\documentclass[12pt,journal,final,letterpaper,onecolumn]{article}
\usepackage[top=1.0in, bottom=1.0in, left=1.in, right=1.in]{geometry}
\usepackage{graphicx} % more modern
\usepackage{subfigure}
\usepackage{algorithm}
\usepackage{algorithmic}
\usepackage{hyperref}
\usepackage{amsmath,  amssymb, amsthm}

\usepackage[authoryear,colon, sort&compress]{natbib}

\usepackage{color}

\usepackage{setspace}
\newcommand{\beq}{\vspace{0mm}\begin{equation}}
\newcommand{\eeq}{\vspace{0mm}\end{equation}}
\newcommand{\beqs}{\vspace{0mm}\begin{eqnarray}}
\newcommand{\eeqs}{\vspace{0mm}\end{eqnarray}}
\newcommand{\barr}{\begin{array}}
\newcommand{\earr}{\end{array}}

\newcommand{\given}{\,|\,}

\newcommand{\zv}{\boldsymbol{z}}

\newcommand{\thetav}{\boldsymbol{\theta}}

\newcommand{\E}{\mathbb{E}}

\allowdisplaybreaks

\newtheorem{thm}{Theorem} %[section]
\newtheorem{cor}[thm]{Corollary}

\usepackage{natbib,natbibspacing}
\renewcommand\footnotemark{}
\begin{document}
\title{\vspace{-.mm} Frequency of Frequencies Distributions and\\\vspace{-.mm} Size Dependent Exchangeable Random Partitions} 
%\title{Split or Merge? Mixed Poisson Processes and\\ Exchangeable Random Partitions}
%\title{Mixed Poisson Processes and\\ the Representation of Cluster Structures} 
\author{ Mingyuan~Zhou$^\dagger$, Stefano Favaro$^*$, 
%\thanks{\emph{Address for correspondence}: Department of Information, Risk, and Operations Management, 2110 Speedway Stop B6500, Austin, TX 78712, USA.
%\emph{Email:} \texttt{mingyuan.zhou@mccombs.utexas.edu}. M. Zhou is also  with the Department of Statistics and Data Sciences, University of Texas at Austin. S. G. Walker is with the Department of Mathematics and Department of Statistics and Data Sciences, University of Texas at Austin.
%% The author's work was supported by the Department of Information, Risk, and Operations Management, McCombs %School of Business and the Division of Statistics and Scientific Computation at the University of Texas at Austin.}
%}
and Stephen~G~Walker$^\dagger$\\
\texttt{\footnotesize{mingyuan.zhou@mccombs.utexas.edu, stefano.favaro@unito.it, s.g.walker@math.utexas.edu}}\vspace{2mm}
\\ 
%Department of Information, Risk, and Operations Management\\
 $ ^\dagger$The University of Texas at Austin, Austin, TX 78712, USA\\ %\vspace{-2mm}
%$^*$Universit\`a di 
$^*$University of Torino  and Collegio Carlo Alberto,  10134 Torino, Italy %, Austin, TX 78712, USA\\
%\texttt{mingyuan.zhou@mccombs.utexas.edu} \\%\\
%\thanks{ %\emph{Address for correspondence}: Department of Information, Risk, and Operations Management, 2110 Speedway Stop B6500, Austin, TX 78712, USA.
%%\emph{Email:} \texttt{mingyuan.zhou@mccombs.utexas.edu}.
%%M. Zhou is an assistant professor  in %the Department of Information, Risk, \& Operations Management at 
%%the McCombs School of Business at the University of Texas at Austin, Austin, TX 78712, USA. % and Department of Statistics and Data Sciences, 
%%S. Favaro is an associate professor in the Departimento di Scienze Economico-Sociali e Matematico-Statistiche at Universit\`a di Torino, 10134 Torino, Italy.
%%S. G. Walker is a professor in the Departments of Mathematics and Statistics \& Data Sciences at
%%the University of Texas at Austin, Austin, TX 78712, USA. 
%\emph{Emails:} \texttt{mingyuan.zhou@mccombs.utexas.edu, stefano.favaro@unito.it, s.g.walker@math.utexas.edu}.
%%% The author's work was supported by the Department of Information, Risk, and Operations Management, McCombs School of Business and the Division of Statistics and Scientific Computation at the University of Texas at Austin.}
%}
}

\maketitle

\vspace{-3mm}

%\begin{spacing}{1.45}
\begin{abstract}
%A  fundamental problem in ecology is to measure species diversity, which is usually conducted via the analysis of a random sample of species, where %a random number of observations are categorized into different species, with the count of each species 
%both the number of distinct species and the size of each species are random. This motivates us to
 
Motivated by the fundamental problem of modeling the frequency of frequencies (FoF) distribution, this paper introduces the concept of a cluster structure to define a probability function that governs the joint distribution of a random count and its exchangeable random partitions. A cluster structure, naturally arising from a completely random measure mixed Poisson process, allows the probability distribution of the random partitions of a subset of a population to be dependent on the population size, a distinct and motivated feature that makes it more flexible than a partition structure. This allows it to model an entire FoF distribution whose structural properties change as the population size varies. A FoF vector can be simulated by drawing an infinite number of Poisson random variables, or by a stick-breaking construction with a finite random number of steps. A generalized negative binomial process model is proposed to generate a cluster structure, where in the prior the number of clusters is finite and Poisson distributed, and the cluster sizes follow a truncated negative binomial distribution. We propose a simple Gibbs sampling algorithm to extrapolate the FoF vector of a population given the FoF vector of a sample taken without replacement from the population. We illustrate our results and demonstrate the advantages of the proposed models through the analysis of real text, genomic, and survey data.  
 \vspace{5mm}
 
\end{abstract}
%\end{spacing}

\newpage
%\input{Introduction}
%\vspace{-6mm}
%\begin{spacing}{1.5}

\vspace{-4mm}\section{Introduction}\vspace{-3mm}

%A fundamental problem in scientific studies is to 
Characterizing a finite population whose individuals are partitioned into different classes is a fundamental research topic  in physical, biological, environmental, and social sciences. One common problem is to estimate certain quantities of a sample  taken from the population. For example, to disseminate survey data to the public, the government statistical agency has the responsibility to  %limit the size and resolution of the microdata, for the purpose of controlling 
assess the risk for  the disclosed microdata records to be matched to specific individuals of the surveyed population, based on the size and resolution of the microdata, while making them informative enough to be useful for education, research, business, and social welfare \citep{bethlehem1990disclosure,fienberg1998confidentiality,skinner2002measure,skinner2008assessing,manrique2012estimating}. 
%For example, to disseminate survey data for public use,  it is common to disclose a microdata consisting of a subset of $m$ records randomly taken from the whole dataset with $n$ individual records. The government statistical agency has the responsibility to limit the size and resolution of the microdata, for the purpose of controlling the risk for  the disclosed microdata records to be matched to specific individuals of the surveyed population, while making them informative enough to be useful for education, research, business, and social welfare \citep{bethlehem1990disclosure,fienberg1998confidentiality,skinner2002measure,manrique2012estimating}.

 %where the sample size $m$ is often chosen to be much smaller than the population size $n$ to control the disclosure  risk \citep{bethlehem1990disclosure,fienberg1998confidentiality,skinner2002measure, % skinner2008assessing,
%manrique2012estimating}. 

%Thus, the disclosure risk of a microdata need to be estimated based on its size (number of records) and number of classes. %If the sample size $i$ is small in comparison to the population size $n$, one may solve 
%The problem of estimating the interested sample quantities could be simply solved by  taking multiple random samples %with/without replacement
% from the observed population. However, the problem remains if one wishes to %build a statistical model to 
%provide parametric forms to measure these sample quantities. 

In practice, one may not observe the population but only a sample taken from it. This  brings another problem often more challenging to solve: to %estimate the structure of the partitions of 
predict how the $n$ individuals of a finite population %(or a sample with a larger size) 
are partitioned into different classes,  on observing the partitions of a sample of $m<n$ individuals randomly taken from this population. % (or the larger sample).
 For example, in high-throughput sequencing, one is often interested in estimating how many more new genomic sequences not found in the current sample would be detected  if  the sequencing depth is increased  \citep{wang2009rna,liu2014rna,sims2014sequencing}. To address this problem, one may define an appropriate procedure to extrapolate the random partitions of the population from the sample. %Given a random sample, one may construct statistical models to fit this particular sample and estimate the model parameters. The problem is these model parameters may be dependent on sample size and hence may not be directly used to describe the population. 
One may also consider constructing a statistical model to fit the random partitions of the observed sample, with the assumption that the same  model parameters inferred from the sample also apply to %the samples of different sizes and 
the population. The size-independent assumption, however, could considerably  limit the flexibility of the selected statistical model. In addition, it could be restrictive to assume that the individuals of a random sample taken %uniformly at random
 without replacement from a finite subpopulation are partitioned in the same way as those of  a random sample  taken %uniformly at random
  without replacement from a larger population 
 to which the subpopulation belongs. 
  %that consists of multiple subpopulations.

%
%%As the full distribution describing the population structure could be %not only challenging to model but also 
%%difficult to quickly interpret, 
%For easy  interpretation, 
%simple consistent statistics that can be measured from a sample, such as species richness and evenness \citep{Fisher1943, simpson1949measurement,hill1973diversity, bunge1993estimating}, are also commonly introduced to characterize the structural property of the sample, and hence that of the population. % if the sample statistics are consistent as the sample size varies. 
%%Consistent sample statistics are appealing, % and could be easy to construct,  
%A common problem, however, is to quantify the uncertainty of these  statistics. 
%

To address all these problems under a coherent statistical framework, we will construct nonparametric Bayesian models to describe both the exchangeable  random partitions of the population and those of a random sample taken without replacement from the population. The distribution of the random partitions of a  sample will be constructed to be dependent on the population %(sample) 
size, which is motivated by our observation that given the model parameters, the structural property of a sample's random partitions  could strongly depend on both the size of the sample and that of the population. 
%the size of the pratio of the sample size $i$ to the population size $n$.

%\subsection{Organization of the paper}
The layout of the paper is as follows:
%The remainder of the paper is organized as follows: 
In Section \ref{sec:pre} we provide some background information. %In Section \ref{sec:FoF} we discuss frequency of frequencies (FoF) distributions, and %further motivate the paper by revealing 
%reveal the limitations of existing methods in modeling them. % that are often designed to model their tails using a straight line in logarithmic space. 
%In Section \ref{sec:species} we discuss existing size independent species sampling models, provide motivations to study size dependent  models to accurately model FoF distributions, and briefly describe % provide all the necessary preliminary notation and a description of 
%normalized random measures. 
In Section  \ref{sec:species}, we discuss frequency of frequencies (FoF) distributions and %\ref{sec:CountMixture}
introduce the new model for constructing  size dependent species sampling models. In Section \ref{sec:gNBP}  we apply the theory in Section \ref{sec:species} % \ref{sec:CountMixture}  
 to the generalized negative binomial process and provide   the asymptotics on both the number and sizes of clusters. We  
 present real data applications in Section \ref{sec:results}.  % end the paper with a brief conclusion and 
We conclude the paper in Section \ref{sec:conclusion} and provide %some additional background information and
  the proofs % of theorems and corollaries
   in Appendix E.

\vspace{-3.5mm}\subsection{Notation and preliminaries}\label{sec:pre}\vspace{-1.5mm}

\textbf{Frequency of frequencies.}
Consider a finite population with $n$ individuals from $K$ different classes, and let $z_i\in\{1,\ldots,K\}$ denote the class individual $i$ is assigned to, let $n_k=\sum_{i=1}^n \delta(z_i=k)$ denote the number of individuals in class $k$, and let $m_i=\sum_{k=1}^K \delta(n_k=i)$ denote the number of classes having $i$ individuals in this finite population, where $\delta(x)=1$  if the condition $x$ is satisfied and $\delta(x)=0$ otherwise. Thus, by definition, we have $$K = \sum_{i=1}^\infty m_i, ~~~n=\sum_{i=1}^\infty i m_i  $$ almost surely (a.s.), and since $m_i=0$ a.s. for all $i\ge n+1$, it is also common to use $\sum_{i=1}^n$ to replace the infinite sum  $\sum_{i=1}^\infty$ in the above equation. %  express the above equations as $K = \sum_{i=1}^n m_i,n=\sum_{i=1}^n i m_i$.  
 For example, we may represent $(z_1,\ldots,z_{14})=(1,2,3,4,5,5,6,6,6,6,7,7,7,7)$ as $(n_1,\ldots,n_7)=(1,1,1,1,2,4,4)$, or $\{m_1,m_2,m_4\}=\{4, 1, 2\}$ and $m_i=0$ for $i\notin \{1,2,4\}$. Since $m_i$ represents the frequency of the classes appearing $i$ times, we refer the count vector  $\mathcal{M}= \{m_i\}_i$ %  (m_1,\ldots,m_n)$ 
 as the frequency of frequencies (FoF) vector, the distribution of which is commonly referred to as the FoF distribution \citep{good1953population}. \vspace{2mm}

%\vspace{-4mm}\section{Bayesian Species Sampling Models}\label{sec:species}\vspace{-1mm}
\noindent\textbf{Exchangeable partition probability functions.}
Assuming the population size $n$ is given,  one may define a probability distribution %use a species sampling model %we will construct a Bayesian species sampling model 
to partition the $n$ individuals into exchangeable random partitions, and hence generate a FoF vector by defining each partition  as a class.
 Let $[m]:=\{1,\ldots,m\}$ denote a subset of the set $[n]:=\{1,\ldots,n\}$, where $m\le n$. For a random partition $\Pi_m=\{A_1,\ldots,A_l\}$ of the set $[m]$, 
where there are $l$ clusters and each individual $i\in[m]$ belongs to one and only one set $A_k$ from $\Pi_m$, we denote $P(\Pi_m\,|\,n)$ as the marginal partition probability for $[m]$ when it is known the population size is~$n$. Note that $P(\Pi_m\given n)=P(z_1,\ldots,z_m\given n)$ %, where $z_{1:m} = \{z_1,\ldots,z_m\}$, 
if %$z_{i}\in\{1,\ldots,l\}$ represents that
 individual $i$ belongs to $A_{z_i}$. %
%We are interested in 
%modeling the distribution of the random partition of %the set $[n]:=\{1,\ldots,n\}$ given the random partition of 
%a subset $[m]:=\{1,\ldots,m\}$ of the set $[n]:=\{1,\ldots,n\}$, where $m\le n$. More specifically, for a random partition $\Pi_m=\{A_1,\ldots,A_l\}$ of the set $[m]$, 
%where there are $l$ clusters and each element $i\in[m]$ belongs to one and only one set $A_k$ from $\Pi_m$, we will first construct the marginal partition probability for $[m]$ when it is known the population size is~$n$, denoted as $P(\Pi_m\,|\,n)$. %As further developed in \citep{pitman1995exchangeable,csp}, if %the probability for an arbitrary partition $\Pi_m$ 
%

If $P(\Pi_m\,|\,n)$ depends only on the number and sizes of the $(A_k)$, regardless of their order, and the population size $n$, then it is referred to in this paper as a size dependent exchangeable partition probability function (EPPF) of~$\Pi_m$. % in a finite population of size $n$. 
If $P(\Pi_m\,|\,m) = P(\Pi_m\,|\,n)$ for all $n\ge m$, then it is referred to as a size independent EPPF. Typical examples of size independent EPPFs include the Ewens sampling formula \citep{ewens1972sampling,Antoniak74}, Pitman-Yor process 
 \citep{perman1992size,pitman1997two}, and those governed by normalized random measures with independent increments (NRMIs)
\citep{regazzini2003distributional,BeyondDP}. %See \citet{muller2004nonparametric}, \citet{BeyondDP} and \citet{Muller2013} for reviews. 
We provide a review on size independent EPPFs in Appendix C. See \citet{csp} for a detailed treatment of EPPFs.\vspace{2mm}% can be found in . \\ 

\noindent\textbf{Completely random measures.} Let us denote $G$ as a completely random measure \citep{Kingman,PoissonP} defined on the product space $\mathbb{R}_+\times \Omega$, where $\mathbb{R}_+=\{x:x>0\}$ and $\Omega$ is a complete separable metric space. It assigns independent infinitely divisible random variables $G(A_j)$  to disjoint Borel sets $A_j\subset \Omega$, with Laplace transforms 
\beq\label{eq:Laplace0}
%L_{a,c,G_0(A)}(s)
\E\left[e^{-\phi\,G(A)}\right] = \exp\bigg\{- \int_{\mathbb{R}_+\times A} (1-e^{-\phi r})\nu(drd\omega)\bigg\},
\eeq
where $\nu(drd\omega)$ is the L\'evy measure. A random draw from $G$ can be expressed as 
$$
G=\sum_{k=1}^Kr_k\delta_{\omega_k}, ~K\sim\mbox{Poisson}(\nu^+),~(r_k,\omega_k)\stackrel{iid}\sim \pi(drd\omega),
$$
where $r_k$ is the weight of atom $\omega_k$, $\nu^+ = \nu(\mathbb{R}_+\times \Omega)$, and $ \nu(drd\omega) = \nu^+\pi(drd\omega)$. % \citep{Wolp:Clyd:Tu:2011}. 
The completely random measure $G$ is well defined if  $\int_{\mathbb{R}_+\times \Omega}\min\{1,r\} \nu(drd\omega)<\infty$, even if the Poisson intensity $\nu^+ $ is infinite.
 In this paper, we consider homogenous completely random measures where the L\'evy measure can be written as $ \nu(drd\omega) =\rho(dr)G_0(d\omega)$, where $G_0$ is a finite and continuous base measure over~$\Omega$.

The generalized gamma process  $G\sim\mbox{g}\Gamma\mbox{P}(G_0,a,1/c)$ of \citet{brix1999generalized}, %is a homogenous completely random measure defined on the product space $\mathbb{R}_+\times \Omega$, 
where $a< 1$ is a discount parameter and $1/c$ is a scale parameter, %, and $G_0$ is a finite and continuous base measure over $\Omega$.
 %The L\'{e}vy measure of the generalized gamma process
%$G\sim{{}}\mbox{g}\Gamma\mbox{P}(a,1/c,G_0)$
%can be expressed as 
is defined with the L\'{e}vy measure as %, where
\beqs\label{eq:LevyGGP}
\nu(drd\omega) =\rho(dr)G_0(d\omega) = \frac{1}{\Gamma(1-a)}r^{-a-1}e^{-cr}\,dr\, G_0(d\omega).  %\,G_0(d\omega).
\eeqs
A detailed description on the generalized gamma process is provided in Appendix D.

\vspace{-3mm}\section{Bayesian modeling of frequency of frequencies}\label{sec:species}

%A fundamental problem in scientific studies is to characterize a population whose individuals are partitioned into different classes. 
%\subsection{Modeling the tail of a frequency of frequencies distribution}\vspace{-1mm}

%Note that the population size $n$ could also be interpreted as the sample size, if the population could be considered as a sample taken without replacement from a larger-size population. Hence, at the outset, we will not make a distinction between these two  interpretations, and whether $n$ is referred to as the population or sample size will  be clear from the context. 
\vspace{-2mm}\subsection{Frequency of frequencies distributions}\label{sec:FoF}\vspace{-2mm}

\begin{figure}[!tb]
 
 \begin{center}
\includegraphics[width=0.76\columnwidth]{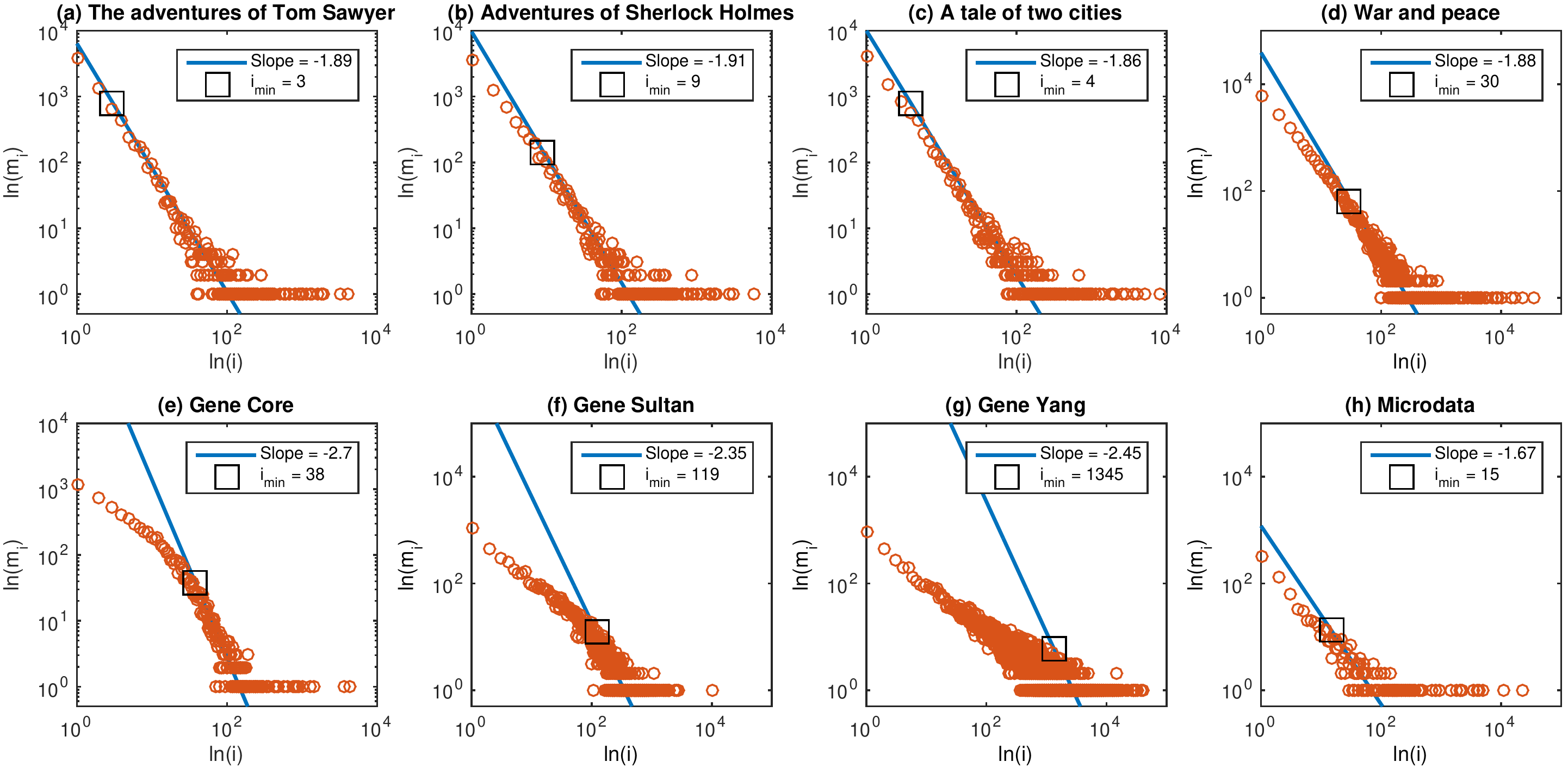}
\end{center}
\vspace{-7.5mm}
\caption{ \label{fig:FoF} \small The log-log plots of the frequency of frequencies (FoF) vectors for (a) the words in ``The Adventures of Tom Sawyer'' by Mark Twain,  (b) the words in ``The Adventures of Sherlock Holmes'' by Arthur Conan Doyle, (c) the words in ``A Tale of Two Cities'' by Charles Dickens, (d) the words in ``War and Peace'' by Leo Tolstoy and translated by Louise and Aylmer Maude,  (e) the RNA sequences studied in \citet{Core19122008}, (f) the RNA sequences studied in \citet{Sultan15082008}, (g) the RNA sequences studied in \citet{yang2010global}, and (h) the microdata provided in Table A.6 of \citet{greenberg1990geographic}. 
For each subfigure,  
a least squares line with the slope fixed as $-\alpha$ is fitted to  $\{[\ln i, \ln(m_i)]\}_{i:i\ge i_{\min},m_i\ge 3}$, where $i_{\min}$ is a %power-law 
lower cutoff point  and $\alpha$ is a scaling parameter estimated using the software provided for \citet{clauset2009power}.
%with the intercept  of the fitted line set as $\left[\sum_{i \in I}\ln(m_i) +\alpha \sum_{i\in I}\ln i \right]/|I|
%$, where $I=\{i:i\ge i_{min}, m_i\ge 3 \}$.
% of the least squares regression line fitting the points $\{[\ln(i),\ln(m_i)]\}_{i:m_i\ge 10}$ of the FoF vector is displayed in the legend. 
%For each subfigure,  the slope of the least squares regression line fitting the points $\{[\ln(i),\ln(m_i)]\}_{i:m_i\ge 10}$ of the FoF vector is displayed in the legend. 
}\vspace{-2mm}
\end{figure}

The need to model the distributions of the class sizes $\{n_k\}_k$, or the FoF vector, arises in a wide variety of settings. For example, in computational linguistics and natural language processing, if we let $n_k$ denote the frequency of the $k$th most frequent word in a text corpus, then $\ln(n_k)$ and $\ln(k)$ would be approximately linearly related according to Zipf's law \citep{zipf1949human}. Alternatively, if we let $m_i$ denote the frequency of the words appearing $i$ times, then $\ln(m_i)$ often appears to follow a straight line as a function of $\ln(i)$, as shown in Figures \ref{fig:FoF}(a)-(d) for the words of four different novels.
%The word FoF distributions commonly exhibit a power-law behavior. 
For many other  natural and artificial phenomena, 
the FoF distributions also exhibit %power-law
similar behavior in their tails, such as those on the number of citations of scientific papers, the degrees of proteins in a protein-interaction network, and the peak gamma-ray intensity of solar flares, to name a few; see   \citet{newman2005power} and \cite{clauset2009power} for reviews. In addition, we  find that the tails of the FoF distributions for the genomic sequences in high-throughput sequencing data and the classes of the microdata also often exhibit %power-law 
similar behaviors. For example, in Figure~\ref{fig:FoF}  are the FoF vectors for the words of four different novels\footnote{\href{https://www.gutenberg.org/ebooks/}{https://www.gutenberg.org/ebooks/}}, the RNA sequences of three different RNA-seq samples\footnote{\href{http://bowtie-bio.sourceforge.net/recount/}{http://bowtie-bio.sourceforge.net/recount/}} provided by \citet{frazee2011recount}, and the classes  of a microdata consists of 87,959 household records, shown in Table A.6 of \citet{greenberg1990geographic}. 

%
%; this phenomenon is commonly known as the power-law behavior, which can often be characterized by %the scaling parameter $\alpha$ that is the absolute value of 
%the slope of the fitted least squares regression line \citep{newman2005power}.   See Appendix B on how to estimate a power-law lower cutoff point and a power-law scaling parameter to characterize the power-law behavior.

 \begin{figure}[!tb]
 
 \begin{center}
\includegraphics[width=0.93\columnwidth]{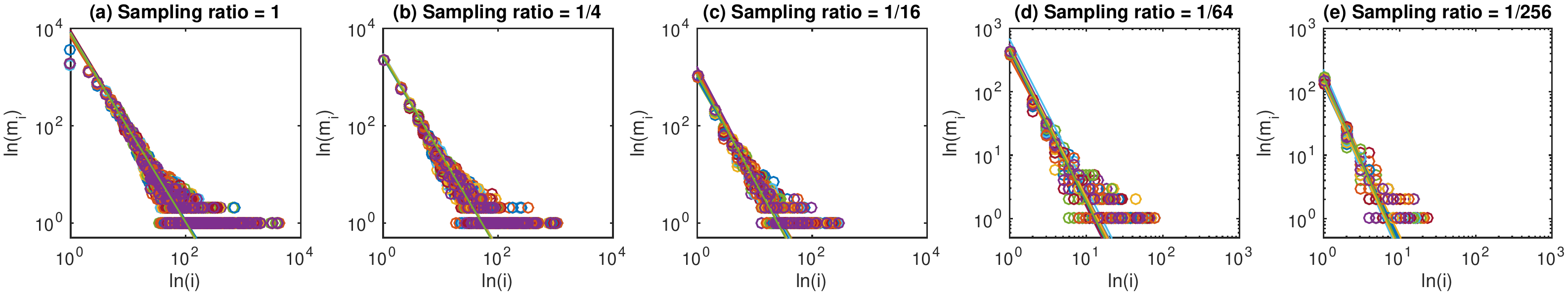}
\end{center}
\vspace{-7.5mm}
\caption{ \label{fig:tomsawyer} \small The log-log plots of the frequency of frequencies (FoF) vectors for the words in the novel ``The Adventure of Tom Sawyer'' by Mark Twain. Each subfigure consists of 20 FoF vectors displayed in different colors.  (a) The 20 FoF vectors, with one curve coming from all the words and each of the other 19 curves coming from a sample of words taken with replacement from the novel,  % each of which comes from a sample of  words taken with replacement from the novel, 
with a sampling ratio of 1; (b)-(e) The 20 FoF vectors, each of which comes from a sample of  words taken without replacement from the novel, with the sampling ratios of 1/4, 1/16, 1/64, and 1/256, respectively. For each FoF vector, a straight line fitting the points $\{[\ln(i),\ln(m_i)]\}_{i:i\ge i_{\min},m_i\ge 3}$ with slope $-\alpha$,
%$\{[\ln(i),\ln(m_i)]\}_{i:m_i\ge 10}$ 
is also plotted, where both the lower %power-law 
cutoff point $i_{\min}$ and scaling parameter $\alpha$ are estimated using the software provided for \citet{clauset2009power}.
}

\begin{center}
\includegraphics[width=0.58\columnwidth]{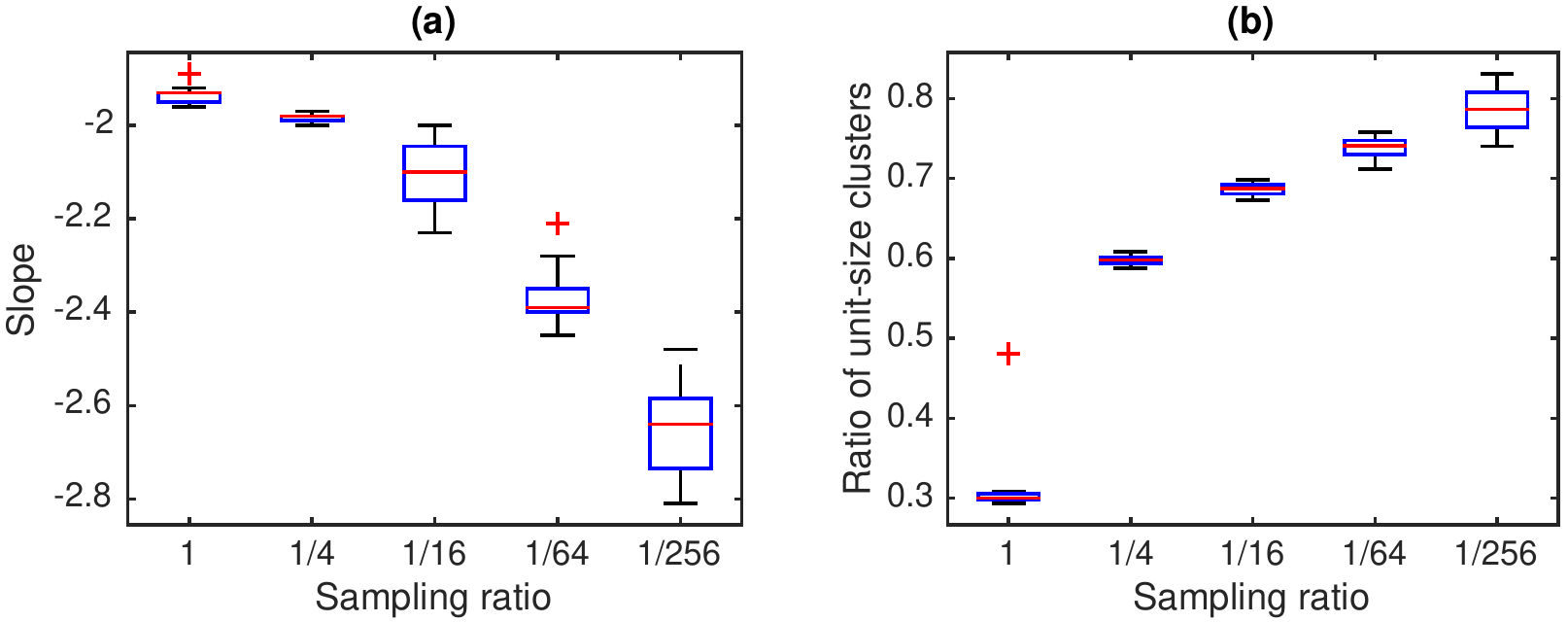}
\end{center}
\vspace{-7.5mm}
\caption{ \label{fig:tomsawyer_alpha} \small Box plots of (a) the slopes of the fitted lines
%stimated power-law scaling parameters 
and (b) the ratios of the clusters of size one %, and (c) Simpson's indices of diversity 
for the FoF vectors in the log-log plots shown in Figure \ref{fig:tomsawyer}. For each sampling ratio, the box plot in each subfigure is based on the corresponding 20 FoF vectors used in Figure \ref{fig:tomsawyer}. 
%For (a), each slope is the slope of the least squares regression line that fits the points $\{[\ln(i),\ln(m_i)]\}_{i:m_i\ge 10}$  of a FoF vector.  (b) Box plots of Simpson's indices of diversity for various sampling ratios.  (c) Box plots of Simpson's indices of diversity for various sampling ratios. For each sampling ratio, the box plot is based on 20 indices, each of which is calculated based on a FoF vector used in Figure \ref{fig:tomsawyer}.
\vspace{-2mm}
 }
\end{figure}

%To elaborate these points, %on the necessity to estimate $\alpha$, 
To illustrate how the characteristics of the FoF vector of a sample are related to the size of the sample, 
we show  in Figure \ref{fig:tomsawyer}(a) the FoF distribution for all the words in the novel ``The Adventures of Tom Sawyer'' by Mark Twain on the logarithmic scale, and also plot the FoF distributions for % the words obtained by sampling 
$1/4$, $1/16$, $1/64$, and $1/256$ of the  words taken without replacement from the novel, in Figures \ref{fig:tomsawyer}(b)-(e), respectively. We further show in Figure \ref{fig:tomsawyer_alpha}(a) the box plots of the slopes of the least squares regression lines fitted to the tails of these FoF vectors, and show in Figure \ref{fig:tomsawyer_alpha}(b) the box plots of  the ratios of unit-size clusters (clusters of size one).  In addition, we provide Figures  \ref{fig:sultan}-\ref{fig:sultan_alpha} in Appendix A 
  as the analogous plots to Figures  \ref{fig:tomsawyer}-\ref{fig:tomsawyer_alpha} for the FoF vectors for a high-throughput sequencing sample for the human transcriptome from a B cell line, as studied in \citet{Sultan15082008}. Note that to estimate the lower cutoff point and slope of the regression line, we use the software provided for  \citet{clauset2009power}, as described in detail in Appendix B. % See Appendix B for more details. %on how to estimate a power-law lower cutoff point and a power-law scaling parameter to characterize the power-law behavior.

%; this phenomenon is commonly known as the power-law behavior, which can often be characterized by %the scaling parameter $\alpha$ that is the absolute value of 
%the slope of the fitted least squares regression line \citep{newman2005power}.   See Appendix B on how to estimate a power-law lower cutoff point and a power-law scaling parameter to characterize the power-law behavior.

It is clear from Figures \ref{fig:tomsawyer}-\ref{fig:tomsawyer_alpha} and \ref{fig:sultan}-\ref{fig:sultan_alpha}  that the slope of the fitted straight line %(negative of the estimated power-law scaling parameter)   %fitted to the points
 and the ratio of unit-size clusters tend to decrease and increase, respectively, as the subsampling ratio decreases. 
 Therefore, %as shown in Figures \ref{fig:tomsawyer}-\ref{fig:sultan_alpha},
 for a sample taken without replacement from a population, its estimated scaling parameter often clearly depends on the sample size. 
   %given the scaling parameter $\alpha$ estimated from a random sample taken without replacement from a finite population, it would be unclear on how to estimate $\alpha$ and hence also the FoF distribution of the finite population. As will be shown in Section \ref{sec:results},  ignoring the size dependence of $\alpha$ %on the sample size 
 % may lead to a poor fitting, let alone prediction, of the population FoF vector.   
 Moreover, it seems that a FoF distribution in some case could be more accurately described with a decreasing concave curve %that is concave down
  than  with a straight line, such as those for the RNA  sequences shown in Figures \ref{fig:FoF}(e)-(g) and Figure  \ref{fig:sultan} in Appendix A.
 % also commonly exhibit power-law behaviors. 
All these empirical observations  motivate us to model the FoF distribution with % more sophisticated 
%more flexible
%statistical models 
a statistical model that could model  %n%ot just the tail, but  
the entire FoF distribution of a finite population, and more importantly, could 
take both the population and sample sizes into consideration, 
providing a principled way to extrapolate the FoF vector of a finite population given a random sample taken without replacement from the population.

\vspace{-3mm}\subsection{Structure of the model}\label{sec:CountMixture}%\vspace{-3mm}

%Moreover, 
As discussed in Section \ref{sec:FoF} and shown in Figures \ref{fig:tomsawyer}-\ref{fig:tomsawyer_alpha}  and \ref{fig:sultan}-\ref{fig:sultan_alpha} in Appendix A, the structural property of a FoF distribution can  strongly depend on $n$. Hence to use the same set of model parameters $\thetav$ to describe the FoF distributions for various  sample sizes, we intend to construct a model that describe the distribution  $P(\Pi_m\,|\,n,\thetav)$, %where $\thetav$ consists of all the model parameters, 
meaning that the EPPF and hence the FoF distribution for a sample of size $m$, taken without replacement from a population of size $n$, depends  not only on the model parameters $\thetav$, but also on the population size $n$. %Note we often omit $\thetav$ for brevity  when writing the conditional distributions. 
To develop this theme, and to allow the mathematics to proceed in a neat way, and without forcing any restrictions, we first make $n$ a random object within the model.

Here we describe how the random allocations of individuals to classes are distributed based on the independent random jumps of a completely random measure. %the generalized gamma process. 
With a random draw from a completely random measure expressed as $G = \sum_{k=1}^{K} r_k \delta_{\omega_k}$, by introducing a categorical latent  variable $z$ with
$
P(z=k\,|\,G) =r_k/G(\Omega), \notag %~z=1,\cdots,K,
$ 
when a population of size $n$ is observed  %$\xv=(x_1,\cdots,x_m)$
we have
%\begin{align}
\beqs\label{eq:f_G_N}
p(\zv\,|\,G,n)= \prod_{i=1}^n \frac{r_{z_i}}{\sum_{k=1}^{K} r_k}
=  \left(\sum_{k=1}^{K} r_k\right)^{-n}%\prod_{k=1}^{K} n_k!}
\prod_{k=1}^{K} {r_k^{n_k}} , 
\eeqs
where  $\zv=(z_1,\ldots,z_n)$ is a sequence of categorical random variables  indicating the class memberships, $n_k = \sum_{i=1}^n  \delta(z_{i}=k)$ is the number of data points assigned to category $k$, and $n=\sum_{k=1}^{K} n_k$.  A random partition $\Pi_n$ of $[n]$ is defined by the ties between the $(z_i)$. So at this point, (\ref{eq:f_G_N}) is standard.
Now (\ref{eq:f_G_N}) exhibits a lack of identifiabilty in that the scale of the $(r_k)$ is arbitrary; the model is the same if we set $\widetilde{r}_k=\kappa\,r_k$ for any $\kappa>0$. Hence, the total mass 
$\sum_{k=1}^K r_k$ is unidentified. 
Additionally, %for reasons outlined in Section \ref{sec:size-dependent}, we want, having marginalized out $G$, for $n$ to remain, and for us to have $p(\zv\,|\,n)$ to remain. 
for the standard models, when $G$ is integrated out, $n$ disappears and we have $p(\zv)$ depending solely on the model parameters $\thetav$.

%As has been previously mentioned, 
We solve both these issues by %allowing $n$ to depend on $G$ via we 
 linking the population size $n$ to the total random mass of $G$ with a Poisson distribution, allowing $n$ to depend on $G$ via 
\beq
p(n\,|\,G)=\mbox{Poisson}\big[ G(\Omega)\big].\label{Poisson}
\eeq 
Since the $n$ data points are clustered according to the normalized random probability measure $G/G(\Omega)$, we have the equivalent sampling mechanism given by
\beq
p(n_k\,|\,G)=\mbox{Poisson}( r_k)\quad\mbox{independently for}\quad k=1,2,\ldots\,, \notag
\eeq 
and, since $n=\sum_k n_k$, we obviously recover (\ref{Poisson}). We note here then that the prior model is for $p(n,G)$ and, consequently, $p(G\,|\,n)$ means $G$ depends on $n$; $i.e.$, for each $n$ we will have a different random measure for $G$.

Therefore, we link directly the cluster sizes $(n_k)$ to the weights $(r_k)$ with  independent Poisson distributions, which is in itself an appealing intuitive feature. %\citep{BNBP_PFA_AISTATS2012}. 
The mechanism to generate a sample of arbitrary size is now well defined and $G$ is no longer scaled freely.  The new construction also allows $G(\Omega)=0$, for which $n= 0$ a.s.  Allowing $G(\Omega)=0$ with a nonzero probability relaxes the requirement of $\nu^+=\infty$ ($i.e.$, $K=\infty$ a.s.), a necessary condition to normalize a completely random measure \citep{regazzini2003distributional,BeyondDP}. For us we will not necessarily be assuming that $K=\infty$ a.s. In fact our model is such that $K=0\iff n=0$, which is coherent, and, moreover, $P(K=0\,|\,n>0)=0$.  

%We solve both these issues by allowing $n$ to depend on $G$ via 
%$$p(n\,|\,G)=\mbox{Poisson}[G(\Omega)],$$
%from which we have independently
%$$p(n_k\,|\,G)=\mbox{Poisson}(r_k).$$
%We note here then that the prior model is for $p(n,G)$ and, consequently, $p(G\,|\,n)$ means $G$ depends on $n$; $i.e.$, for each $n$ we will have a different random measure for $G$.

%Linking $n$ to $G(\Omega)$ with a Poisson distribution %not only introduces dependence between the normalized random probability measure and the sample size, %be dependent on the sample size, 
%but also 
%makes the scale of $G$ become identifiable.
 With $G$ marginalized out from the $G$ mixed Poisson process, the joint distribution of  $n$ and its exchangeable random partition $\Pi_n$ is called an exchangeable cluster probability function (ECPF), which further leads to a FoF distribution that is shown to be an infinite product of Poisson distributions. %, %, which is in a fully factorized form 
On observing a population of size $n$, we are interested in the EPPF $P(\Pi_n\,|\,n,\thetav)$ and, marginalizing over $n-m$ elements, we would consider $P(\Pi_m\,|\,n,\thetav)$. Note that distinct from a partition structure of \citet{kingman1978random,kingman1978representation} that requires $P(\Pi_m\,|\,n,\thetav)=P(\Pi_m\,|\,m,\thetav)$ for all $n>m$, we no longer have or require this condition % in this paper. %$P(\Pi_m\,|\,n,\thetav)=P(\Pi_m\,|\,m,\thetav)$ for $n>m$ 
for exchangeable  random partitions % ECPF and  EPPF
 generated under a $G$ mixed Poisson process, which will be  referred to as a cluster structure. 
%in  a cluster structure. 
%With  the completely random measure $G$ marginalized out, 

We provide in Section \ref{sec:prop} %Section \ref{sec:CountMixture}
 the general form for both $p(\zv,n) = P(\Pi_n,n\given \thetav)$ and $p(\zv\,|\,n)=P(\Pi_n\,|\,n,\thetav)$, and make connections to previous work in Section \ref{sec:2.4} by letting $G$ be drawn from the gamma process. % \citep{ferguson73}. 
 We provide  in Section \ref{sec:gNBP} the specific case when $G$ is drawn from the generalized gamma process $G\sim\mbox{g}\Gamma\mbox{P}(G_0,a,1/c)$ and the asymptotics on the number and sizes of clusters as $n\rightarrow \infty$.  
In Section \ref{sec:results} we use MCMC methods to extrapolate the FoF vector of the population from a random sample taken without replacement from it. 

% and estimate the posterior values of Simpson's index of diversity, using real text and genomic data. % data and next generation sequencing data. %  sequence frequency count data. 

%
%We work at a fundamental level with a normalized completely  random measure. Hence, the total (random) mass  is unidentified and consequently arbitrary. We take this opportunity to use it to model the, prior to observation, random population size $n$.  More specifically, we model  $n$ as a Poisson random variable, %from a Poisson distribution, 
%the  mean of which is parameterized by the total random mass of a completely random measure $G$ over a complete and separable metric space $\Omega$. The total random mass $G(\Omega)$ 
%is used to normalize $G$ to obtain a random probability measure $G(\cdot)/G(\Omega)$.
%Linking $n$ to $G(\Omega)$ with a Poisson distribution %not only introduces dependence between the normalized random probability measure and the sample size, %be dependent on the sample size, 
%%but also 
%makes the scale of $G$ become identifiable.

\vspace{-3mm}
\subsection{Properties of the model}\label{sec:prop}
\begin{figure}[!tb]
\begin{center}
\includegraphics[width=0.7\columnwidth]{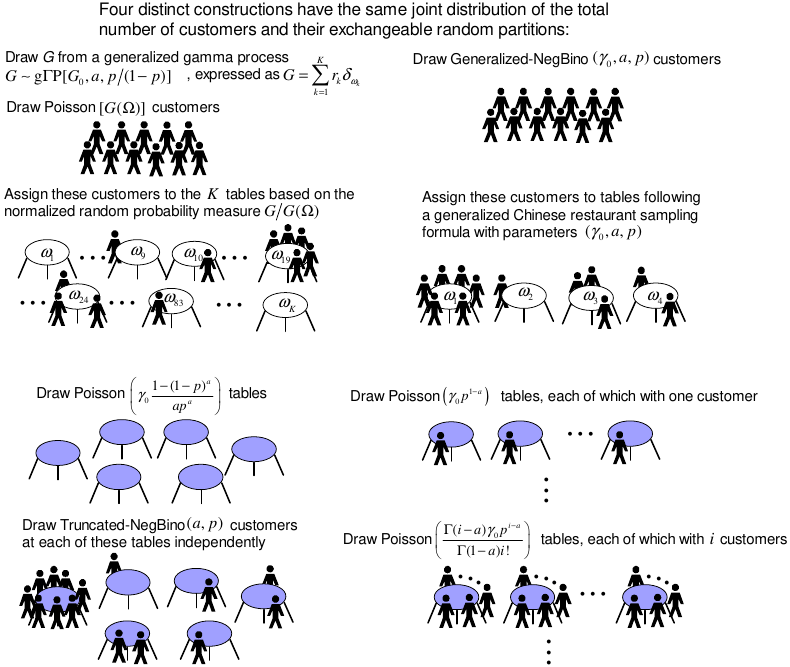}
\end{center}
\vspace{-6.5mm}
\caption{ \label{fig:gNBPdraw} \small The cluster structure of the generalized negative binomial process %count-mixture model
 can be either constructed by assigning $\mbox{Poisson}[G(\Omega)]$ number of customers to tables following a normalized generalized gamma process $G/G(\Omega)$, where $G\sim\mbox{g}\Gamma\mbox{P}[G_0,a,p/(1-p)]$, or constructed by assigning $n\sim\mbox{gNB}(\gamma_0,a,p)$ number of customers to tables following a generalized Chinese restaurant sampling formula $\zv\sim$~$\mbox{gCRSF}(n,\gamma_0,a,p)$, where $\gamma_0=G_0(\Omega)$. A equivalent cluster structure can be generated by first drawing $\mbox{Poisson}\big(\gamma_0\frac{1-(1-p)^a}{ap^a}\big)$ number of tables, and then drawing  $\mbox{TNB}(a,p)$ number of customers independently at each table. Another equivalent one can be generated by drawing $\mbox{Poisson}\big(\frac{\Gamma(i-a)\gamma_0 p^{i-a}}{\Gamma(1-a)i!}\big)$ number of tables, each of which with $i$ customers, for $i\in\{1,2,\ldots\}$.
}\vspace{-2mm}
\end{figure}

\vspace{-2mm}
A key insight of this paper is that %the clustering structure, introduced by 
a completely random measure mixed Poisson process produces a cluster structure that is identical in distribution to ($i$) the one produced  %, is equivalent in distribution to the one  introduced 
by assigning the total random count of the Poisson process into exchangeable random partitions, using  the random probability measure normalized from that completely random measure, ($ii$) the one produced by assigning the total (marginal) random count $n$ of the mixed Poisson process into exchangeable random partitions using an EPPF of $\Pi_n$, and ($iii$) the one produced by constructing a FoF vector, %the $i$th element of which representing the number of distinct clusters of size $i$  is 
the $i$th element of which is generated from a Poisson distribution parameterized by a specific function of $i$. %generating a Poisson random number of clusters for every possible cluster sizes. 
 For example, when the generalized gamma process $G\sim\mbox{g}\Gamma\mbox{P}[G_0,a,p/(1-p)]$ is used as the completely random measure in this setting, our key discoveries are summarized in 
Figure \ref{fig:gNBPdraw}, %. These points will be made clearer in the following discussions.  %
which will be discussed further in Section \ref{sec:gNBP}.

%
%We note that  \citet{BNBP_PFA_AISTATS2012} and \citet{NBP2012} have explored related ideas to mix a gamma or beta process with a negative binomial process, and use that hierarchical process for mixture modeling of grouped data. Yet the authors marginalized neither  the beta nor gamma process %due to technical difficulties 
%and relied on finite truncation for inference. 
%The techniques developed here to model a random count vector serve as the foundation for  \citet{NBP_CountMatrix} to construct a family of nonparametric Bayesian priors for infinite random count matrices, and for \citet{BNBP_EPPF} to define a prior distribution that describes the random partition of a count vector into a latent random count matrix. 

%We will discuss at the end of the paper that the ideas and techniques developed in this paper serve as the foundation for \citet{NBP_CountMatrix} to develop priors for random count matrices and for \citet{BNBP_EPPF} to develop a size dependent EPPF for the beta-negative binomial process.

In Theorem \ref{thm:compoundPoisson}, we establish  the marginal model for the $(n_k)$ with $G$ marginalized out. We provide the L\'evy measure, ECPF, EPPF, FoF distribution, stick-breaking construction, and prediction rule in Corollaries \ref{cor:compoundPoisson}-\ref{thm:predict}. The proofs are provided in  Appendix E. 

\begin{thm}[Compound Poisson Process]\label{thm:compoundPoisson}
 It is that the $G$ mixed Poisson process is also a compound Poisson process; a random draw of which can be expressed as 
 %\begin{align}
$$X(\cdot)=\sum_{k=1}^{l} n_k \,\delta_{\omega_k}(\cdot)\quad\mbox{with  }~l\sim\emph{\mbox{Poisson}}\left[G_0(\Omega)\int_{0}^\infty(1-e^{-r})\rho(dr)\right],$$
and independently
$$P(n_k=j)=\frac{ {\int_{0}^\infty r^j e^{-r} \rho(dr)} }{{j!}\int_{0}^\infty(1-e^{-r})\rho(dr)}~~\mbox{for}~~j=1,2,\ldots\notag$$
 %(n_k,\omega_k) \sim \tilde{\nu}(dnd\omega)/\tilde{\nu}^+, 
 %\end{align} 
 where $\int_{0}^\infty(1-e^{-r})\rho(dr) %\le \int_{0}^\infty\min\{1,s\}\rho(ds) 
<\infty$ is a condition required for the characteristic functions of $G$ to be well defined, $\omega_k\stackrel{iid}{\sim} g_0$, and $g_0(d\omega)=G_0(d\omega)/G_0(\Omega)$. % is the base distribution. 
 \end{thm}
 
 \begin{cor} % [L\'evy measure]
 \label{cor:compoundPoisson}
 The L\'evy measure of the $G$ mixed Poisson process can be expressed as 
 $$
 \nu(dnd\omega) = \sum_{j=1}^\infty\int_{0}^\infty \frac{ r^j e^{-r}  }{j!} \rho(dr)~\delta_j(dn)G_0(d\omega).
 $$
% %\begin{align}
%$$X(\cdot)=\sum_{k=1}^{l} n_k \,\delta_{\omega_k}(\cdot)\quad\mbox{with }l\sim\emph{\mbox{Poisson}}\left[G_0(\Omega)\int_{0}^\infty(1-e^{-r})\rho(dr)\right],$$
%and independently
%$$P(n_k=j)=\frac{ {\int_{0}^\infty r^j e^{-r} \rho(dr)} }{{j!}\int_{0}^\infty(1-e^{-r})\rho(dr)}~~\mbox{for}~~j=1,2,\ldots\notag$$
% %(n_k,\omega_k) \sim \tilde{\nu}(dnd\omega)/\tilde{\nu}^+, 
% %\end{align} 
% where $\int_{0}^\infty(1-e^{-r})\rho(dr) %\le \int_{0}^\infty\min\{1,s\}\rho(ds) 
%<\infty$ is a condition required for the characteristic functions of $G$ to be well defined, $\omega_k\stackrel{iid}{\sim} g_0$, and $g_0(d\omega)=G_0(d\omega)/G_0(\Omega)$. % is the base distribution. 
 \end{cor}

The compound Poisson representation dictates %in the prior 
the model to have a Poisson distributed finite number of  clusters, whose sizes follow a positive discrete distribution. % $(\pi_j)$.
 The mass parameter $\gamma_0=G_0(\Omega)$ has a linear relationship with the expected number of clusters, % and hence the expected number of samples, 
 but has no direct impact on the cluster-size distribution in the prior.  % , including both the cluster-number and cluster-size distributions, specified by the count-mixture model.
 Note that a draw from $G$ contains $K< \infty$ or $K=\infty$ atoms a.s., but only $l$ of them would be associated with nonzero counts if $G$ is mixed with a Poisson process. 
Since the cluster indices are unordered and exchangeable, 
without loss of generality, in the following discussion,  we relabel %the indexes of the 
the atoms with nonzero counts % in $\mathcal{D}_n=\{\omega_k\}_{k:n_k>0}$ 
in order of appearance from $1$ to $l$  %$l:=\,|\,\mathcal{D}_n|=L(\Omega)$ 
and then $z_i\in\{1,\ldots,l\}$ for $i=1,\ldots,n$,  with $n_k>0$ if and only if $1\le k \le l$ and $n_k=0$ if $k>l$. %Note that $l$ is a simplified notation of $\,|\,\mathcal{D}_m|$, which depends on %is a function of 
%the sample size $m$.

\begin{cor}[Exchangeable Cluster/Partition Probability Functions]\label{thm:ECPF}
The model 
has a fully factorized  exchangeable cluster probability function (ECPF) as
$$p(\zv,n\,|\,\gamma_0,\rho) = %\E_{G} [f(\zv,n\,|\,G)] =
 \frac{\gamma_0^l}{n!} \exp\left\{\gamma_0\int_{0}^\infty(e^{-r}-1)\rho(dr)\right\}
\prod_{k=1}^l \int_0^\infty r^{n_k} e^{-r} \rho(dr),$$
the marginal distribution for the population size $n=X(\Omega)$ has probability generating function
$$\E[t^{n}\,|\,\gamma_0,\rho] = \exp\left\{ \gamma_0 \int_{0}^\infty (e^{-(1-t)r}-1)\rho(dr)\right\}$$ 
 and probability mass function 
$\left. p_N(n\,|\,\gamma_0,\rho)=\frac{d^n (\E[t^n\,|\,\gamma_0,\rho])} { n! d t^n } \right |_{t=0},$
and an exchangeable partition probability function (EPPF) of $\Pi_n$ as
$$p(\zv\,|\,n,\gamma_0,\rho) = {p(\zv,n\,|\,\gamma_0,\rho) }\big/{p_N(n\,|\,\gamma_0,\rho)}.\notag$$ 
\end{cor}

The proof of this is straightforward given the representation in Theorem \ref{thm:compoundPoisson} and given the one-to-many-mapping combinatorial coefficient taking $(n_1,\ldots,n_l,l)$ to $(z_1,\ldots,z_n,n)$ is
$$\frac{l!}{n!}\,\prod_{k=1}^l n_k!\,.$$

\begin{cor}[Frequency of Frequencies Distribution] \label{cor:m_i}
Let  $\mathcal{M}=\{m_{i}\}_i$ be the frequency of frequencies (FoF) vector, where $m_{i}=\sum_{k=1}^{l}\delta(n_k=i)$ is the number of distinct types of size~$i$,  $\sum_{i=1}^\infty m_{i}=l$, and $\sum_{i=1}^\infty im_{i}=n$. For the $G$ mixed Poisson process, we  can generate  a random sample of 
$\mathcal{M}$ by drawing each of its element independently as 
\beq
m_i\sim\emph{\mbox{Poisson}}\left(m_i; \frac{\gamma_0\int_0^\infty r^{i} e^{-r} \rho(dr)}{i!}\right)
\eeq
for $i\in\{1,2,\ldots\}$. Alternatively, we may first draw $$l\sim\emph{\mbox{Poisson}}\left(\gamma_0\int_{0}^\infty(1-e^{-r})\rho(dr)\right) $$ as the total number of distinct clusters (species) with nonzero counts, then draw $m_i$ sequentially %, %$i.e.$, the number of species of size $i$,    
 %\in\{1,2,\ldots\}$
 using a stick-breaking construction as
\beq
m_{i}\,|\,l, m_1,\ldots,m_{i-1} \sim 
%\mbox{Binomial}\left(\ell -\sum_{t=1}^{i-1}m_{t} , \frac{\frac{\Gamma(i-a)\gamma_{0}p^{i-a}}{\Gamma(1-a)(i)!}}{  \gamma_0\frac{1-(1-p)^a}{ap^{a}} - \sum_{t=1}^{i-1} \frac{\Gamma(t-a)\gamma_{0}p^{t-a}}{\Gamma(1-a)t!}  }  \right)
\emph{\mbox{Binomial}}\left(l-\sum_{t=1}^{i-1}m_{t} , \frac{\frac{\int_0^\infty r^{i} e^{-r} \rho(d r)}{i!}}{  \sum_{t=i}^{\infty} \frac{\int_0^\infty r^{t} e^{-r} \rho(d r)}{t!} } 
 \right) 
\eeq
for $i = 1,2,\ldots$ until $l = \sum_{t=1}^i m_{i}$, and further let $m_{i+\kappa}=0$ for all $\kappa\in\{1,2,\ldots\}$. %, where $\kappa$ can be interpreted as the size of the largest partition in a random sample of size $\sum_{i=1}^{\kappa} im_i =n$. 

\end{cor}
%The proof for this Corollary is provided in Appendix E.

\begin{cor}[Prediction Rule]\label{thm:predict} 
Let $l^{-i}$ represent the number of clusters in $\zv^{-i}:=\zv\backslash z_i$ and $n_k^{-i}:=\sum_{j\neq i} \delta(z_j=k)$. We can express the prediction rule of the  model as
\beq%\label{eq:PredictRuleG}
P(z_{i} = k\,|\,\zv^{-i},n,\gamma_0,\rho) \propto
\begin{cases}\vspace{2mm}
\displaystyle\frac{\int_{0}^\infty r^{n_k^{-i}+1} e^{-r} \rho(d r)}{\int_0^\infty r^{n_k^{-i}}e^{-r}  \rho(d r)} , & \emph{\mbox{for }} k=1,\ldots,l^{-i};\\
\displaystyle \gamma_0\int_0^\infty re^{-r}\rho(dr), & \emph{\mbox{if } }k=l^{-i}+1.
\end{cases}\notag
\eeq
This prediction rule can be used to simulate an exchangeable random partition of $[n]$ via Gibbs sampling.
\end{cor}

%The proofs for both Corollaries are provided in Appendix E. 
\subsection{Related work}\label{sec:2.4}
To make connections to previous work, 
let us first consider the special case that $G$ %\sim\Gamma\mbox{P}[G_0,p/(1-p)]$
 is a gamma process with L\'evy measure $\nu(drd\omega) = r^{-1}e^{-p^{-1}(1-p)r}drG_0(d\omega)$, which is a special case of the generalized gamma process $G\sim\mbox{g}\Gamma\mbox{P}[G_0,a,p/(1-p)]$ with $a=0$. This $G$ mixed Poisson process is defined as the negative binomial process  $X\sim\mbox{NBP}(G_0,p)$  in \citet{NBP2012}. For $X\sim\mbox{NBP}(G_0,p)$, with Corollary \ref{cor:compoundPoisson}, the L\'evy measure can be expressed as 
$
\nu(dnd\omega) = \sum_{j=1}^\infty j^{-1}{p^j} \delta_j(dn) G_0(d\omega).
$ 
With Corollary \ref{thm:ECPF}, 
%the ECPF can be expressed as
we have the ECPF 
$
p(\zv,n\given \gamma_0,p) = ({n!})^{-1} p^{n}(1-p)^{\gamma_0}\gamma_0^l \prod_{k=1}^{l} \Gamma(n_k)\notag
$
and probability mass function (PMF)  $p_N(n\given \gamma_0,p)=\frac{\Gamma(n+\gamma_0)}{\Gamma(\gamma_0)} p^n(1-p)^{\gamma_0}$, which is the PMF of the negative binomial (NB) distribution $n\sim\mbox{NB}(\gamma_0,p)$.
Thus the EPPF for $X$ %\sim\mbox{NBP}(G_0,p)$ 
can be expressed~as 
\begin{align}\label{eq:CRPEPPF}
p(\zv\given \gamma_0) &=\frac{p(\zv,n\given \gamma_0,p)}{p_N(n\given \gamma_0,p)}=  \frac{\Gamma(\gamma_0)\gamma_0^l}{\Gamma(n+\gamma_0)} \prod_{k=1}^{l} \Gamma(n_k),
\end{align}
which is the EPPF of the Chinese restaurant process (CRP) \citep{aldous:crp}, % with concentration parameter $\gamma_0$, 
a variant of the widely used Ewens sampling formula \citep{ewens1972sampling,PolyaUrn}. 

%While the CRP does not directly define a prior distribution over a FoF vector, 
For the CRP, multiplying  its EPPF $p(\zv\given \gamma_0)$  by the PMF of $n\sim\mbox{NB}(\gamma_0,p)$ leads to the ECPF, and as in Corollary \ref{cor:m_i}, further multiplying its ECPF with the combinatorial coefficient  ${n!}/[{\prod_{i=1}^n(i!)^{m_i}m_i!}]$ leads to 
the distribution of a 
%the %PMF of the 
%joint distribution of the  population size $n$ and 
FoF vector $\mathcal{M}=\{m_i\}_i$ as
\begin{align} %\label{main_samp}
p(\mathcal{M},n\,|\,\gamma_0,p) &=\left\{\prod_{i=1}^\infty \mbox{Poisson}\left(m_i; \gamma_0\frac{p^i}{i}\right)\right\} \times \delta\left(n=\sum_{i=1}^\infty i m_i\right),
%\left\{\prod_{i=n+1}^\infty \mbox{Poisson}\left(0; \frac{\gamma_0\int_0^\infty s^{i} e^{-s} \rho(dr)}{i!}\right)\right\}.
\notag
\end{align}
which can be generated by simulating countably infinite Poisson random variables, or %, as in Corollary \ref{cor:m_i}, using a stick-breaking constructing with a finite number of steps that 
using a stick-breaking construction that first draws $l\sim\mbox{Poisson}[-\gamma_0\ln(1-p)]$ number of of nonempty clusters, and then draws $m_i$ sequentially 
\beq
m_{i}\,|\,l, m_1,\ldots,m_{i-1} \sim 
%\mbox{Binomial}\left(\ell -\sum_{t=1}^{i-1}m_{t} , \frac{\frac{\Gamma(i-a)\gamma_{0}p^{i-a}}{\Gamma(1-a)(i)!}}{  \gamma_0\frac{1-(1-p)^a}{ap^{a}} - \sum_{t=1}^{i-1} \frac{\Gamma(t-a)\gamma_{0}p^{t-a}}{\Gamma(1-a)t!}  }  \right)
{\mbox{Binomial}}\left(l-\sum_{t=1}^{i-1}m_{t} , \frac{i^{-1}{p^i}}{ -\ln(1-p) - \sum_{t=1}^{i-1} t^{-1}{p^t} } 
 \right) 
\eeq
for $i = 1,2,\ldots$ until $l = \sum_{t=1}^i m_{i}$, and further lets $m_{i+\kappa}=0$ for all $\kappa\in\{1,2,\ldots\}$.

The EPPF of the widely used Piman-Yor process  \citep{%perman1992size,pitman1997two,
csp},  %given the mass parameter $\gamma_0$ and discount parameter $a\in[0,1)$, the
 with  mass parameter $\gamma_0$ and discount parameter $a\in[0,1)$, can be  expressed as
\begin{align}
P(\zv\given \gamma_0,a) &= \frac{\Gamma(\gamma_0)}{\Gamma(n+\gamma_0)} \prod_{k=1}^{l} \frac{\Gamma(n_k-a)}{\Gamma(1-a)}[\gamma_0+(k-1)a]\notag.
\end{align}
However, unless $a=0$, it is unclear whether the Pitman-Yor process can be related to a FoF vector whose countably infinite elements simply follow the Poisson distributions. % Poisson random variables. % Poisson distributions. 
There exists the class of Gibbs-type EPPF that provides a generalization of the EPPF induced by the Pitman-Yor process.  See \citet{gnedin2006exchangeable}  for details and  \citet{de2015gibbs} for a Bayesian nonparametric treatment.
%and the ECPF is the above EPPF multiplied by the negative binomial distribution %$n\sim\mbox{NB}(\gamma_0,p)$, with 
%
%The Chinese restaurant process and Pitman-Yor process do not define a prior on FoF distribution, directly. The negative binomial process is a special case of our model. It is ability to model FoF vector is previously unknown and is revealed in this paper. But its parameterization is relatively restrictive. 
%

Note that the ideas of mixing multiple group-specific Poisson processes with a gamma process, or mixing multiple group-specific negative binomial (NB) processes with a gamma or beta process have been exploited in \citet{NBP2012} to construct priors for mixed-membership modeling, and in \citet{NBP_CountMatrix} to construct priors for  random count matrices. When the number of groups reduces to one, the %gamma-Poisson or
 NB process in \citet{NBP2012} and \citet{NBP_CountMatrix} %, also related to the one in \citet{lo1982bayesian} and \citet{InfGaP},
  becomes a special case of the generalized NB process to be thoroughly investigated in Section  \ref{sec:gNBP}. %The EPPF governed by the negative binomial process is constrained to be independent on the population size, whereas the EPPF governed by the generalized negative binomial process is not subject to this constraint.
   Following the hierarchical construction in \citet{NBP2012} and \citet{NBP_CountMatrix}, the proposed generalized NB process or other completely random measure mixed Poisson processes may also be extended to a multiple group setting to construct more sophisticated nonparametric Bayesian priors for both mixed-membership modeling and random count matrices. 
   
    %In the next section 
    Below we will study a particular process: the generalized NB process, whose ECPF and FoF distribution both have simple analytic expressions and whose exchangeable random partitions can not only be simulated via Gibbs sampling using the above prediction rule, but also be sequentially constructed using a recursively calculated prediction rule.

\vspace{-4mm}\section{Generalized negative binomial process}\label{sec:gNBP} \vspace{-3mm}
In the following discussion, 
we study the generalized NB process (gNBP) model where $G\sim\mbox{g}\Gamma\mbox{P}[G_0,a,p/(1-p)]$  with $a<0$, $a=0$, or $0<a<1$. 
Here we apply the results in Section~3 to this specific case.
Using (\ref{eq:LevyGGP}), we have
$\int_{0}^\infty r^{n}e^{-r}\rho(dr)= {\frac{\Gamma(n-a)}{{\Gamma(1-a)} }p^{n-a}}$ %\quad \mbox{and} \quad
and $\int_0^\infty(1-e^{-r})\rho(dr) = \frac{1-(1-p)^a}{ap^{a}}.$
Marginalizing out $G(\Omega)$ from 
$n\,|\, \lambda\sim\mbox{Poisson}[G(\Omega)]$ with $G\sim{{}}\mbox{g}\Gamma\mbox{P}[\gamma_0,a,p/(1-p)]$, leads to a generalized NB distribution; $i.e.$, $n\sim\mbox{gNB}(\gamma_0,a,p)$, with shape parameter $\gamma_0$, discount parameter $a<1$, and probability parameter $p$. 
Denote by $\sum_{*}$ as the summation over all sets of positive integers $(n_1,\ldots,n_l)$ with ${\sum_{k=1}^l n_k = n}$.  As derived in Appendix F, the PMF of the generalized NB distribution can be expressed as
\beq\label{eq:f_M0}
p_N(n\,|\,\gamma_0,a,p)
 = 
{p^n}e^{-{\gamma_0}\frac{1-(1-p)^a}{ap^a}} \sum_{l=0}^n \gamma_0^l p^{-al} \frac{S_a(n,l)}{n!},
\eeq 
where  $S_a(n,l)$, as defined in detail in  Appendix F, multiplied by $a^{-l}$ are %Toscano's formula or 
generalized Stirling numbers \citep{charalambides2005combinatorial,csp}.
Marginalizing out $G$ in the generalized gamma process mixed Poisson process 
\beq\label{eq:gGaPP0}
X\,|\,G\sim\mbox{PP}(G)\quad\mbox{and}\quad G\sim{{}}\mbox{g}\Gamma\mbox{P}\left[G_0,a, {p}/{(1-p)}\right]
\eeq %\eeqs %in (\ref{eq:gGaPP}) 
leads to a generalized NB process
$
X\sim\mbox{gNBP}(G_0,a,p),
$
such that for each $A\subset \Omega$, $X(A)\sim\mbox{gNB}(G_0(A),a,p)$. This process is also a compound Poisson process as
\begin{align}\label{eq:GNBPdraw}
X(\cdot)=\sum_{k=1}^{l} n_k\delta_{\omega_k}(\cdot),~l\sim\mbox{Poisson}\Big(\gamma_0\frac{1-(1-p)^a}{ap^a}\Big),~n_k  \stackrel{iid}{\sim} \mbox{TNB}(a,p),~\omega_k \stackrel{iid}{\sim} g_0,
\end{align}
 where $\mbox{TNB}(a,p)$ denotes a truncated NB distribution, with PMF
\beqs\label{eq:TNB}
p_U(u\,|\,a,p)= \frac{\Gamma(u-a)}{u!\Gamma(-a)}\frac{p^u(1-p)^{-a}}{1-(1-p)^{-a}},~u=1,2,\ldots.
\eeqs
Note that $\lim_{a\rightarrow 0}\frac{1-(1-p)^a}{ap^a} = -\ln(1-p)$ and $\lim_{a\rightarrow 0}\mbox{TNB}(a,p)$ becomes the logarithmic distribution with parameter $p$ \citep{Fisher1943,LogPoisNB,johnson2005univariate}.
The L\'evy measure of the gNBP can be expressed as 
$
\nu(dnd\omega) = \sum_{j=1}^\infty \frac{\Gamma(j-a)}{j!\Gamma(1-a)}p^{j-a} \delta_j(dn) G_0(d\omega).
$ 

The ECPF of the gNBP %generalized NB process 
model is given by
\beqs \label{eq:f_Z_M}
p(\zv,n\,|\,\gamma_0,a,p) 
=\frac{1}{n!}e^{-\gamma_0\frac{1-(1-p)^a}{ap^{a}}}
\gamma_0^{l{}} p^{n-al{}}
\prod_{k=1}^{l{}} \frac{\Gamma(n_k-a)}{{\Gamma(1-a)} },
\eeqs 
which is fully factorized and will be used as the likelihood to infer $\gamma_0$, $a$, and $p$.
The EPPF of $\Pi_n$ %of the generalized NB process 
is the ECPF in (\ref{eq:f_Z_M}) divided by the marginal distribution of $n$ in (\ref{eq:f_M0}), given by
\begin{align} \label{eq:EPPF}
p(\zv\,|\,n,\gamma_0,a,p) = 
\frac{\gamma_0^{l}  p^{-al}}{
\sum_{\ell=0}^n \gamma_0^\ell p^{-a\ell} S_a(n,\ell)}
\prod_{k=1}^{l{}}\frac{\Gamma(n_k-a)}{\Gamma(1-a)}.
\end{align}
We define  the EPPF in (\ref{eq:EPPF})  %in (\ref{eq:EPPF}) % probability function
as the generalized Chinese restaurant sampling formula (gCRSF),  and we denote a random draw under this EPPF as %given by 
$$\zv\,|\,n\sim{\mbox{gCRSF}}(n,\gamma_0,a,p).$$
%As derived in the Appendix, 
The conditional distribution of the number of clusters in a population of size $n$ %conditioning on the sample size $m$ 
can be expressed as
\begin{align}\label{eq:f_L2_0}
p_L(l\,|\,n,\gamma_0,a,p) =\frac{1}{l!}\sum_{*}\frac{n!}{\prod_{k=1}^l n_k!}p(\zv\,|\,n,\gamma_0,a,p)= \frac{\gamma_0^{l}  p^{-al}S_a(n,l)}{
\sum_{\ell=0}^n \gamma_0^\ell p^{-a\ell} S_a(n,\ell)}.
\end{align}
Recall that $m_{i}=\sum_{k=1}^{l}\delta(n_k=i)$ represents the number of distinct types of size $i$, with $\sum_{i=1}^\infty m_{i}=l$ and $\sum_{i=1}^\infty im_{i}=n$. With Corollary \ref{cor:m_i}, 
%we can express the  conditional distribution  of $\mathcal{M}$ in a sample of size $n$ as
%\begin{align}\label{main_samp}
%p(\boldsymbol{m}\,|\,n,\gamma_{0},a,p)&=\frac{n!}{\prod_{i=1}^n (i!)^{m_i}m_i!} p(\zv\,|\,n,\gamma_0,a,p)\notag\\
%&=
%\frac{n!}{\sum_{l=0}^{n}\gamma_{0}^{l}p^{-a l} S_{a}(n,l)}\prod_{i=1}^{n}\frac{1}{m_{i}!}\left(\frac{\Gamma(i-a)\gamma_{0}p^{-a}}{\Gamma(1-a)i!}\right)^{m_{i}}. %\notag\\
%%&=
%\end{align}
we can express the joint distribution of $n$ and $\mathcal{M}$, under the constraint that $n=\sum_{i=1}^\infty im_{i}$, as
\begin{align} %\label{main_samp}
 p(\mathcal{M},n\,|\, \gamma_{0},a,p) %=e^{-\gamma_0\frac{1-(1-p)^a}{ap^{a}}}
% \prod_{i=1}^{n}\frac{1}{m_{i}!}\left(\frac{\Gamma(i-a)\gamma_{0}p^{i-a}}{\Gamma(1-a)i!}\right)^{m_{i}}\notag\\
&=\left\{\prod_{i=1}^{\infty}\mbox{Poisson}\left(m_i; \frac{\Gamma(i-a)\gamma_{0}p^{i-a}}{\Gamma(1-a)i!}\right)\right\}
%\left\{\mbox{Poisson} 
%\left(0; \sum_{i=n+1}^\infty \frac{\Gamma(i-a)\gamma_{0}p^{i-a}}{\Gamma(1-a)i!}\right) \right\} \notag\\
%&
\times 
\delta\left(n = \sum_{i=1}^\infty i m_i\right) \label{eq:SpeciesSeries}, %\\
%&=\left\{\prod_{i=1}^{n}\mbox{Poisson}\left(m_i; \frac{\Gamma(i-a)\gamma_{0}p^{i-a}}{\Gamma(1-a)i!}\right)\right\}\left\{\mbox{Poisson}\left(0; \gamma_0\frac{1-(1-p)^a}{ap^{a}} - \sum_{i=1}^n \frac{\Gamma(i-a)\gamma_{0}p^{i-a}}{\Gamma(1-a)i!}\right) \right\},
\end{align}
%\begin{align} %\label{main_samp}
% p(\mathcal{M},n\,|\, \gamma_{0},a,p) %=e^{-\gamma_0\frac{1-(1-p)^a}{ap^{a}}}
%% \prod_{i=1}^{n}\frac{1}{m_{i}!}\left(\frac{\Gamma(i-a)\gamma_{0}p^{i-a}}{\Gamma(1-a)i!}\right)^{m_{i}}\notag\\
%&=\left\{\prod_{i=1}^{n}\mbox{Poisson}\left(m_i; \frac{\Gamma(i-a)\gamma_{0}p^{i-a}}{\Gamma(1-a)i!}\right)\right\}\left\{\mbox{Poisson} 
%\left(0; \sum_{i=n+1}^\infty \frac{\Gamma(i-a)\gamma_{0}p^{i-a}}{\Gamma(1-a)i!}\right) \right\} \notag\\
%&\times \delta\left(n = \sum_{i=1}^\infty i m_i\right) \label{eq:SpeciesSeries}, %\\
%%&=\left\{\prod_{i=1}^{n}\mbox{Poisson}\left(m_i; \frac{\Gamma(i-a)\gamma_{0}p^{i-a}}{\Gamma(1-a)i!}\right)\right\}\left\{\mbox{Poisson}\left(0; \gamma_0\frac{1-(1-p)^a}{ap^{a}} - \sum_{i=1}^n \frac{\Gamma(i-a)\gamma_{0}p^{i-a}}{\Gamma(1-a)i!}\right) \right\},
%\end{align}
where we apply the fact that
$\sum_{i=1}^n \frac{\Gamma(i-a)}{i!\Gamma(-a)}{p^i(1-p)^{-a}} = 1-(1-p)^{-a}$ for $a<1$. 
Thus to generate a cluster structure governed by the generalized negative binomial process, one may draw $m_i\sim\mbox{Poisson}\left(\frac{\Gamma(i-a)\gamma_{0}p^{i-a}}{\Gamma(1-a)i!}\right)$ independently for each $i$, or first draw 
\beq
l\sim\mbox{Poisson}\left(\gamma_0\frac{1-(1-p)^a}{ap^{a}}\right) \label{eq:ell}
\eeq
 number of unique partitions (species), and then draw $m_i$ %, $i.e.$, the number of species of size $i$,   
 for $i\ge1$ using 
\beq \label{eq:Bino}
m_{i}\,|\,l, m_1,\ldots,m_{i-1} \sim 
%\mbox{Binomial}\left(\ell -\sum_{t=1}^{i-1}m_{t} , \frac{\frac{\Gamma(i-a)\gamma_{0}p^{i-a}}{\Gamma(1-a)(i)!}}{  \gamma_0\frac{1-(1-p)^a}{ap^{a}} - \sum_{t=1}^{i-1} \frac{\Gamma(t-a)\gamma_{0}p^{t-a}}{\Gamma(1-a)t!}  }  \right)
\mbox{Binomial}\left(l -\sum_{t=1}^{i-1}m_{t} , \frac{\frac{\Gamma(i-a)p^{i}}{i!}}{  \sum_{t=i}^{\infty} \frac{\Gamma(t-a)p^{t}}{t!}  } 
 \right) 
\eeq
 until $l = \sum_{t=1}^im_{t}$. Note that in the prior, $\E[m_i]=\left(\frac{\Gamma(i-a)\gamma_{0}p^{i-a}}{\Gamma(1-a)i!}\right)$ 
 %$\ln(\E[m_i])  = (i-a)\ln(p) + \ln \left(\frac{\Gamma(i-a)\gamma_{0} }{\Gamma(1-a)i!}\right)$ 
 and hence, using the property of the gamma function, we have
 $$\ln(\E[m_i]) ~\sim ~ - (a+1)\ln(i) + \ln(p) i $$
 as $i\rightarrow \infty$. Thus if $p\rightarrow 1$,  we may consider $a+1$ as a power-law scaling parameter.

Note that if $a\rightarrow 0$, we recover from (\ref{eq:SpeciesSeries}) the logarithmic series of \citet{Fisher1943}, as also discussed in \citet{Sampling_NB_1950} and  \citet{watterson1974models},  and we  recover from (\ref{eq:EPPF}) the EPPF for the CRP, as shown in \eqref{eq:CRPEPPF}. %a variant of the Ewens sampling formula  which is the EPPF of the Chinese restaurant process (CRP) \citep{PolyaUrn,aldous:crp}.
 When $a\neq 0$, we generalize CRP by making the EPPF be dependent on the population size $n$. This generalization differs from those in \citet{ishwaran2003generalized} and \citet{cerquetti2008generalized}, where the EPPFs are independent of $n$.

The prediction rule for  the EPPF in (\ref{eq:EPPF})   can be expressed as % can be expressed as
\beq\label{eq:PredictRule}
%p(z_{m+1} = k\,|\,\zv,m,\gamma_0,a,p)
P(z_{i} = k\,|\,\zv^{-i},n,\gamma_0,a,p)  \propto
\begin{cases}
n_k^{-i} -a, &  {\mbox{for }} k=1,\ldots,l^{-i};\\ %\frac{n_k-a}{m+\gamma_0p^{-a}-al}, 
\gamma_0 p^{-a}, & {\mbox{if } }k=l^{-i}+1. %\frac{\gamma_0 p^{-a}} {m+\gamma_0p^{-a}-al}, 
\end{cases}
\eeq
This prediction rule can be used in a Gibbs sampler to simulate an exchangeable random partition $\zv\,|\,n\sim{\mbox{gCRSF}}(n,\gamma_0,a,p)$ of $[n]$. As it is often unclear how many Gibbs sampling iterations are required to generate an unbiased sample from this EPPF, below we present a sequential construction for  this EPPF to directly generate  an unbiased sample.

Marginalizing out $z_n$ from (\ref{eq:EPPF}), we have
\begin{align}
{p(z_{1:n-1}\,|\,n,\gamma_0,a,p)}%&=\frac{j!f(z_{1:j}|j,\gamma_0,\rho)f_J(j\,|\,\gamma_0,\rho)}{m!f_M(m\,|\,\gamma_0,\rho)} p^{i} \prod_{i=j}^{m-1}(\gamma_0p^{-a}+ i- a l_{(i)})\\
~~=&~~~p(z_{1:n-1}\,|\,n-1,\gamma_0,a,p) \notag\\
&\times \frac{\sum_{\ell =0}^{n-1} \gamma_0^\ell p^{-a\ell}S_a(n-1,\ell)}{\sum_{\ell=0}^n \gamma_0^\ell p^{-a\ell }S_a(n,\ell)}\left[\gamma_0p^{-a}+ (n-1)- a l_{(n-1)}\right],\notag
\end{align}
where $z_{1:i}:=\{z_1,\ldots,z_i\}$, $l_{(i)}$ denotes the number of partitions in $z_{1:i}$, and $l_{(n)}=l$.
Further marginalizing out $z_{n-1},\ldots,z_{i+1}$, we have
\begin{align}
{p(z_{1:i}\,|\,n,\gamma_0,a,p)}
&=p(z_{1:i}\,|\,i,\gamma_0,a,p)\frac{\sum_{\ell=0}^{i} \gamma_0^\ell p^{-a\ell}S_a(i,\ell)}{\sum_{\ell=0}^n \gamma_0^\ell p^{-a\ell }S_a(n,\ell)}   R_{n,\gamma_0,a,p}(i,l_{(i)}) \notag\\
&= \frac{R_{n,\gamma_0,a,p}(i,l_{(i)}) \gamma_0^{l_{(i)}}  p^{-al_{(i)}}}{\sum_{\ell=0}^n \gamma_0^\ell p^{-a\ell}S_a(n,\ell)}  \prod_{k\,:\,n_{k,(i)}>0}\frac{\Gamma(n_{k,(i)}-a)}{\Gamma(1-a)} ,\label{eq:SizeEPPF}
\end{align}
where $n_{k,(i)}:=\sum_{j=1}^i \delta(z_j=k)$; $R_{n,\gamma_0,a,p}(i,j)= 1$ if $i=n$  and is recursively calculated for $i=n-1,n-2,\ldots,1$ with
\beq\label{eq:R}
R_{n,\gamma_0,a,p}(i,j) = R_{n,\gamma_0,a,p}(i+1,j)(i-a j) +  R_{n,\gamma_0,a,p}(i+1,j+1)\gamma_0p^{-a}.
\eeq
We name (\ref{eq:SizeEPPF}) as a size-dependent EPPF as its distribution on an exchangeable random partition of $[i]$ is a function of the population size $n$.
Note that if $a=0$, the EPPF becomes the same as that 
of the Chinese restaurant process and no longer depends on $n$. %the population size. %,  as  one may show that it 
%then $$\frac{\sum_{l=0}^{i} \gamma_0^lp^{-al}S_a(i,l)}{\sum_{l=0}^n \gamma_0^lp^{-al}S_a(n,l)} = \frac{\sum_{l=0}^{i} \gamma_0^l |s(i,l)|}{\sum_{l=0}^n \gamma_0^l|s(n,l)|}  = \frac{\Gamma(i+\gamma_0)}{\Gamma(n+\gamma_0)}$$ and $R_{n,\gamma_0,a=0,p}(i,l) = \frac{\Gamma(n+\gamma_0)}{\Gamma(i+\gamma_0)}$, and hence ${p(z_{1:i}\,|\,n,\gamma_0,a=0,p)}
%= p(z_{1:i}\,|\,i,\gamma_0,a=0,p)$. Thus when $a=0$, the EPPF becomes independent of the population size,  which is a well-known property for the Chinese restaurant process. 

In Appendix F, we show the sequential prediction rule of the generalized Chinese restaurant sampling formula that constructs $\Pi_{i+1}$ from $\Pi_i$ in a population of size $n$ by assigning element $(i+1)$ to $A_{z_{i+1}}$, and show the predictive distribution of $z_{i+1\,:\,n}$ 
%$p(z_{i+1\,:\,n}\,|\,z_{1:i},n,\gamma_0,a,p) $.
%
given $z_{1:i}$, the population  size $n$, and model parameters. % the model parameters  $\gamma_0$, $a$ and $p$ p(z_{i+1\,:\,n}\,|\,z_{1:i},n,\gamma_0,a,p) 

In summary, 
a draw from the generalized NB process (gNBP) represents a cluster structure with a Poisson distributed finite number of clusters, whose sizes 
follow a truncated NB distribution. Marginally, the population  size 
follows a generalized NB distribution. 
These three count distributions and the prediction rule are determined by a discount, a probability, and a mass parameter, which together with~$i$ are used to parameterize the Poisson rate for the random number of clusters of size $i$  for the FoF distribution. %For the FoF distribution, the number of clusters of size $i$ follows a Poisson distribution parameterized by a function of $i$ and the three model parameters. 
These parameters are convenient to infer using the fully factorized ECPF. Since $P(\Pi_m\,|\,n)= P(\Pi_m\,|\,m)$ is often not true for $n>m$,  the EPPF of the gNBP, which is derived by applying Bayes' rule on the ECPF and the generalized NB distribution, generally violates the addition rule required in a partition structure and hence is dependent on the population  size.  This size dependent EPPF is  referred to as the generalized Chinese restaurant sampling formula. To generate an exchangeable random partition of $[n]$ under this EPPF, we show we could use either a Gibbs sampler or a recursively-calculated sequential prediction rule.

We conclude this section by investigating the large $n$ asymptotic behavior of both the number of clusters $p_L(l\,|\,n,\gamma_0,a,p)$ shown in   \eqref{eq:f_L2_0} and the sizes of clusters $p(\mathcal{M}\,|\, n, \gamma_{0},a,p) = p(\mathcal{M},n\,|\, \gamma_{0},a,p) / p_N(n\,|\, \gamma_{0},a,p)$, which can be obtained with (\ref{eq:SpeciesSeries}) and (\ref{eq:f_M0}). An interesting question to answer is if we fix the model parameters $\gamma_0$, $a$, and $p$, where $0<\gamma_0<\infty$, $a<1$, and $0<p<1$, and assume the population size $n$ is given, how   $l_{(n)}$, the cluster number, and $M_{i,n}$, the number of clusters of size $i$, would behave as the population size $n$ approaches infinity.  We summarize our findings in Table \ref{tab:asymptotics} and provide the details in Appendices \ref{sec_asym_1} and  \ref{sec_asym_2}. Table \ref{tab:asymptotics} characterizes three asymptotic regimes according to the choice of the parameter $a$, that is $a\in(0,1)$, $a=0$, and $a\in\{-1,-2,\ldots\}$. 

For $a=0$ the distribution \eqref{eq:f_L2_0} coincides with the distribution of the number of clusters in a sample of size $n$ from a Dirichlet process. Hence, the large $n$ asymptotic behavior of $l_{(n)}$ is known from  \citet{hollander73} whereas the large $n$ asymptotic behavior of $M_{i,n}$ is known from \citet{ewens1972sampling}.

For any $a\in(0,1)$ the number of clusters minus one, $l_{(n)}-1$, converges weakly to $\mbox{Poisson}[\gamma_0/(ap^a) ]$, whereas $M_{i,n}$ converges weakly to $\mbox{Poisson}\left(\frac{\Gamma(i-a) \gamma_{0}p^{-a}}{\Gamma(1-a) i!}\right)$. Note that, for any $a\in(0,1)$, $a\frac{\Gamma(i-a)}{\Gamma(1-a) i!}$ is a proper probability distribution over the natural numbers, that is   $a\frac{\Gamma(i-a)}{\Gamma(1-a) i!}\in(0,1)$ for any $i\geq1$ and $\sum_{i=1}^\infty a\frac{\Gamma(i-a)}{\Gamma(1-a) i!}=1$. In other terms, for large $n$ the number $M_{i,n}$ of clusters of size $i$ becomes a proportion $a\frac{\Gamma(i-a)}{\Gamma(1-a) i!}$ of $l_{(n)}-1$, and such a proportion decreases with the index $i$. 
It is also interesting to notice that the logarithmic of $\frac{\Gamma(i-a) \gamma_{0}p^{-a}}{\Gamma(1-a) i!}$ can be approximated by  
$$-(a+1)\ln( i) + C$$
when $i$ is large, where the coefficient $C=\ln\left(\frac{ \gamma_{0}p^{-a}}{\Gamma(1-a) }\right)$ is not related to the index $i$. Thus we may consider $a+1$ as a power-law scaling parameter as $n\rightarrow \infty$.

\begin{table}[!tb]\small \caption{Large $n$ asymptotic regimes with respect to the parameter $a$. %We denote by $P_{c}$ a Poisson random variable with parameter $c$.
}\label{tab:asymptotics}
\begin{center}\small
\begin{tabular}{c*{8}{c}r} $a$ & Distinct types $l_{(n)}$ & Distinct types $M_{i,n}$ \\[0.2cm] 
\hline\hline\\
$(0,1)$ & $\displaystyle l_{(n)}\rightarrow 1+\mbox{Poisson}\left(\frac{\gamma_{0}}{ap^{a}}\right)$ & $\displaystyle M_{i,n}\rightarrow \mbox{Poisson}\left(\frac{\Gamma(i-a) \gamma_{0}p^{-a}}{\Gamma(1-a) i!}\right)$ %$\sim \left(\frac{ \gamma_{0}p^{-a}}{\Gamma(1-a) }\right) l^{-a-1} $
\\[0.4cm] 
0 & $\displaystyle\frac{l_{(n)}}{\log n}\rightarrow \gamma_{0}$ & $\displaystyle M_{i,n}\rightarrow \mbox{Poisson}\left(\frac{\gamma_{0}}{i}\right)$ \\[0.4cm] 
$-a\in\{1,2,\ldots\}$ & $\displaystyle\frac{l_{(n)}}{n^{\frac{-a}{1-a}}}\rightarrow\frac{(\gamma_{0}p^{-a})^{\frac{1}{1-a}}}{-a}$ & $\displaystyle M_{i,n}\rightarrow  \mbox{Poisson}\left(
{ \frac{\Gamma(i-a)\gamma_{0}p^{-a}}{\Gamma(1-a){i!}}}\right)$
%\frac{\Gamma(l-a+2)\gamma_{0}p^{-a}}{\Gamma(l+1)}\right)$
\end{tabular}
\vspace{-5mm}
\end{center}
\end{table}

Finally, for any $a\in\{-1,-2,\ldots\}$ the number of clusters rescaled by $n^{-a/(1-a)}$ converges weakly to the constant $\frac{(\gamma_{0}p^{-a})^{\frac{1}{1-a}}}{-a}$, whereas $M_{i,n}$ converges weakly to $\mbox{Poisson}\left(\frac{\Gamma(i-a) \gamma_{0}p^{-a}}{\Gamma(1-a) i!}\right)$. Note that, differently from the case $a\in(0,1)$, for any $a\in\{-1,-2,\ldots\}$, $\sum_{i=1}^\infty a\frac{\Gamma(i-a)}{\Gamma(1-a) i!}=+\infty$, that is $a\frac{\Gamma(i-a)}{\Gamma(1-a) i!}$ is not a probability distribution over the natural numbers. In particular, $a\frac{\Gamma(i-a)}{\Gamma(1-a) i!}$ is a constant when $a=-1$ and  increases with the index $i$ when $a\in\{-2,-3,\ldots\}$.

\vspace{-3mm}\section{Illustrations}\label{sec:results}\vspace{-2mm}

Species abundance data of a population 
is usually represented with a FoF vector as $\mathcal{M}=\{m_i\}_i$, where $m_i$ denotes the number of species that have been observed $i$ times in the population. %of size $i$, % for $i\in\{1,2,\ldots\}$, 
As discussed before, this data can also be converted into a sequence of cluster indices $\zv=(z_1,\ldots,z_n)$ or a cluster-size vector $(n_1,\ldots,n_l)$, where $n_k$ is the number of individuals in cluster $k$, $n = \sum_i  im_i=\sum_{k=1}^l n_k$ is the size of the population  and $l=\sum_i m_i$ is the number of distinct clusters in the population. For example, we may represent $\{m_1, m_2, m_3\}=\{2,1,2\}$ as $\zv=(1,2,3,3,4,4,4,5,5,5)$ or $(n_1,\ldots,n_5)=(1,1,2,3,3)$.
 For %a sample of 
 species frequency counts, we use (\ref{eq:f_Z_M})  as the likelihood for the model parameters $\thetav=\{\gamma_0,a,p\}$. With appropriate priors imposed on $\thetav$, we use MCMC to obtain posterior samples $\thetav^{(j)}=\{\gamma_0^{(j)}, a^{(j)}, p^{(j)}\}$. % and then calculate $S_{\thetav^{(j)}}$.
The details of MCMC update equations are provided in  Appendix I.

%\vspace{-2mm}\subsection{Extrapolation of the FoF vector}
To understand the structural properties of the population, one often has to make a choice between taking more but smaller size samples and taking fewer but larger size samples. For example, in high-throughput sequencing, to increase the number of detected sequences given a fixed budget, one may need to decide whether to reduce the sequencing depth per sample to allow collecting more biological replicates \citep{sims2014sequencing}. These motivate us to consider the fundamental problem of extrapolating the FoF vector of a sample, taken without replacement from the population, to reconstruct the FoF vector of the population.  This extrapolation problem %has already been 
is readily answered under our framework by $p(z_{i+1\,:\,n}\,|\,z_{1:i},n,\gamma_0,a,p)$ in (\ref{eq:extrapolate}), which shows the joint distribution of the cluster indices of the unobserved $n-i$ individuals of the population given the observed clusters indices $(z_1,\ldots,z_i)$  of the sample of size $i$, the population size $n$, and the model parameters. To reconstruct $(z_{i+1},\ldots,z_{n})$, one can either use (\ref{eq:PredictRule}) to sequentially construct the vector from $z_{i+1}$ to $z_n$, or randomly initialize the vector and then use (\ref{eq:PredictRulej}) in a Gibbs sampling algorithm. For a population with tens of thousands or millions of individuals, we prefer the second method as it is often more computationally efficient.

We consider   the novel ``The Adventures of Tom Sawyer'' by Mark Twain, with a total of $n=77,514$ words from $l=7,772$ terms; the novel ``The Adventures of Sherlock Holmes'' by Arthur Conan Doyle, with a total of  $n=106,007$ words from $l=7,896$ terms; the high-throughput sequencing dataset studied in \citet{Sultan15082008}, with a total of $n=418,650$ sequences from $l=6,712$ unique sequences; the high-throughput sequencing dataset studied in \citet{Core19122008}, with a total of $n=125,794$ sequences from $l=7,124$ unique sequences; and the mircodata provided in Table A.6 of \citet{greenberg1990geographic}, with a total of $n=87,959$ household records from $l=929$ groups. We randomly take  $1/32$,  $1/16$, $1/8$, $1/4$, or  $1/2$ of the individuals without replacement from the population to form a sample $(z_1,\ldots,z_i)$,  where $i$ is the sample size, from which we use Gibbs sampling to simulate the indices of the remaining individuals $(z_{i+1},\ldots,z_n)$, where $n$ is the population size. In each Gibbs sampling iteration, we draw $T=5$ times the indices in $\{z_{i+1},\ldots, z_n\}$ in a random order  using (\ref{eq:PredictRulej}) and then sample the model parameters $\gamma_0$, $a$, and $p$ once. %We set $T=5$ in this paper. 

For comparison, we  consider using the software provide for  \citet{clauset2009power} to estimate a lower cutoff point $i_{\min}$ and a scaling parameter $\alpha$ from a random sample taken without replacement from the finite population, and then find  $-\alpha_h$,  the slope of the least squares line fitting the first $i_{\min}-1$ FoF points of the random sample on the log-log plot. We then fit a straight line to the population FoF points $\{\ln i, \ln(m_i)\}_{i<i_{\min}}$, with $-\alpha_h$ as the slope and $[\sum_{i\in I_h}(\ln(m_i)+\alpha_h \ln(m_i)]/|I_h|$ as the intercept, where $I_h=\{i:1\le i < i_{\min},\, m_i>=1\}$, and another straight line to the population FoF points $\{\ln i, \ln(m_i)\}_{i\ge i_{\min}}$, with $-\alpha$ as the slope and $[\sum_{i\in I_t}(\ln(m_i)+\alpha \ln(m_i)]/|I_t|$ as the intercept, where $I_t=\{i: i \ge i_{\min},\, m_i>=3\}$. We emphasize that this least squares (LS) procedure is merely used as a baseline, which refits the population FoF points under the assumption that $i_{\min}$, $\alpha_h$, and $\alpha$ all
%the power-law lower cutoff point and scaling parameter for the tail of the sample FoF vector, and the slope of the straight line that fits the beginning part of the sample FoF vector 
all stay  unchanged as the sample size varies; it may fit the tail well, but may perform poorly in fitting the center part of a FoF distribution. 
%, and hence %It is used for comparison, but it does not work in practice
%is impractical to be used to predict the FoF vector of a population given  that of a random sample. % taken from it. 

We also make comparisons with the Pitman-Yor  process  \citep{perman1992size,pitman1997two,csp}, a widely used nonparametric Bayesian prior with a size independent EPPF that $P(\Pi_m\,|\, \gamma_0,a, m)=P(\Pi_m\,|\, \gamma_0,a, n)$ for all $n\ge m$, where $\gamma_0$ and $a$ are the concentration and discount parameters, respectively, for the Pitman-Yor process. We describe a Gibbs sampling algorithm in Appendix I, using data augmentation techniques developed in \citet{teh2006bayesian}.  In addition, we also consider the Chinese restaurant process.

\begin{figure}[!tb]

\begin{center}
\includegraphics[width=0.88\columnwidth]{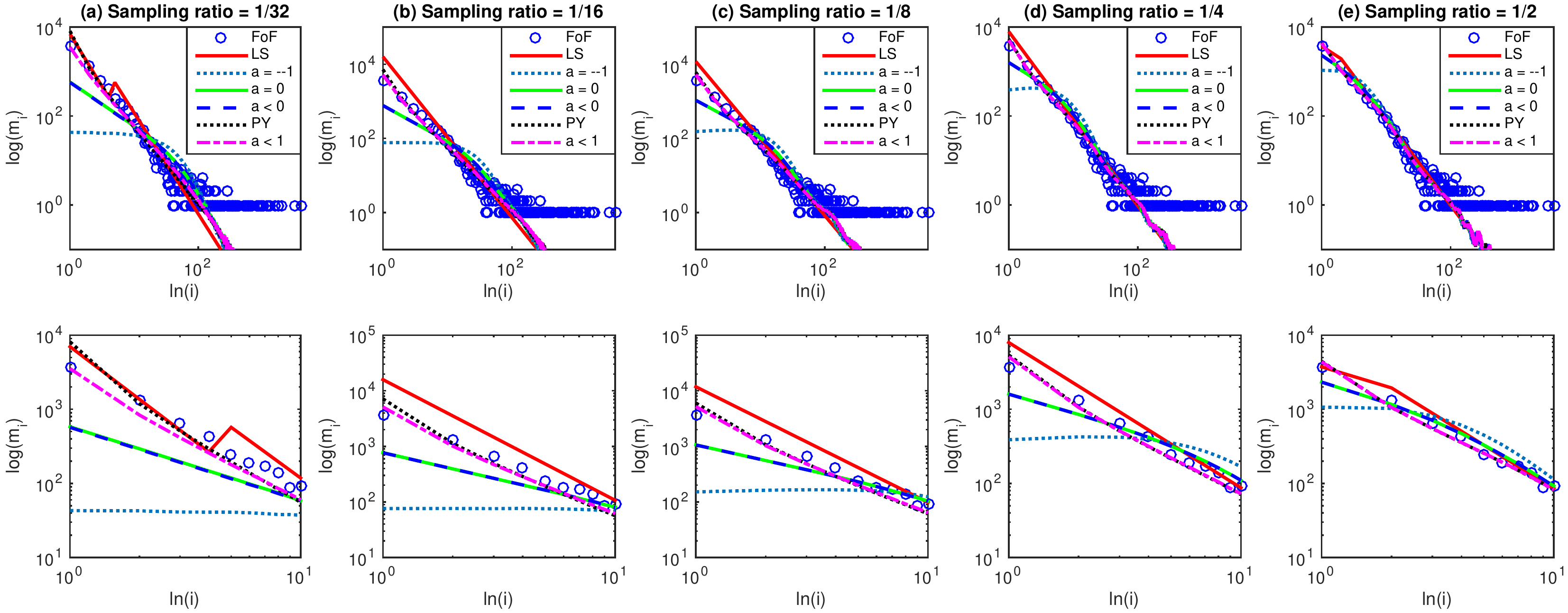}
\end{center}
\vspace{-6mm}
\caption{ \label{fig:tomsawyer_FoF_extrapolate} \small The posterior means of the population FoF vectors extrapolated from  sample FoF vectors for ``The Adventures of Tom Sawyer'' by Mark Twain, using the least squares (LS) refitting procedure, the Chinese restaurant process, the Pitman-Yor (PY) process, and the generalized negative binomial process (gNBP), whose discount parameter is set as $a=-1$, $a=0$, %$a=0.5$, 
$a\in(-\infty,0)$, %$a\in[0,1)$, 
or $a\in(-\infty,1)$. Each sample is taken without replacement from the population with a sampling ratio of $1/32$, $1/16$, $1/8$, $1/4$, or $1/2$. The performance of the Chinese restaurant process is found to be almost identical to the gNBP with $a=0$, and hence omitted for brevity. 
%\citet{Sultan15082008}. 
}%\vspace{-2mm}
%\end{figure}

 \begin{center}
\includegraphics[width=0.6\columnwidth]{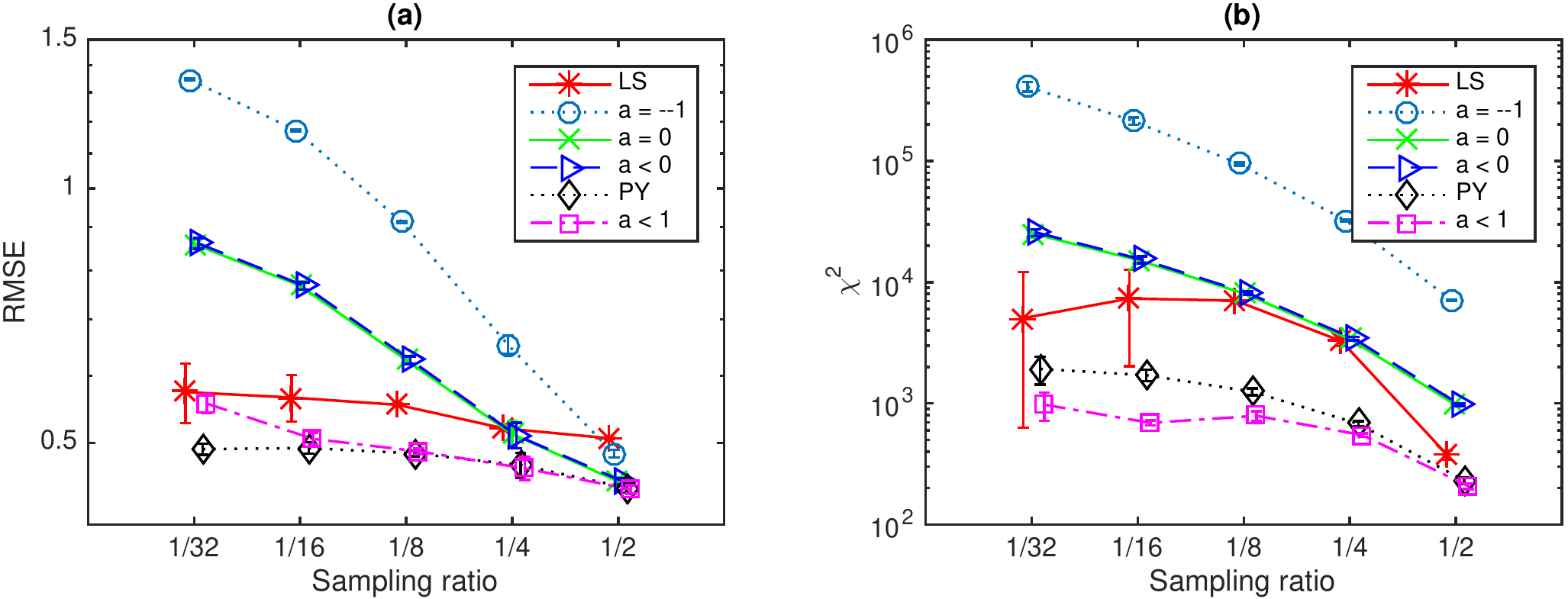}
\end{center}
\vspace{-6mm}
\caption{ \label{fig:tomsawyer_extrapolate} \small (a) RMSEs and (b) chi-squared ($\chi^2$) test statistics 
for the extracted FoF vectors shown in Figure \ref{fig:tomsawyer_FoF_extrapolate}.
% of the population extrapolated from that of a sample, which is  taken without replacement from the population, with a sampling ratio of $1/32$, $1/16$, $1/8$, $1/4$, or $1/2$. The discount parameter is set as $a=-1$, $a=0$, $a=0.5$,  $a\in[0,1)$, or $a\in(-\infty,1)$.
}

%\begin{figure}[!tb]

\end{figure}

\begin{figure}[!tb]
 
\begin{center}
\includegraphics[width=0.88\columnwidth]{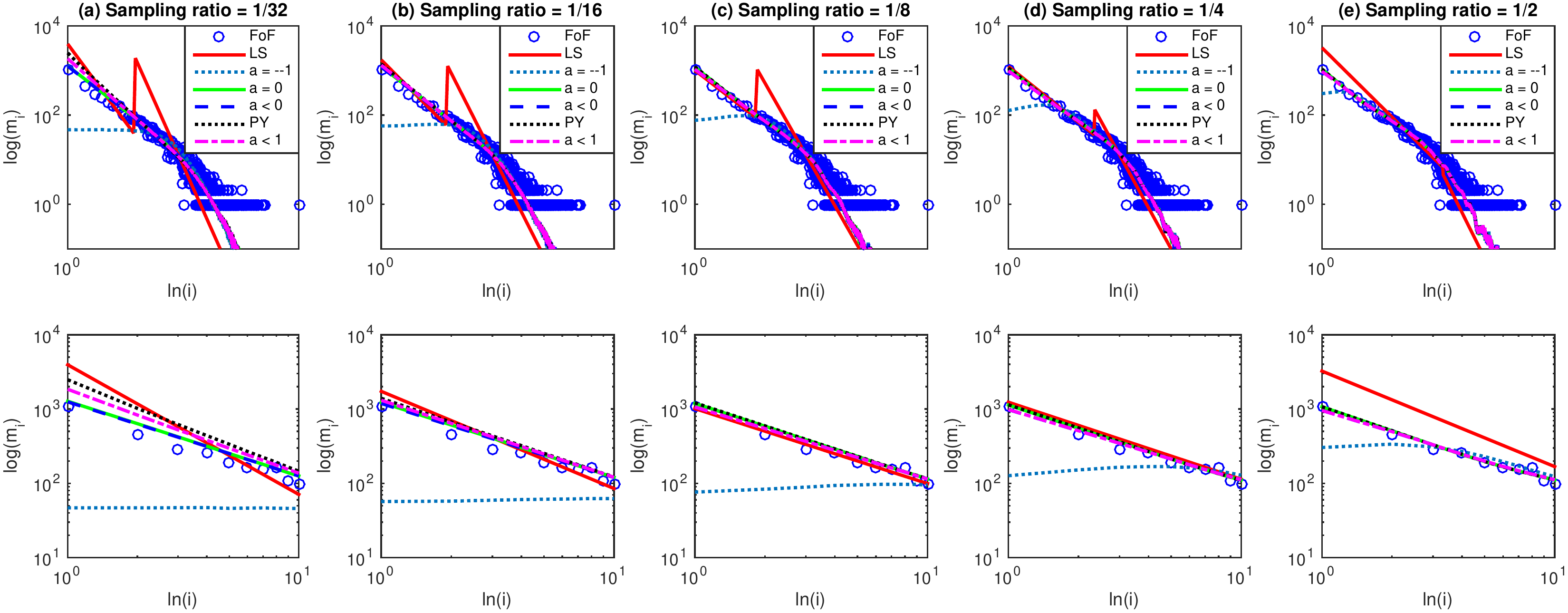}
\end{center}
\vspace{-6mm}
\caption{ \label{fig:sultan_FoF_extrapolate} \small Analogous plots to Figure \ref{fig:tomsawyer_FoF_extrapolate} for a RNA-seq  data studied in \citet{Sultan15082008}. 
}%\vspace{-2mm}
%\end{figure}

 \begin{center}
\includegraphics[width=0.6\columnwidth]{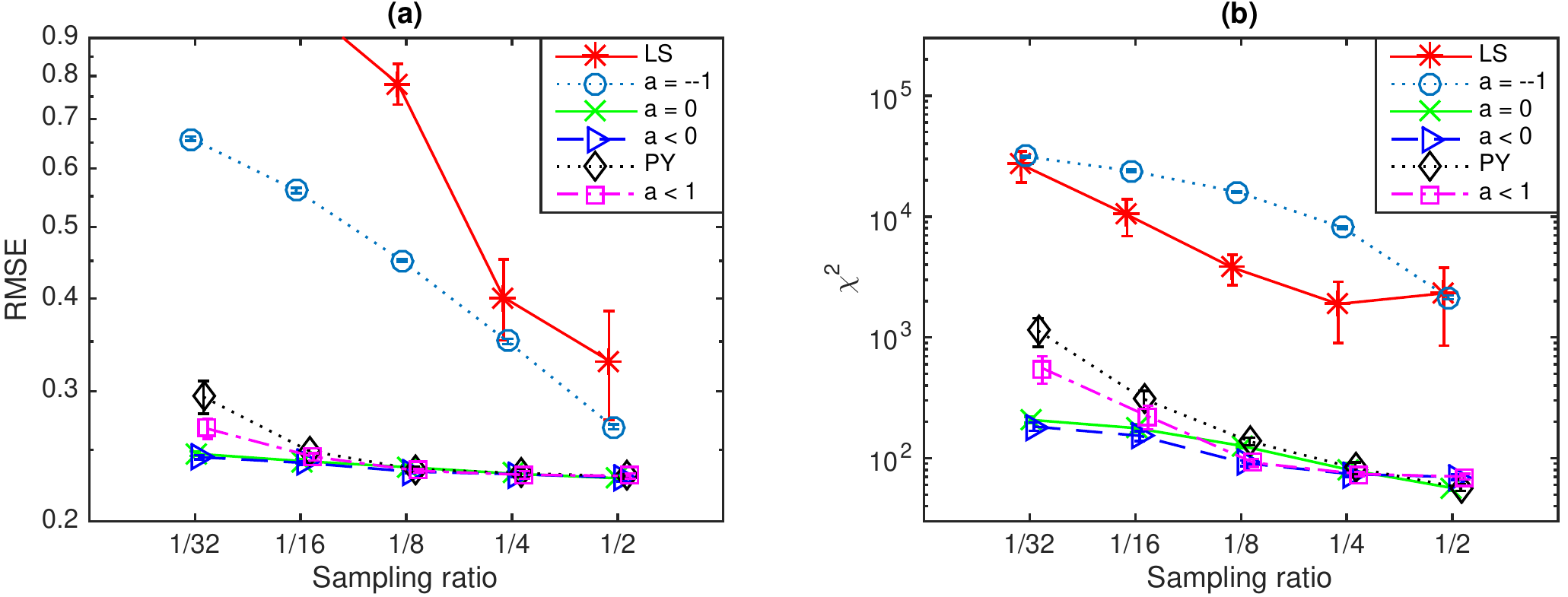}
\end{center}
\vspace{-6mm}
\caption{ \label{fig:sultan_extrapolate} \small 
Analogous plots to Figure \ref{fig:tomsawyer_extrapolate} for a RNA-seq  data studied in \citet{Sultan15082008}. 
%(a) RMSE and (b) chi-squared ($\chi^2$) test statistic 
%for the FoF vector of the population extrapolated from that of a sample, which is  taken without replacement from the population, with a sampling ratio of $1/32$, $1/16$, $1/8$, $1/4$, or $1/2$. The discount parameter is set as $a=-1$, $a=0$, $a=0.5$,  $a\in[0,1)$, or $a\in(-\infty,1)$.
}

%\begin{figure}[!tb]

\end{figure}

 For all MCMC based algorithms, we consider 1000  iterations and collect the last 500 samples, for each of which we convert the cluster index vector $(z_1,\ldots,z_n)$ to a population FoF vector, and take the average of all the 500 collected vectors, denoted by $\widehat{\mathcal{M}}=(\hat{m}_1,\ldots,\hat{m}_n)$, as the posterior mean of the population FoF vector, given the sample $(z_1,\ldots,z_i)$ and the population size $n$. 
Using the observed  population FoF vector $\mathcal{M}$, we measure the extrapolation  performance using the root mean squared error (RMSE), defined as
%\beq
%\mbox{RMSE} = \sqrt{\frac{\sum_{i:1\le m_i\le 100} \left[\ln(m_i) - \ln(\hat{m}_i)\right]^2}{ \sum_{i} \delta(1\le m_i\le 100)}}
%\eeq
\beq
\mbox{RMSE} = \sqrt{\frac{\sum_{i=1}^{100} \delta(m_i>0)\left[\ln(m_i) - \ln(\hat{m}_i)\right]^2}{ \sum_{i=1}^{100} \delta(m_i>0)}}
\eeq
and the chi-squared test statistic, defined as
\beq
\chi^2 =  \frac{(\sum_{i=50}^n m_i-\sum_{i=50}^n\hat{m}_i)^2}{\sum_{i=50}^n\hat{m}_i}+\sum_{i=1}^{49}\frac{(m_i-\hat{m}_i)^2}{\hat{m}_i} .
\eeq
The RMSE and chi-squared test statistic measure the distances between the observed population FoF vector and the extrapolated FoF vector in the logarithmic and original scales, respectively. 
Examining the trace plots of the inferred model parameters, % and performance measurements, 
we find that 1000 MCMC iterations are sufficient for both the Pitman-Yor and generalized NB process, as the Markov chains appear to converge fast and mix well in all experiments. We provide example trace plots for three different datasets in Figures \ref{fig:MCMC_tomsawyer}-\ref{fig:MCMC_microdata} of Appendix A.

\begin{figure}[!tb]
 
\begin{center}
\includegraphics[width=0.88\columnwidth]{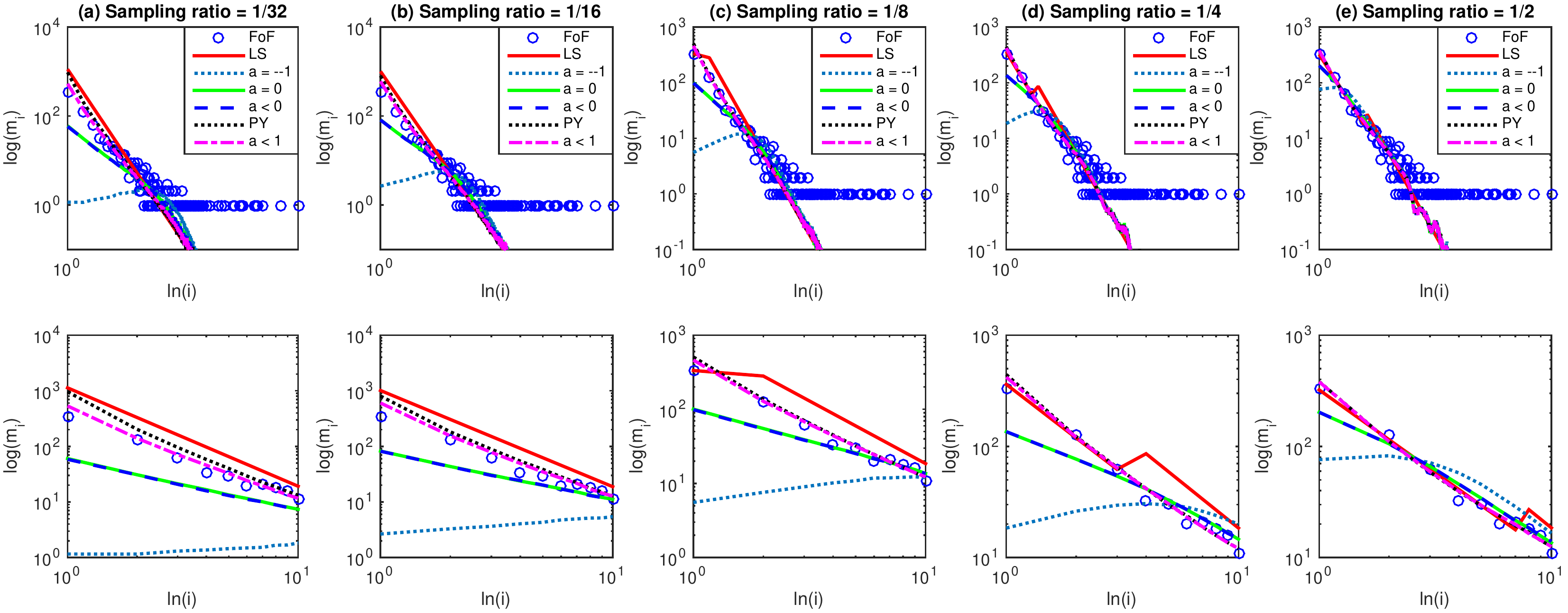}
\end{center}
\vspace{-6mm}
\caption{ \label{fig:microdata_FoF_extrapolate} \small Analogous plots to Figure \ref{fig:tomsawyer_FoF_extrapolate} for the microdata provided in Table A.6 of \citet{greenberg1990geographic}. 
}%\vspace{-2mm}
%\end{figure}

 \begin{center}
\includegraphics[width=0.6\columnwidth]{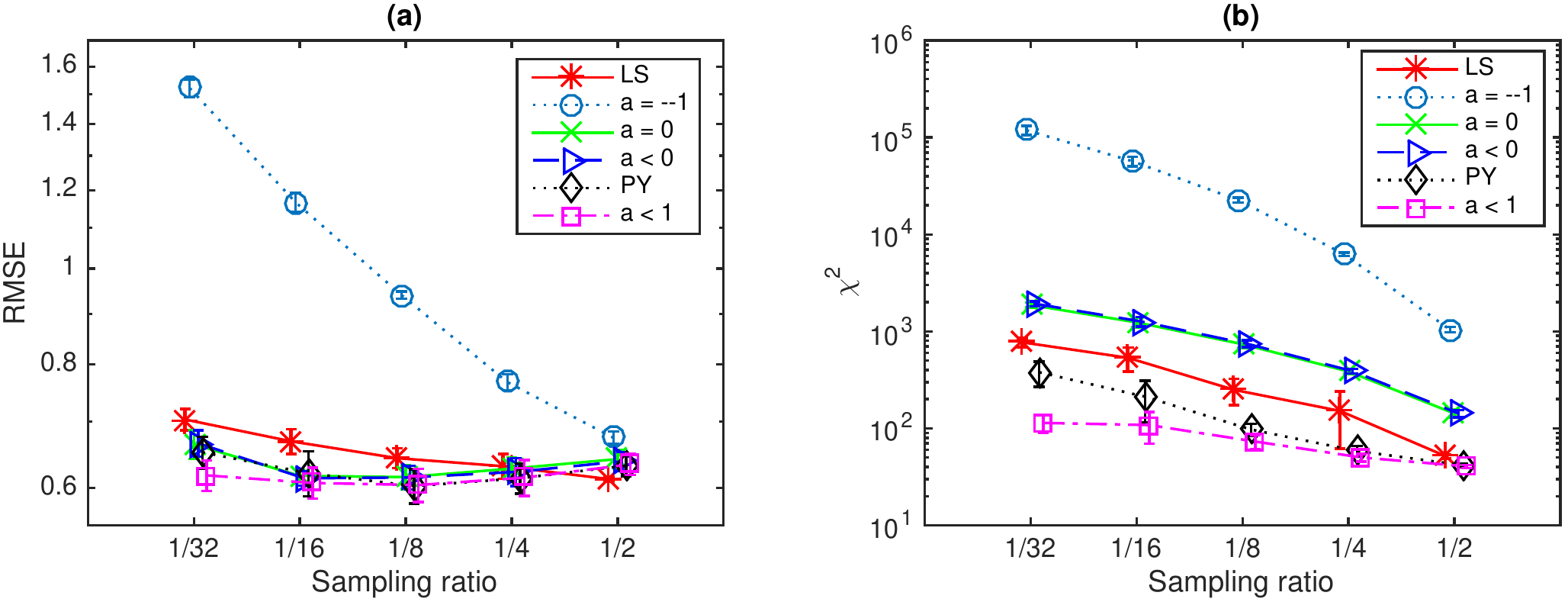}
\end{center}
\vspace{-6mm}
\caption{ \label{fig:microdata_extrapolate} \small 
Analogous plots to Figure \ref{fig:tomsawyer_extrapolate} for the microdata provided in Table A.6 of \citet{greenberg1990geographic}. 
%(a) RMSE and (b) chi-squared ($\chi^2$) test statistic 
%for the FoF vector of the population extrapolated from that of a sample, which is  taken without replacement from the population, with a sampling ratio of $1/32$, $1/16$, $1/8$, $1/4$, or $1/2$. The discount parameter is set as $a=-1$, $a=0$, $a=0.5$,  $a\in[0,1)$, or $a\in(-\infty,1)$.
}

%\begin{figure}[!tb]

\end{figure}

Shown in Figure \ref{fig:tomsawyer_FoF_extrapolate} are the posterior means of the population FoF vectors extrapolated from  sample FoF vectors for ``The Adventures of Tom Sawyer'' by Mark Twain, using least squares (LS) lines fitted to the population FoF points on the log-log plots, using the Pitman-Yor process, or using  the generalized negative binomial process under various settings of the discount parameter $a$. 
 Shown in Figure \ref{fig:tomsawyer_extrapolate}  are the corresponding  RMSEs and chi-squared test statistics. 
 Note that the slopes of these LS lines are estimated from the sample FoF vectors, whereas the intercepts are obtained by refitting these straight lines to the population FoF vectors. % under the LS criterion. 
 Thus the LS procedure is appropriate for fitting the data but impractical for out-of-sample prediction. The results of the Chinese restaurant process are almost identical to these of the generalized negative binomial process with $a=0$, and hence are omitted from these figures. 
 % under various settings for the extrapolated population FoF vectors for ``The Adventures of Tom Sawyer''. 
Figures \ref{fig:sultan_FoF_extrapolate}-\ref{fig:sultan_extrapolate} are analogous plots to Figures \ref{fig:tomsawyer_FoF_extrapolate}-\ref{fig:tomsawyer_extrapolate}
for a high-throughput RNA-seq data studied in \citet{Sultan15082008}, and Figures \ref{fig:microdata_FoF_extrapolate}-\ref{fig:microdata_extrapolate} are analogous plots to Figures \ref{fig:tomsawyer_FoF_extrapolate}-\ref{fig:tomsawyer_extrapolate}
for a microdata. In Appendix A, we also provide corresponding  Figures \ref{fig:holmes_FoF_extrapolate}-\ref{fig:holmes_extrapolate} for ``The Adventures of Sherlock Holmes'' by Arthur Conan Doyle, %are analogous plots to Figures \ref{fig:tomsawyer_FoF_extrapolate}-\ref{fig:tomsawyer_extrapolate}
 and Figures \ref{fig:core_FoF_extrapolate}-\ref{fig:core_extrapolate}  for a high-throughput RNA-seq data studied in \citet{Core19122008}.
% are analogous plots to Figures \ref{fig:tomsawyer_FoF_extrapolate}-\ref{fig:tomsawyer_extrapolate}

As shown in Figures \ref{fig:tomsawyer_FoF_extrapolate}-\ref{fig:microdata_extrapolate} and Figures \ref{fig:holmes_FoF_extrapolate}-\ref{fig:core_extrapolate} of Appendix A, the LS refitting procedure,  impractical for real applications, consistently underperforms both the Pitman-Yor process and the gNBP with $a<1$, and may  perform poorly if the population FoF vector appears to follow a decreasing concave curve. The gNBP with $a=-1$ appears to strongly discourage the frequencies of small-size clusters. Although it has poor performance for all the data considered in the paper, it shows that $a=-1$ or even smaller values could be used  for certain applications that favor the population FoF vector to follow a concave shape. Both the gNBP with $a=0$, %whose performance, as expected,  is 
with almost identical performance to that of the Chinese restaurant process, and the gNBP with $a<0$ perform well on both RNA-seq genomic data, each of whose population FoF vectors clearly follows a decreasing concave curve, but clearly underperform both the Pitman-Yor process and gNBP with $a<1$ on the other three datasets, whose population FoF vectors more closely follow decreasing straight  lines. The Pitman-Yor process performs well for all datasets, but in general clearly underperforms the gNBP with $a<1$.
%For ``The Adventures of Tom Sawyer'', as shown in Figures \ref{fig:tomsawyer_FoF_extrapolate}-\ref{fig:tomsawyer_extrapolate}, the population FoF vector could be well modeled by fixing $a=0.5$, inferring $a$ from $[0,1)$, or inferring $a$ from $(-\infty,1)$; for the RNA-seq data, as shown in Figures \ref{fig:sultan_FoF_extrapolate}-\ref{fig:sultan_extrapolate}, the population FoF vector could be well modeled by fixing $a=0$, inferring $a$ from $(-\infty,0)$, or inferring $a$ from $(-\infty,1)$; whereas for the microdata, as shown in Figures \ref{fig:microdata_FoF_extrapolate}-\ref{fig:microdata_extrapolate}, the population FoF vector could be well modeled by inferring $a$ from $[0,1)$ or inferring $a$ from $(-\infty,1)$. 
In addition to the five datasets, we have also examined the other three datasets shown in Figure \ref{fig:FoF}. % and two additional novels. 
Our observations on all these datasets % shown  in Figure \ref{fig:FoF} 
consistently suggest that  choosing the gNBP, with $a$ vary freely within $(-\infty,1)$, achieves the performance that is either the best or close to the best, which is hence recommended as the preferred choice, if there is no clear prior information on how the population FoF vector is distributed. % if there is no clear prior information on the range of $a$. 

\vspace{-4mm}\section{Conclusions} \label{sec:conclusion}
%\normalsize
We propose an infinite product of Poisson density functions to model the entire frequency of frequencies (FoF) distribution of a population consisting of a random number of individuals, and propose a size dependent exchangeable random partition function to model the FoF distribution of a population whose number of individuals is given.  We first present a general framework that uses a completely random measure mixed Poisson process to support a FoF distribution,  and then focus on studying the generalized negative binomial process constructed by mixing the generalized gamma process with the Poisson process. Our asymptotic analysis shows how the generalized negative binomial process can adjust its discount parameter to model different %power-law
tail behaviors for %the tails of the
the FoF distributions. 
%on the number and sizes of clusters show that  
%shows that 
On observing  a single sample taken without replacement from a population, we propose a simple Gibbs sampling algorithm to extrapolate the FoF vector of the population from the FoF vector of that sample. The performance  of the algorithm is demonstrated in estimating  FoF vectors for text corpora, high-throughput sequencing data, and microdata, where a population typically consists of tens of thousands or millions of individuals. Since various kinds of  statistics commonly used to characterize the properties of a population can often be readily calculated given the population FoF vector, being able to accurately model the FoF distributions of big datasets brings new opportunities to advance the state-of-the-art of a wide array of real discrete data applications, % involving millions of individuals, 
such as  making comparisons between different text corpora, finding a good compromise between the depth and coverage of high-throughput sequencing for genomic data, estimating entropy in a nonparametric Bayesian manner, and assessing disclosure risk for microdata.
%A size dependent species model is introduced to  allow flexible modeling of species abundance frequency count data. We gain this flexibility with a simple model and consequently posterior inference via MCMC is also simple. The paper provides a framework to jointly model a single random count and its exchangeable random partition. 
%It is natural to extend the same framework  to mixture modeling, where the usual task is to partition a set of data points into exchangeable clusters, where both the number and sizes of clusters are unknown and need to be inferred.  The techniques developed here to model a random count vector also serve as the foundation for  \citet{NBP_CountMatrix} to construct a family of nonparametric Bayesian priors for infinite random count matrices, and for \citet{BNBP_EPPF} to define a prior distribution that describes the random partition of a count vector into a latent random count matrix. 

%%
\vspace{-4mm}
\section*{Acknowledgements} 
\vspace{-3mm}
The authors thank the Associate Editor and three anonymous referees, whose invaluable
comments and suggestions have helped us to improve the paper substantially. 
M. Zhou  thanks  Lawrence Carin, Fernando A. Quintana, Peter M\"uller  for their comments on an earlier draft of this paper, and thanks 
Xiaoning Qian and Siamak Zamani Dadaneh for discussions on high-throughput sequencing count data.
%and
% thanks 
%Texas Advanced Computing Center (TACC) for computational support. 
S. G. Walker is supported by  the U. S. National Science Foundation through grant DMS-1506879.  S. Favaro is supported by the European Research Council (ERC) through StG N-BNP 306406.
%The authors are grateful to the Editor, an Associate Editor and two anonymous referees for their invaluable comments and suggestions that help significantly improve the paper.  
%M. Zhou thanks  Lawrence Carin, Fernando A. Quintana and  Peter M\"uller  for their helpful comments and suggestions on an early draft of this paper. The authors thank Xiaoning Qian for his helpful comments on high-throughput sequencing data. %helpful 
%%%discussions. 
%\newpage
%\small
%\singlespacing

%\end{spacing}

\begin{spacing}{1.03}
\small
\vspace{-2mm}
\bibliographystyle{plainnat}
\bibliography{References052016.bib}

%\bibliography{/Users/zhoum/Dropbox/WorkingPapers/DeepPoGamma/References102014.bib,References052016.bib}
\end{spacing}

\normalsize

\newpage
\appendix
\numberwithin{figure}{section}
\numberwithin{equation}{section}
\begin{center}
\Large{Frequency of Frequencies Distributions and Size Dependent Exchangeable Random Partitions: Supplementary Material}
\end{center}
\section{Additional figures}

\begin{figure}[!h]
 
 \begin{center}
\includegraphics[width=0.88\columnwidth]{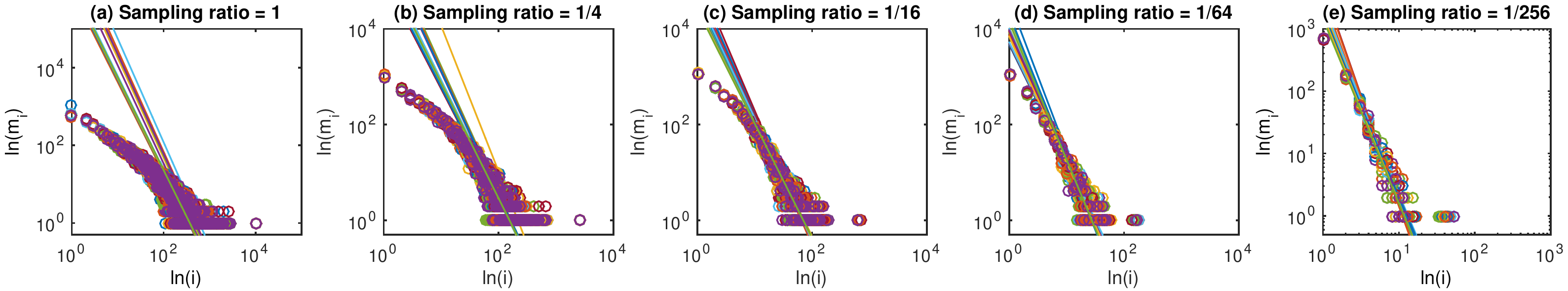}
\end{center}
\vspace{-6mm}
\caption{ \label{fig:sultan} \small Analogous plots to Figure \ref{fig:tomsawyer} for the frequency of frequencies (FoF) vectors for the RNA sequences of a high-throughput sequencing sample %of the human transcriptome from a B cell line 
studied in \citep{Sultan15082008}. 
}

\begin{center}
\includegraphics[width=0.6\columnwidth]{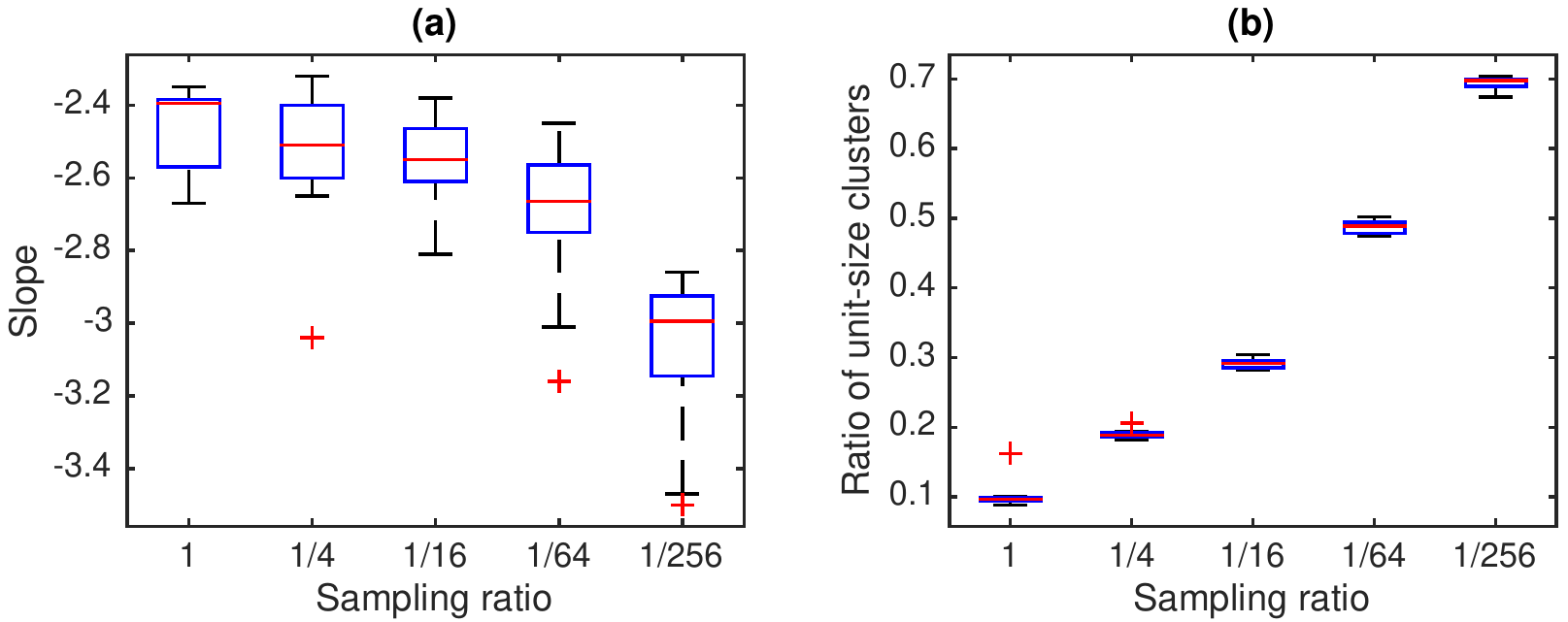}
\end{center}
\vspace{-6mm}
\caption{ \label{fig:sultan_alpha} \small Analogous plots to Figure \ref{fig:tomsawyer_alpha} for the frequency of frequencies (FoF) vectors for the RNA sequences of a high-throughput sequencing sample %of the human transcriptome from a B cell line 
studied in \citep{Sultan15082008}. 
}

\end{figure}

\begin{figure}[!h]
 
\begin{center}
\includegraphics[width=0.65\columnwidth]{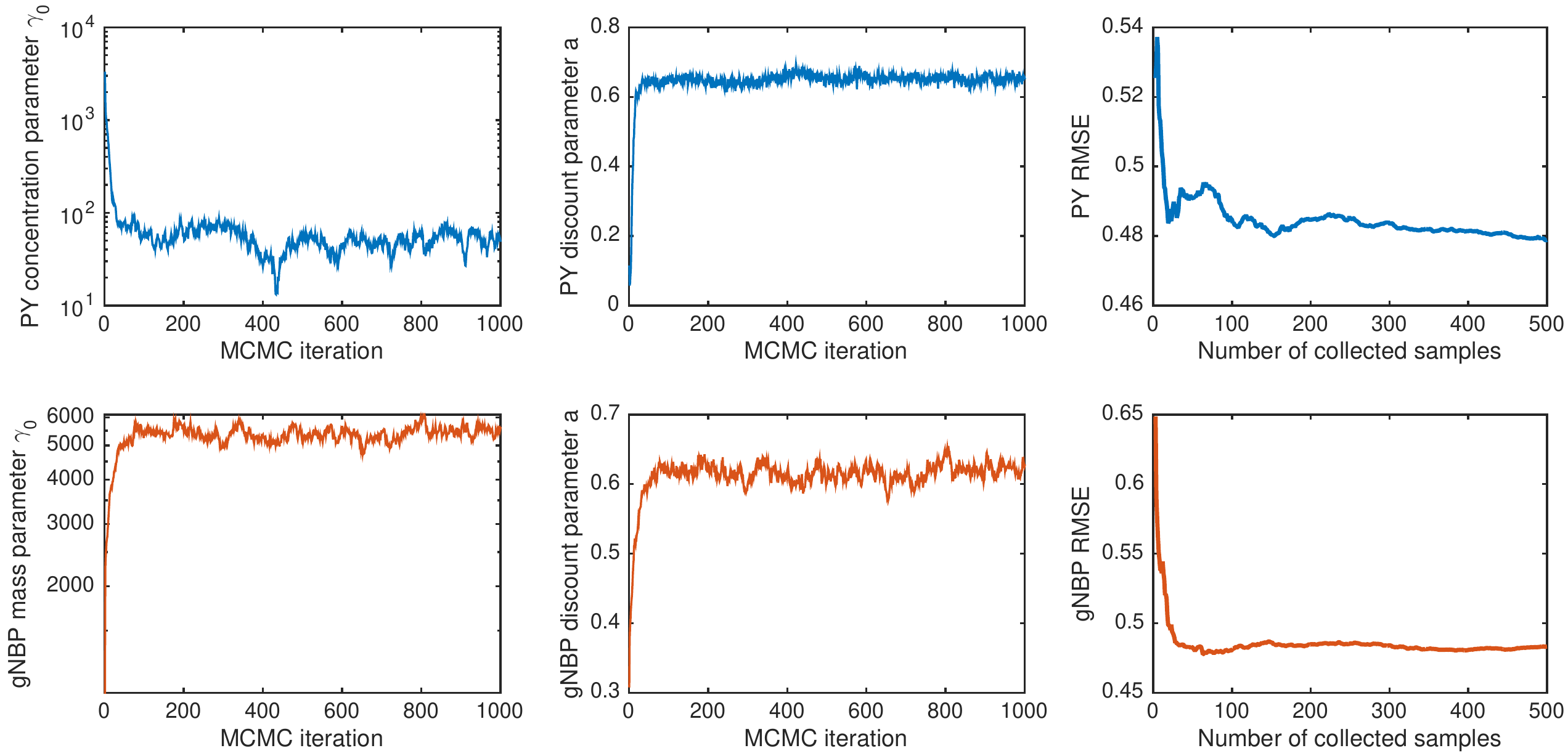}
\end{center}
\vspace{-6mm}
\caption{ \label{fig:MCMC_tomsawyer} \small For  ``The Adventures of Tom Sawyer'' by Mark Twain, with a sampling ratio of  $1/8$, the trace plots in the first row %from the left to right 
are for the concentration parameter $\gamma_0$, discount parameter $a$, and RMSE, respectively, for the Pitman-Yor (PY) process; the trace plots in the second row %from the left to right 
are  for the mass parameter $\gamma_0$, discount parameter $a$, and RMSE, respectively, for the generalized negative binomial process (gNBP) with $a$ varying freely within $(-\infty,1)$.
}%\vspace{-2mm}
\end{figure}
\begin{figure}[!h]
\begin{center}
\includegraphics[width=0.65\columnwidth]{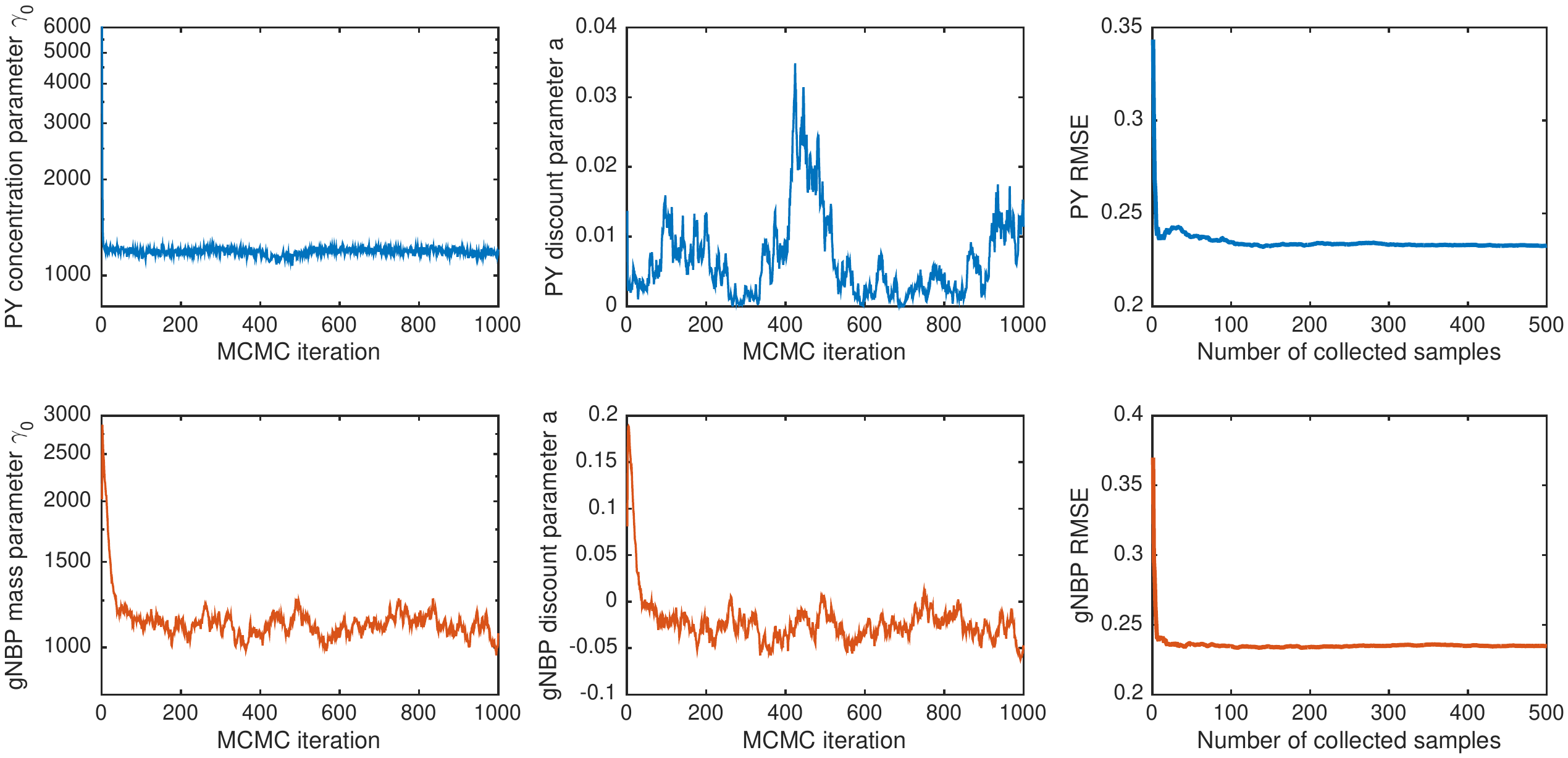}
\end{center}
\vspace{-6mm}
\caption{ \label{fig:MCMC_sultan} \small Analogous plots to Figure \ref{fig:MCMC_sultan} for a RNA-seq data studied in \citet{Sultan15082008}, with a sampling ratio of  $1/8$. 
}
\end{figure}
\begin{figure}[!h]
\begin{center}
\includegraphics[width=0.65\columnwidth]{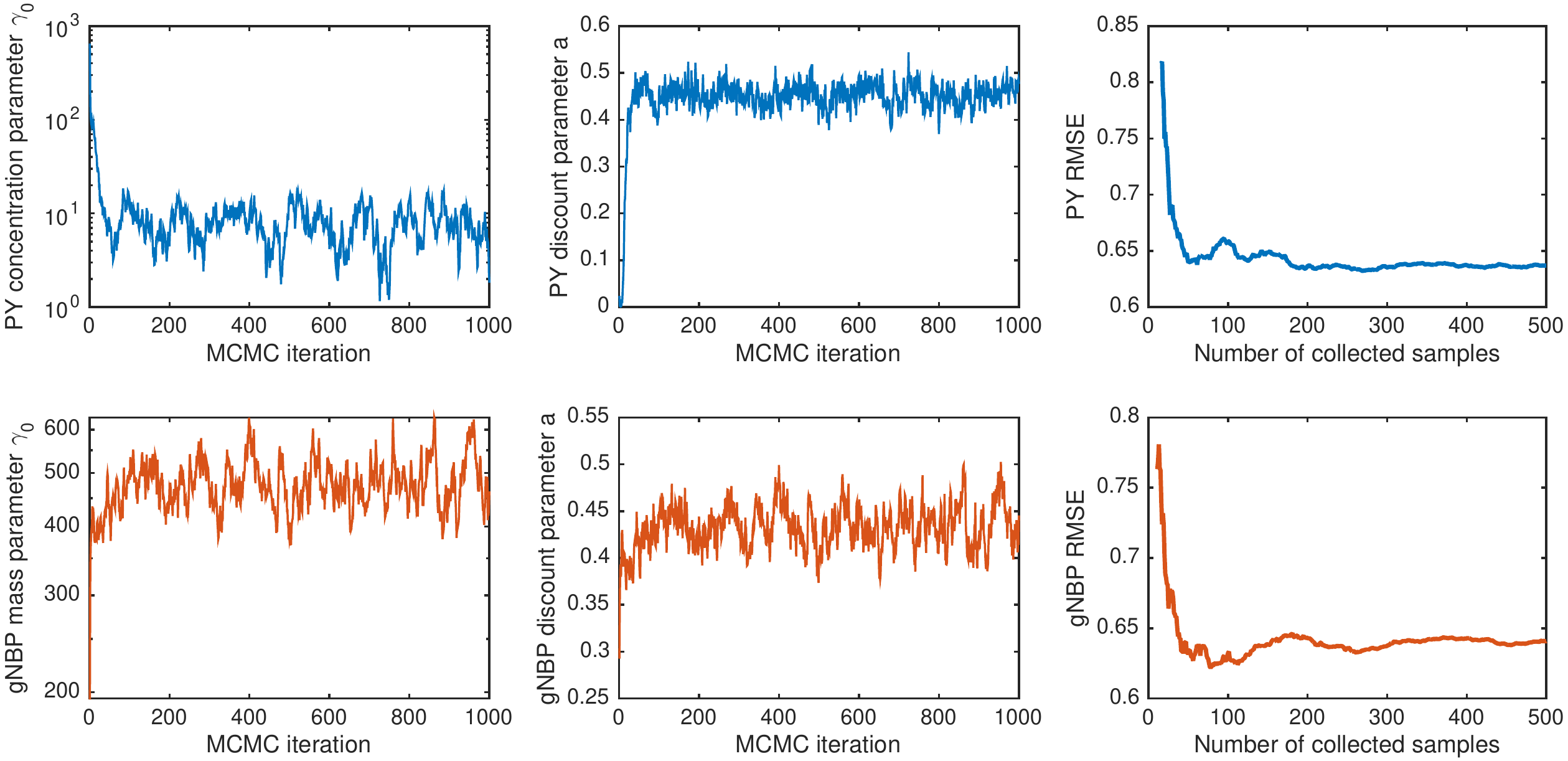}
\end{center}
\vspace{-6mm}
\caption{ \label{fig:MCMC_microdata} \small Analogous plots to Figure \ref{fig:MCMC_tomsawyer} for the Microdata provided in Table A.6 of \citet{greenberg1990geographic}, with a sampling ratio of  $1/8$. 
}

\end{figure}

\begin{figure}[!h]
 
\begin{center}
\includegraphics[width=0.88\columnwidth]{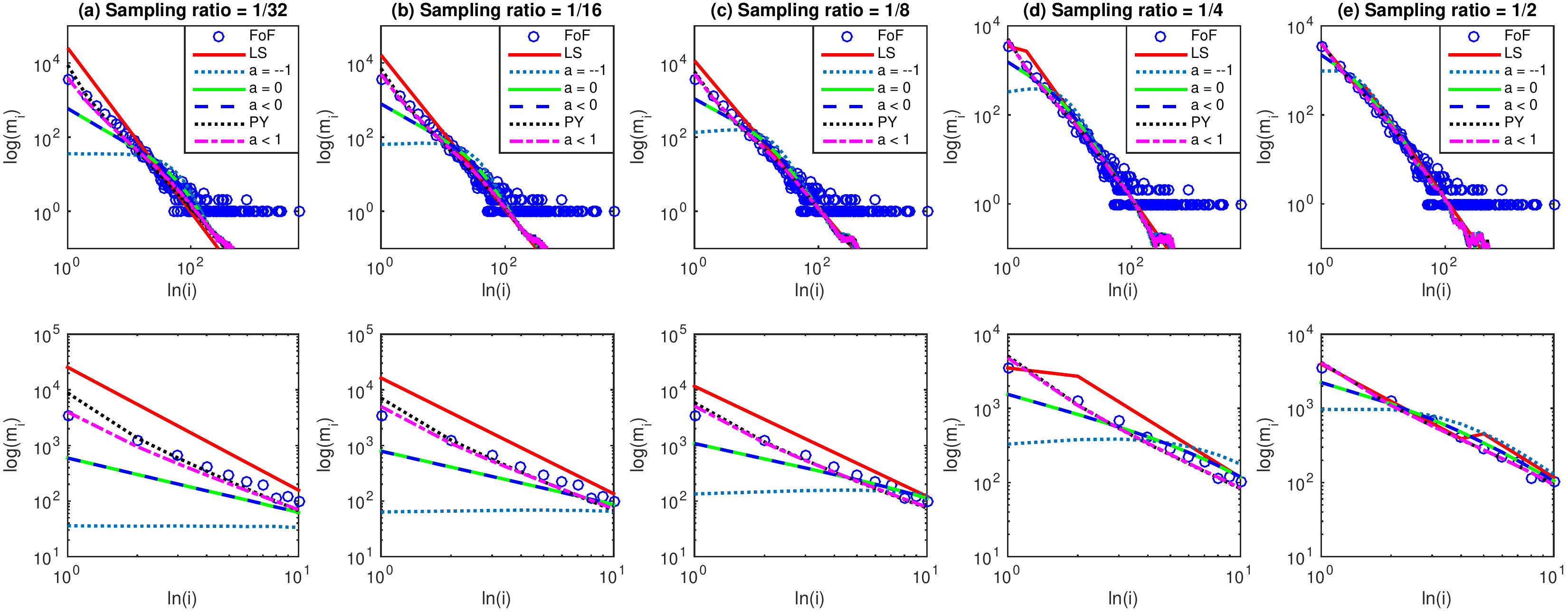}
\end{center}
\vspace{-6mm}
\caption{ \label{fig:holmes_FoF_extrapolate} \small Analogous plots to Figure \ref{fig:tomsawyer_FoF_extrapolate} for the novel ``The Adventures of Sherlock Holmes'' by Arthur Conan Doyle. 
}%\vspace{-2mm}
\end{figure}

\begin{figure}[!h]
 \begin{center}
\includegraphics[width=0.6\columnwidth]{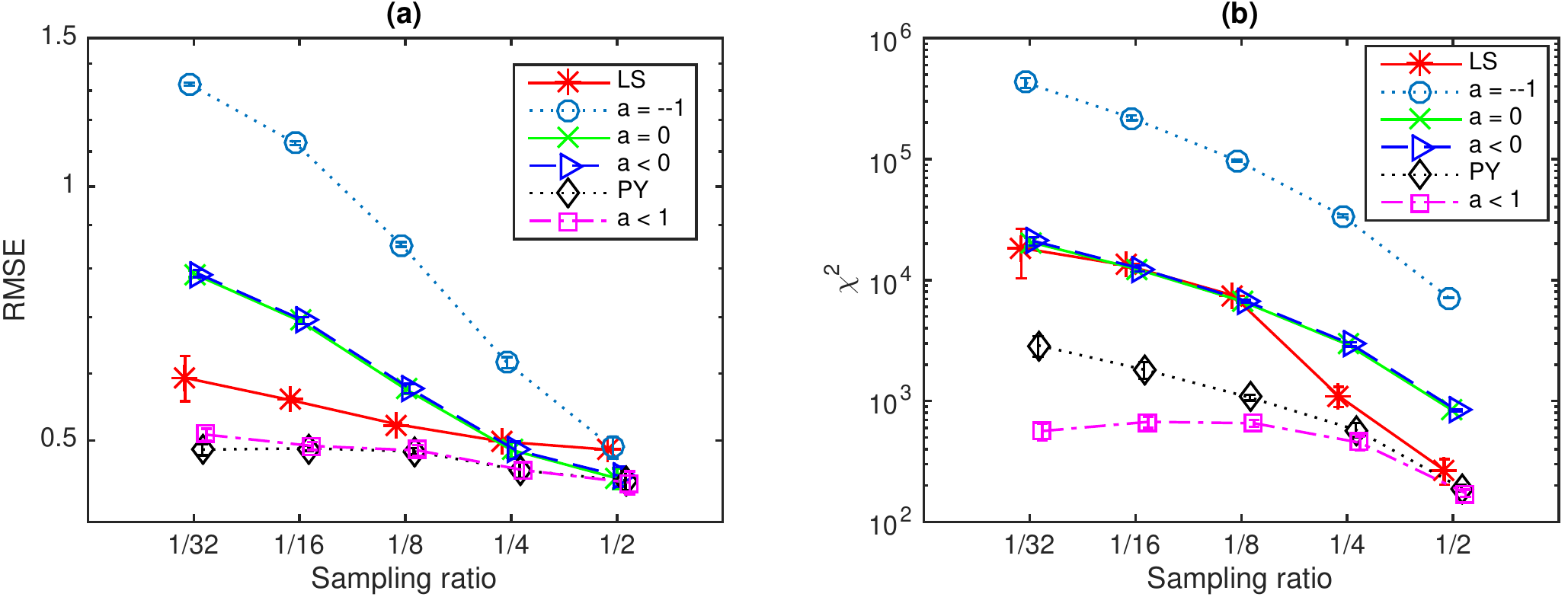}
\end{center}
\vspace{-6mm}
\caption{ \label{fig:holmes_extrapolate} \small 
Analogous plots to Figure \ref{fig:tomsawyer_extrapolate} for the novel ``The Adventures of Sherlock Holmes'' by Arthur Conan Doyle. 
%(a) RMSE and (b) chi-squared ($\chi^2$) test statistic 
%for the FoF vector of the population extrapolated from that of a sample, which is  taken without replacement from the population, with a sampling ratio of $1/32$, $1/16$, $1/8$, $1/4$, or $1/2$. The discount parameter is set as $a=-1$, $a=0$, $a=0.5$,  $a\in[0,1)$, or $a\in(-\infty,1)$.
}

%\begin{figure}[!tb]

\end{figure}

\begin{figure}[!h]
 
\begin{center}
\includegraphics[width=0.88\columnwidth]{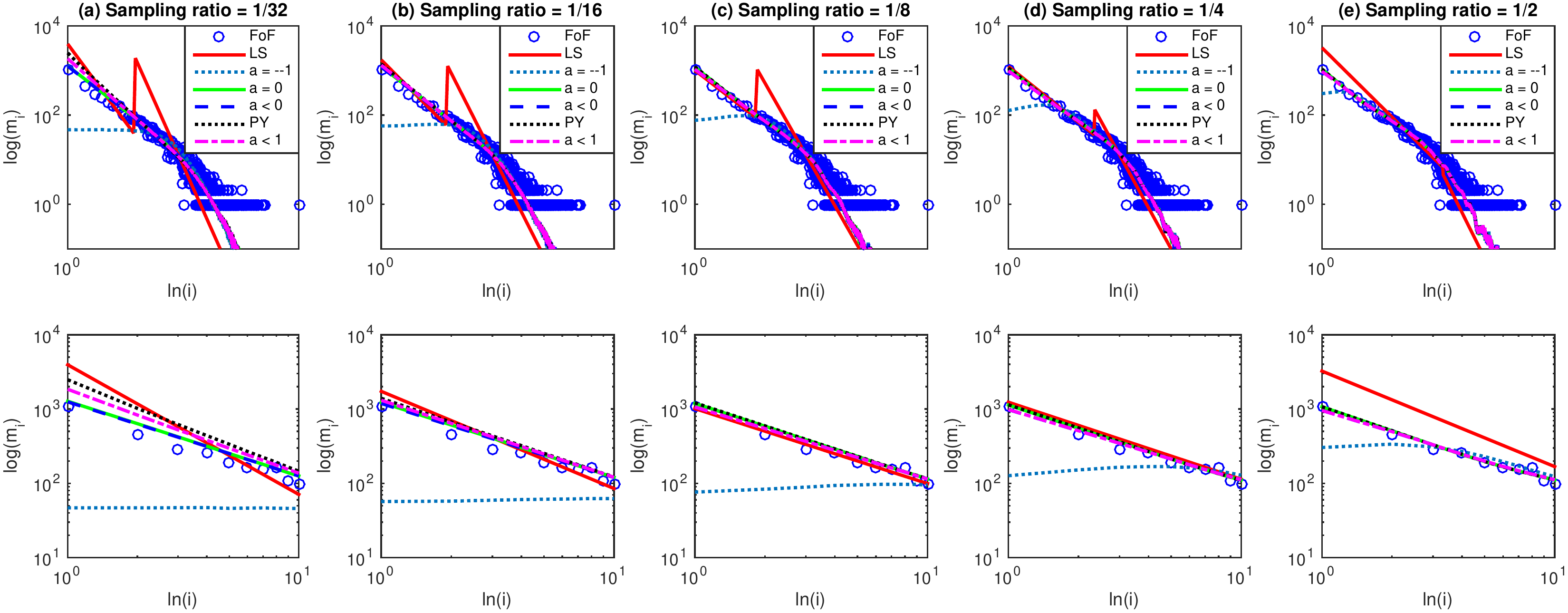}
\end{center}
\vspace{-6mm}
\caption{ \label{fig:core_FoF_extrapolate} \small Analogous plots to Figure \ref{fig:tomsawyer_FoF_extrapolate} for a RNA-seq  data studied in \citet{Core19122008}. 
}%\vspace{-2mm}
\end{figure}

\begin{figure}[!h]
 \begin{center}
\includegraphics[width=0.6\columnwidth]{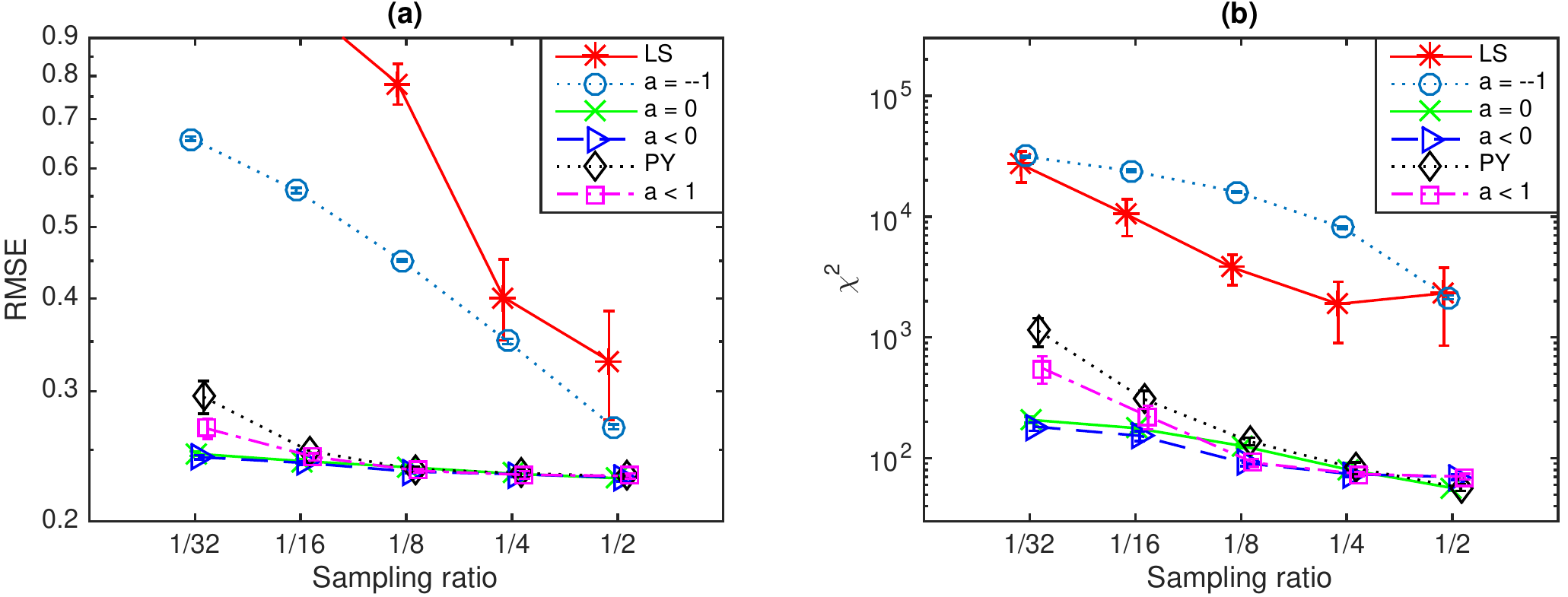}
\end{center}
\vspace{-6mm}
\caption{ \label{fig:core_extrapolate} \small 
Analogous plots to Figure \ref{fig:tomsawyer_extrapolate} for a RNA-seq  data studied in \citet{Core19122008}. 
%(a) RMSE and (b) chi-squared ($\chi^2$) test statistic 
%for the FoF vector of the population extrapolated from that of a sample, which is  taken without replacement from the population, with a sampling ratio of $1/32$, $1/16$, $1/8$, $1/4$, or $1/2$. The discount parameter is set as $a=-1$, $a=0$, $a=0.5$,  $a\in[0,1)$, or $a\in(-\infty,1)$.
}

%\begin{figure}[!tb]

\end{figure}

\section{Characterizing the tails of FoF distributions % power-law distributions
%Estimation of power-law scaling parameter
}

% \vspace{-2mm}\subsection{Modeling  the tail of a 
 %frequency of frequencies distribution}
As in \citet{newman2005power}, to model the tail of a FoF distribution that follows a power law, one may  define a probability mass function for the class sizes as $$P(n_k=i) ={i^{-\alpha}}\big/{\zeta(\alpha,i_{\min}) }, ~~i\in\{i_{\min},i_{\min}+1,\ldots\},$$ where $i_{\min}$ is the cutoff integer  which one considers as  the starting point for the power law, and $\zeta(\alpha,i_{\min})  = \sum_{j=i_{\min}}^\infty j^{-\alpha} $  is the Hurwitz zeta function. Thus, given %the number of clusters with sizes no less than  $i_{\min}$ as
 $K^*=\sum_{i=i_{\min}}^n m_i$, one has $\E[m_i] = K^* P(n_k=i)$ and hence $\ln(\E[m_i]) =  -\alpha \ln(i) + C$ for $i\in\{i_{\min},i_{\min}+1,\ldots\} $, where $C$ is a constant not related to $i$. To estimate the scaling parameter $\alpha$ for  a finite population  of $n$ individuals, a straightforward approach is to plot $\ln(m_i)$ against $\ln(i)$, % for $i\in\{i:m_i>0, i\ge i_{\min}\} $, 
 and then estimate $-\alpha$ using the slope of a straight line fitted to the points on the plot. This simple approach is criticized in \cite{clauset2009power}, who suggest estimating  $\alpha$ by maximizing the likelihood
$%\beq
\mathcal{L}(\alpha) = %- K^* \ln \zeta(\alpha,i_{\min}) - \alpha \sum_{k:n_k\ge i_{\min}} \ln(n_k) =    
- \sum_{i=i_{\min}}^n m_i\left[\ln \zeta(\alpha,i_{\min})+\alpha\ln(i)\right]. \notag
$
For each subfigure in Figure \ref{fig:FoF}, we use the software\footnote{\href{http://tuvalu.santafe.edu/~aaronc/powerlaws/}{http://tuvalu.santafe.edu/$\sim$aaronc/powerlaws/}} provided for \cite{clauset2009power} to estimate both the power-law lower cutoff point $i_{\min}$ and the scaling parameter $\alpha$, and fit a straight line to the FoF points on the loglog plot using $-\alpha$ as the slope and $\left[\sum_{i \in I}\ln(m_i) +\alpha \sum_{i\in I}\ln i \right]/|I|
$, where $I=\{i:i\ge i_{min}, m_i\ge 3 \}$, as the intercept. 
% and using $\left[\sum_{i \in I}\ln(m_i) +\alpha \sum_{i\in I}\ln i \right]/|I|
%$, where $I=\{i:i\ge i_{min}, m_i\ge 3 \}$ as the intercept. 
%; a least squares regression line is fitted to the points in the set $\{[\ln(i),\ln(m_i)]\}_{i:m_i\ge 10}$ for each FoF vector, with its slope displayed in the legend of the corresponding plot. 

\vspace{-2mm}\section{Size independent species sampling models}
%Assuming the population size $n$ is given,  using a Bayesian species sampling model, one may partition the $n$ individuals into exchangeable random partitions, and hence generate a FoF vector by defining each partition  as a class. 
The underlying structure of existing Bayesian species sampling models  is built on Kingman's concept of a partition structure \citep{kingman1978random,kingman1978representation}, which defines a family of %sequence of 
consistent probability distributions for random partitions of a set $[m]:=\{1,\ldots,m\}$. The sampling consistency requires  the  probability distribution of the random partitions of a subset of size $m$ of a set of size $n\geq m$ to be the same for all~$n$.
More specifically, for a random partition $\Pi_m=\{A_1,\ldots,A_l\}$ of the set $[m]$,  
%where there are $l$ clusters and each element $i\in[m]$ belongs to one and only one set $A_k$ from $\Pi_m$, 
such a constraint requires
that $P(\Pi_m\,|\,n)=P(\Pi_m\,|\,m)$ does not depend on  $n$. %, where $P(\Pi_m\,|\,n)$ denotes the marginal partition probability for $[m]$ when it is known the population size is~$n$. 
As further developed in \citet{pitman1995exchangeable,csp}, if %the probability for an arbitrary partition $\Pi_m$ 
$P(\Pi_m\,|\,m)$ depends only on the number and sizes of the $(A_k)$, regardless of their order, then it is called an exchangeable partition probability function (EPPF) of~$\Pi_m$, expressed as  $P(\Pi_m=\{A_1,\ldots,A_l\}\,|\,m)=p_m(n_1,\ldots,n_l)$, where $n_k=|A_k|$. 
The sampling consistency  %of $P(\Pi_m)$ as $m$ varies %
amounts to an addition rule \citep{csp,Gnedin_deletion} for %
the EPPF; that $p_1(1) = 1$ and
\beqs\label{eq:addrule}
p_m(n_1,\ldots,n_l) =  p_{m+1}(n_1,\ldots,n_l,1)+ \sum_{k=1}^l p_{m+1}(n_1,\ldots,n_k+1,\ldots,n_l).
\eeqs
An EPPF of  $\Pi_m$ satisfying this constraint is considered as an EPPF of $\Pi :=(\Pi_1,\Pi_2,\ldots)$. 
For an EPPF of $\Pi$, 
 $\Pi_{m+1}$ can be constructed from  $\Pi_m$ by assigning element $(m+1)$ to $A_{z_{m+1}}$ based on the prediction rule as
\beq\notag
z_{m+1}\,|\,\Pi_m=
\begin{cases} \vspace{3mm}
l+1& \mbox{with probability  }\displaystyle\frac{p_{m+1}(n_1,\ldots,n_l,1)}{p_m(n_1,\ldots,n_l)} , \\ 
k & \mbox{with probability  }\displaystyle\frac{p_{m+1}(n_1,\ldots,n_k+1,\ldots,n_l)}{p_m(n_1,\ldots,n_l)}.\end{cases}
\eeq

%Note that the conventional definition of EPPF often requires that the probability distribution of the random partitions of a subset of size $m$ of a set of size $n\ge m$ to be the same for all $n$, which means $P(\Pi_m\,|\,m) = P(\Pi_m\,|\,n)$ for all $n\ge m$ \citep{csp}. We refer to an EPPF satisifying this constraint as a size independent EPPF.  

A basic  EPPF of $\Pi$ %satisfying this constraint % of $\Pi$
 is the Ewens sampling formula \citep{ewens1972sampling,Antoniak74}. Moving beyond the Ewens sampling formula, 
various  approaches, including the Pitman-Yor process 
 \citep{perman1992size,pitman1997two}, normalized random measures with independent increments (NRMIs)
\citep{regazzini2003distributional}, Poisson-Kingman models \citep{pitman2003poisson}, species sampling \citep{Pitman96somedevelopments}, %,lee2013}, 
  stick-breaking priors \citep{ishwaran2001gibbs}, and Gibbs-type random partitions \citep{gnedin2006exchangeable},   have been proposed to construct more general size independent EPPFs. % under this constraint. % of $\Pi$.  
See \citet{muller2004nonparametric}, \citet{BeyondDP} and \citet{Muller2013} for reviews. 
 %We provide more details on size independent EPPFs %species sampling models
 %in the Appendix. 

Among these approaches,
there has been increasing interest in 
normalized random measures with independent increments (NRMIs)
\citep{regazzini2003distributional}, %,lijoi2005inverseGaussian,lijoi2007controlling,james2009posterior}, 
where a completely random measure \citep{Kingman,PoissonP} with a finite and strictly positive total random mass is normalized to construct a random probability measure. For example, the normalized gamma process is a Dirichlet process \citep{ferguson73}. %, the marginalization of which leads to a variant of the Ewens sampling formula \citep{DP_Mixture_Antoniak}. 
More advanced completely random measures, such as the generalized gamma process of \citet{brix1999generalized}, can be employed to produce more general size-independent exchangeable random partitions %  of $\Pi$ 
\citep{pitman2003poisson,csp,lijoi2007controlling}. However,  the expressions of the EPPF and its associated prediction rule usually involve integrations that are difficult to calculate.

\vspace{-2mm}\section{Completely random measures}\label{sec:preliminary}

In this section we provide the mathematical foundations for an independent increment process with no Gaussian component.  These are pure jump processes and for us will have finite limits so that the process can be normalized by the total sum of the jumps to provide a random distribution function.
The most well known of such processes is the gamma process (see, for example, \citet{ferguson1972representation}) and we will be specifically working with a generalized gamma process in Section \ref{sec:ggp}.

\vspace{-2mm}\subsection{Generalized gamma process}\label{sec:ggp}

%In this section we provide the mathematical foundations for an independent increment process with no Gaussian component.  These are pure jump processes and for us will have finite limits so that the process can be normalized by the total sum of the jumps to provide a random distribution function.
%The most well known of such processes is the gamma process (see, for example, \citet{ferguson1972representation}) and we will be specifically working with a generalized gamma process in Section \ref{sec:ggp}.
%\vspace{-2mm}\subsection{Generalized gamma process}\label{sec:ggp}
The generalized gamma process,   denote by $G\sim\mbox{g}\Gamma\mbox{P}(G_0,a,1/c)$, is a completely random (independent increment) measure defined on the product space $\mathbb{R}_+\times \Omega$, where  $a< 1$ is a discount parameter, $1/c$ is a scale parameter, and $G_0$ is a finite and continuous base measure over a complete separable metric space $\Omega$ \citep{brix1999generalized}.  
It assigns independent infinitely divisible generalized gamma ($\mbox{g}\Gamma$)  distributed random variables $G(A_j)\sim{{}}\mbox{g}\Gamma(G_0(A_j),a,1/c)$ to disjoint Borel sets $A_j\subset \Omega$, %. For each subset $A\subset\Omega$, $G(A)$ follows a generalized gamma distribution,
with Laplace transform given by
\beq\label{eq:Laplace}
%L_{a,c,G_0(A)}(s)
\E\left[e^{-\phi\,G(A)}\right] = \exp\left\{-\frac{G_0(A)}{a}\left[(c+\phi)^a-c^a\right]\right\}.
\eeq
%The L\'{e}vy measure of the generalized gamma process
%$G\sim{{}}\mbox{g}\Gamma\mbox{P}(a,1/c,G_0)$
%can be expressed as
%\beqs\label{eq:LevyGGP}
%\nu(dr d\omega) = \frac{1}{\Gamma(1-a)}r^{-a-1}e^{-cr}\,dr \,G_0(d\omega).
%\eeqs
%The connection between (\ref{eq:Laplace}) and (\ref{eq:LevyGGP}), not given here, is the well known form for the Laplace transform of an infinitely divisible random variable. 
The generalized gamma distribution %distribution $x\sim{{}}\mbox{g}\Gamma\mbox{P}(\gamma,a,1/c)$, %with $\E[x]=\gamma c^{a-1}$, 
 was independently suggested by \citet{tweedie1984index} and \citet{hougaard1986survival} and also studied in \citet{bar1986reproducibility,alen1992modelling}, and \citet{jorgensen1997theory}.
%As $a\rightarrow 0$, since $\lim_{a\rightarrow 0}\frac{1-(1-p)^a}{ap^a} = -\ln(1-p)$ and hence $ %\beq
%L_{a,c,G_0(A)}(s)
%\lim_{a\rightarrow 0}\E[e^{-sx}] = (1+ s/c)^{-\gamma}
%$, the generalized gamma distribution  becomes a gamma distribution  $x\sim\mbox{Gamma}(\gamma,1/c)$, with $\E[x]=\gamma/c$. A generalized gamma distribution scaled with a positive constant $\beta>0$ is distributed as $\beta x\sim\mbox{g}\Gamma\mbox{P}(\gamma\beta^a,a,\beta/c)$.
 
When $a\rightarrow0$, we recover the gamma process \citep{ferguson73,PoissonP},   and if $a=1/2$, we recover the inverse Gaussian process \citep{lijoi2005inverseGaussian}. 
 A draw $G$ from %the generalized  gamma process
$\mbox{g}\Gamma\mbox{P}(G_0,a,1/c)$ %consists of countably infinite points,
 can be expressed as
\beq
G = \sum_{k=1}^{K} r_k \delta_{\omega_k},\notag \eeq
with $K\sim\mbox{Poisson}(\nu^+)$ and $(r_k,\omega_k)\stackrel{i.i.d.}{\sim} \pi(drd\omega)$, 
where $r_k=G(\omega_k)$ is the weight for atom $\omega_k$ %, $G(\Omega)=\sum_{k=1}^{K} r_k$ is the total random mass, 
and $\pi(dr\, ,d\omega)\nu^{+} = \nu(dr\,,d\omega)$. 
Except where otherwise specified, we only 
consider $a<1$ and $c>0$. %or $0<a<1$ and $c=0$
If $0\le a<1$, since the Poisson intensity $\nu^+ = \nu(\mathbb{R}_+\times \Omega) = \infty$ ($i.e.$, $K=\infty$ a.s.) and
  $
 \int_{\mathbb{R}_+\times \Omega} \min\{1, s\} \nu(dr\, d\omega)  %= \gamma_0c^{a-1}
 $
 is finite, a draw from $\mbox{g}\Gamma\mbox{P}(G_0,a,1/c)$ consists of countably infinite atoms. On the other hand, if $a<0$, then $\nu^+=-\gamma_0c^a/a$ and thus $K\sim \mbox{Poisson}(-\gamma_0c^a/a)$ ($i.e.$, $K$ is finite a.s.) and $r_k\stackrel{i.i.d.}{\sim}\mbox{Gamma}(-a,1/c)$. %This process will be seen again in Section \ref{sec:gNBP}.

\vspace{-2mm}\subsection{Normalized random measures % with Independent Increments
}\label{NRMI}

A NRMI model \citep{regazzini2003distributional} is
%Since 
a normalized completely random measure 
$$\widetilde{G}=G/G(\Omega)$$ where 
$G(\Omega)=\sum_{k=1}^{K} r_{k}$ is the total random mass, which is required to be finite and strictly positive. 
Note that the strict positivity of $G(\Omega)$ implies that $\nu^+=\infty$ and hence $K=\infty$  a.s. \citep{regazzini2003distributional,BeyondDP}. 
For MCMC inference, 
following \citet{%nieto2004normalized,
james2009posterior}, %,griffin2011posterior%,favaromcmc
 a specific auxiliary variable  $T>0$, with %(v\,|\,m,\sum_{k=1}^{K} r_k)
 $ %\beq\label{eq:auxiliary}
p_T(t\,|\,n,G(\Omega)) =\mbox{Gamma}[n,1/G(\Omega)]
$, 
can be introduced to yield a fully factorized likelihood, stimulating the development of a number of posterior simulation algorithms including 
\citet{griffin2011posterior,barrios2012modeling}, and \citet{%griffin2011sequential,
 favaromcmc}.  
Marginalizing out $G$ and then $T$ from that fully factorized likelihood leads to an EPPF of $\Pi$ \citep{pitman2003poisson,csp,lijoi2007controlling}.  However, the prediction rule of the EPPF may not be easy to calculate.

 \section{Proofs}
%\vspace{-4mm}\subsection{Proof for Theorem \ref{thm:compoundPoisson}}
\begin{proof}[Proof for Theorem \ref{thm:compoundPoisson}]
%Here we establish the marginal model for the $(n_k)$ with $G$ integrated out. To this end 
Let us consider the process $X_G$, conditional on $G$, given by
$$X_G(A)=\sum \nolimits_{k} n_k\,\delta(\omega_k\in A).$$  
Now it is easy to see that
$$\E[\exp\{-\phi X_G(A)\}\,|\,G]=\exp\{-G(A)(1-e^{-\phi})\},$$
and using the well known result for homogeneous  L\'evy processes, we have
\beq
\E[\exp\{-\lambda G(A)\}]=\exp\left\{-G_0(A)\,\int_0^\infty \left[1-e^{-\lambda r}\right]\,\rho(dr)\right\}.\label{one}
\eeq
Now, the key observation is the following identity:
\beq
1-e^{-(1-e^{-\phi})r}=1-e^{-r}\sum_{j=0}^\infty \frac{r^j}{j!}e^{-\phi j}=(1-e^{-r})-e^{-r}\sum_{j=1}^\infty \frac{r^j}{j!}e^{-\phi j} = \sum_{j=1}^\infty \frac{r^je^{-r}}{j!}(1-e^{-\phi j}).\label{eq:Iden}
\eeq
Let us put this to one side for now and consider the model for $\tilde{X}$ given by
$$\tilde{X}(A)=\sum_{k=1}^{l} n_k\,\delta(\omega_k\in A)$$
with $l\sim\mbox{Poisson}[\gamma G_0(\Omega)]$ for some non-negative $\gamma$ and 
independently $P(n_k=j)=\pi_j$ for some $\pi_j\leq 1$ and $j\in\{1,2,\ldots\}$.
Now given $l$, we have
$$\E[ \exp\{-\phi \tilde{X}(A)\}|l]=\prod_{k=1}^{l} \E [\exp\{-\phi n_k\,\delta(\omega_k\in A)\}]$$
and each of these expectations is given by
$$\psi=\sum_{j=1}^\infty e^{-\phi j}\pi_j.$$
Thus
$$\E[\exp\{-\phi \tilde{X}(A)\}]=%\E[\psi^L]=
\exp\{-\gamma\, G_0(A)\, (1-\psi)\}$$
which is given by
\beq
\exp\left[-\gamma\,G_0(A)\, \left(1-\sum_{j=1}^\infty e^{-\phi j}\,\pi_j\right)\right].\label{two}
\eeq
Comparing (\ref{one}) and (\ref{two}) we see that we have a match when
$$\gamma=\int_0^\infty (1-e^{-r})\,\rho(dr)$$
and
$$\pi_j=\frac{\int_0^\infty r^j\,e^{-r}\,\rho(dr)}{j! \gamma}\,,$$
and note that it is easy to verify that
$$\sum_{j=1}^\infty \pi_j=1.$$
\end{proof}

%\vspace{-4mm}\section{Proof for Corollary \ref{cor:compoundPoisson}}
\begin{proof}[Proof for Corollary \ref{cor:compoundPoisson}]
Using \eqref{eq:Iden} and \eqref{two}, we have
\begin{align}
\E[\exp\{-\phi {X}(A)\}]
&=
\exp\left\{-\gamma\,G_0(A)\, \left[1-\sum_{j=1}^\infty e^{-\phi j}\,\pi_j\right]\right\}\notag\\
&=
\exp\left[-\,G_0(A)\, \int_0^\infty { \bigg(1-e^{-r}- \sum_{j=1}^\infty e^{-\phi j}\,  \frac{r^j\,e^{-r}}{j!}\bigg)} \rho (dr) \right]\notag\\
&=  \exp\left\{-G_0(A)\,  \int_0^\infty \sum_{j=1}^\infty (1-e^{-\phi j })\frac{r^j\,e^{-r}}{j!}\,\rho(dr) \right\}.\notag %\\
% & = \exp\left\{-\int_{\{1,2,\ldots\}\times A}  \, \sum_{j=1}^\infty (1-e^{-\phi j }) \int_0^\infty \frac{r^j\,e^{-r}}{j!}\,\rho(dr) \delta_j(n) \nu(dn d\omega) \right\}.\notag
\end{align}
Substituting the definition of the L\'evy measure $\nu(dnd\omega)$ in Corollary 2 into
 %$\nu(dnd\omega) = \sum_{j=1}^\infty\frac{ {\int_{0}^\infty r^j e^{-r} \rho(dr)} }{{j!}}\delta_j(dn)G_0(d\omega)$ into 
\eqref{eq:Laplace0},  we have
\begin{align}
\E[\exp\{-\phi {X}(A)\}] & = \exp\left\{-\int_{\mathbb{R}_+\times A}  \, \sum_{j=1}^\infty (1-e^{-\phi j }) \int_0^\infty \frac{r^j\,e^{-r}}{j!}\,\rho(dr)~ \delta_j(dn) G_0(d\omega) \right\}\notag\\
& = \exp\left\{-G_0(A) \, \sum_{j=1}^\infty (1-e^{-\phi j }) \int_0^\infty \frac{r^j\,e^{-r}}{j!}\,\rho(dr) \right\}.\notag
\end{align}
The proof is complete by changing the order of the summation and integration.
\end{proof}

%\vspace{-4mm}\section{Proof for Corollary \ref{cor:m_i}}
\begin{proof}[Proof for Corollary \ref{cor:m_i}]
Since $\sum_{i=1}^\infty r^{i}e^{-r}/i! = 1-e^{-r}$, we can express the  joint distribution  of $\mathcal{M}$ and the population size $n$ as
\begin{align} %\label{main_samp}
p(\mathcal{M},n\,|\,\gamma_0,\rho) &= \frac{n!}{\prod_{i=1}^n(i!)^{m_i}m_i!}p(\zv\,|\,n,\gamma_0,\rho) p_N(n\,|\, \gamma_0,\rho)\notag\\ %\E_{G} [f(\zv,n\,|\,G)] =
&=\exp\left\{\gamma_0\int_{0}^\infty(e^{-r}-1)\rho(dr)\right\}
\prod_{i=1}^{n} \left(\frac{\gamma_0\int_0^\infty r^{i} e^{-r} \rho(dr)}{i!}\right)^{m_i} \frac{1}{m_i!}\notag\\
&=\left\{\prod_{i=1}^\infty \mbox{Poisson}\left(m_i; \frac{\gamma_0\int_0^\infty r^{i} e^{-r} \rho(dr)}{i!}\right)\right\} \times \delta\left(n=\sum_{i=1}^\infty i m_i\right).
%\left\{\prod_{i=n+1}^\infty \mbox{Poisson}\left(0; \frac{\gamma_0\int_0^\infty s^{i} e^{-s} \rho(dr)}{i!}\right)\right\}.
\notag
\end{align}
Therefore, we can generate each $m_i$ independently from a Poisson distribution. The stick-breaking construction to generate $\mathcal{M}$ directly follows  the relationships between the Poisson, multinomial, and binomial distributions. 
\end{proof}

%\vspace{-4mm}\section{Proof for Corollary \ref{thm:predict}}

\begin{proof}[Proof for Corollary \ref{thm:predict}]
This follows directly  from Bayes' rule, since $p(z_i\,|\,\zv^{-i},n,\gamma_0,\rho) = \frac{p(z_i,\zv^{-i},n\,|\,\gamma_0,\rho)}{p(\zv^{-i},n\,|\,\gamma_0,\rho)}$, where 
% if we express the ECPF as
%\beq
%f(z_i,\zv^{-i},m\,|\,\gamma_0,\rho)= \frac{\gamma_0^{l^{-i}}}{m!} e^{\gamma_0\int_{0}^\infty(e^{-r}-1)\rho(dr)} 
%\Big(\gamma_0\int_0^\infty re^{-r}\rho(dr)\Big)^{\delta(z_i=l^{-i}+1)} \prod_{k=1}^{l^{-i}} \int_0^\infty r^{n_k^{-i}+\delta(z_i=k)} e^{-r} \rho(dr).\notag
%\eeq
$$p(z_i,\zv^{-i},n\,|\,\gamma_0,\rho)=\,\,\,\,\,\,\,\,\,\,\,\,\,\,\,\,\,\,\,\,\,\,\,\,\,\,\,\hfill$$%\sum_{z_i}f(z_i,\zv^{-i},m\,|\,\gamma_0,\rho) = 
$$n^{-1\,} p(\zv^{-i},n-1\,|\,\gamma_0,\rho)\,\left[\gamma_0\int_0^\infty re^{-r}\rho(dr)\,{\bf 1}(z_i=l^{-i}+1) \,+ \,\sum_{k=1}^{l^{-i}} \frac{\int_0^\infty r^{n_k^{-i}+1} e^{-r} \rho(dr)}{\int_0^\infty r^{n_k^{-i}} e^{-r} \rho(dr)} {\bf 1}(z_i=k)  \right].\notag$$
Marginalizing out the $z_i$ from $p(z_i,\zv^{-i},n\,|\,\gamma_0,\rho)$ we have
\beqs
&p(\zv^{-i},n\,|\,\gamma_0,\rho)=%\sum_{z_i=1}^{l^{-i}+1}f(z_i,\zv^{-i},m\,|\,\gamma_0,\rho)= %\sum_{z_i}f(z_i,\zv^{-i},m\,|\,\gamma_0,\rho) = 
n^{-1}\,p(\zv^{-i},n-1\,|\,\gamma_0,\rho)\left[{\gamma_0\int_0^\infty re^{-r}\rho(dr)+ \sum_{k=1}^{l^{-i}} \frac{\int_0^\infty r^{n_k^{-i}+1} e^{-r} \rho(dr)}{\int_0^\infty r^{n_k^{-i}} e^{-r} \rho(dr)}  }\right].\notag
\eeqs
\end{proof}

\vspace{-4mm}\section{Derivations for the generalized negative binomial process}

Marginalizing out $\lambda$ from  %a generalized gamma-Poisson mixture distribution
$n| \lambda\sim\mbox{Poisson}(\lambda)$ with $\lambda\sim{{}}\mbox{g}\Gamma\mbox{P}[\gamma_0,a,p/(1-p)]$, leads to a generalized NB distribution; $n\sim\mbox{gNB}(\gamma_0,a,p)$, with shape parameter $\gamma_0$, discount parameter $a<1$, and probability parameter $p$. The probability generating function (PGF) is given by
$$ \E[t^n]   = \E[\E[t^n\,|\,\lambda]] 
 = \exp\left\{-\frac{\gamma_0[(1-pt)^a-(1-p)^a)]}{ap^a}\right\}, %=e^{\gamma_0\frac{(1-p)^a}{ap^a}} \sum_{k=0}^\infty \frac{1}{k!} {\left(\frac{-\gamma_0}{ap^a}\right)^k} \sum_{j=0}^\infty \binom{ak}{j}(-pz)^j.
$$
the mean value is $\gamma_0\big[p/(1-p)\big]^{1-a}$ and the variance is $\gamma_0\big[p/(1-p)\big]^{1-a}(1-ap)/(1-p)$. The PGF  was originally presented in \citet{willmot1988remark} and \citet{gerber1992generalized}. With the PGF written  as 
$$\begin{array}{ll}
\E(t^n)  & =\exp\left\{\gamma_0\frac{(1-p)^a}{ap^a}\right\}\sum_{k=0}^\infty \frac{1}{k!} {\left(\frac{-\gamma_0(1-pt)^a}{ap^a}\right)^k}  \\ \\
& =\exp\left\{\gamma_0\frac{(1-p)^a}{ap^a}\right\} \sum_{k=0}^\infty \frac{1}{k!} {\left(\frac{-\gamma_0}{ap^a}\right)^k} \sum_{j=0}^\infty \binom{ak}{j}(-pt)^j,\end{array}$$
we can derive the PMF as 
\beqs\label{eq:f_M}
p_N(n\,|\,\gamma_0,a,p)
 = %e^{{\gamma}\frac{(1-p)^a}{ap^a}}(-p)^m \sum_{k=0}^\infty \frac{1}{k!}{\left(\frac{-\gamma}{ap^a}\right)^k}  \binom{ak}{m} =
\frac{p^n}{n!}e^{{\gamma_0}\frac{(1-p)^a}{ap^a}} \sum_{k=0}^\infty \frac{1}{k!}{\left(-\frac{\gamma_0}{ap^a}\right)^k} \frac{\Gamma(n-ak)}{\Gamma(-ak)}, ~n\in\{0,1,\ldots\}.
%\\
%&= \frac{\Gamma(m-a+\gamma_0p^{-a})}{m!\Gamma(1-a+\gamma_0p^{-a})}\gamma_0e^{-\gamma_0\frac{(1-(1-p)^a)}{ap^{a}}}p^{m-a}
%&=\exp\left({\gamma}\frac{(1-p)^a}{ap^a}\right)(-p)^m  \sum_{k=0}^\infty \frac{\left(\frac{-\gamma}{ap^a}\right)^k}{k!}  (ak)(ak-1)\cdots(ak-m+1)
\eeqs %\end{align}
We can also generate %$n$ from a generalized NB distribution;
$n\sim{{}}\mbox{gNB}(\gamma_0,a,p)$ %which can be generated 
from a compound Poisson distribution, as
$
n=\sum_{k=1}^l n_k$, with the  $(n_k)$ independent from $\mbox{TNB}(a,p)$, and $l\sim\mbox{Poisson}\big(\frac{\gamma_0(1-(1-p)^a)}{ap^a}\big),
$
where $\mbox{TNB}(a,p)$ denotes a truncated NB distribution, with PGF $\E[t^{u}] =  %\sum_{n=1}^\infty  \frac{\Gamma(n-a)}{n!\Gamma(-a)}\frac{p^n(1-p)^{-a}}{1-(1-p)^{-a}} z^n =
\frac{1-(1-pt)^a}{1-(1-p)^a}$ and
PMF
\beqs\label{eq:TNB}
p_U(u|a,p)= \frac{\Gamma(u-a)}{u!\Gamma(-a)}\frac{p^u(1-p)^{-a}}{1-(1-p)^{-a}},~u\in\{1,2,\ldots\}.
\eeqs
Note that as $a\rightarrow 0$, $u\sim\mbox{TNB}(a,p)$ becomes a logarithmic distribution \citep{LogPoisNB} with PMF $p_U(u|p)=\frac{-1}{\ln(1-p)}\frac{p^u}{u}$ %\citep{johnson2005univariate} 
and $n\sim\mbox{gNB}(\gamma_0,a,p)$ becomes a NB distribution; $n\sim\mbox{NB}(\gamma_0,p)$. The truncated NB distribution with $0<a<1$ is the extended NB distribution introduced in \citet{engen1974species}. 

Here we provide a useful identity which will be used later in this section.
Denote by $\sum_{*}$ as the summation over all sets of positive integers $(n_1,\ldots,n_l)$ with ${\sum_{k=1}^l n_k = n}$.  We call $n\sim\mbox{SumTNB}(l,a,p)$ as a sum-truncated NB distributed random variable that can be generated via $n=\sum_{k=1}^l n_k, ~n_k\sim\mbox{TNB}(a,p)$. Using both (\ref{eq:TNB}) %and PGF of the truncated NB distribution 
and $$
\left[\frac{1-(1-pt)^a}{1-(1-p)^a}\right]^{l}= \frac{ \sum_{k=0}^l \binom{l}{k} (-1)^k \sum_{j=0}^\infty\binom{ak}{j} (-pt)^j }{  [1-(1-p)^a]^l},
$$  we may express the PMF of the sum-truncated NB distribution 
as
$$p_N(n|l,a,p) = 
\sum_{*} \prod_{k=1}^l {\frac{\Gamma(n_k-a)}{n_k!\Gamma(-a)} \frac{p^{n_k}(1-p)^{-a}}{1-(1-p)^{-a}}}%\notag\\
=\frac{p^n}{ [1-(1-p)^a]^l} {\sum_{k=0}^l (-1)^k \binom{l}{k}  \frac{\Gamma(n-ak)}{n!\Gamma(-ak)}  },%= \frac{p^{m}}{ [ \frac{1-(1-p)^{a}}{a}]^{l}} \sum_{\sum_{k=1}^l n_k=m} \prod_{k=1}^l \frac{\Gamma(n_k-a)}{n_k!\Gamma(1-a)}
$$ 
leading to the identity %shown  in (\ref{eq:gStirling}).
\begin{align}\label{eq:gStirling}
S_a(n,l) = \frac{n!}{l!}\sum_{*} \prod_{k=1}^l \frac{\Gamma(n_k-a)}{n_k!\Gamma(1-a)}= \frac{1}{l!a^{l}}\sum_{k=0}^l (-1)^k \binom{l}{k}  \frac{\Gamma(n-ak)}{\Gamma(-ak)},
\end{align}
%whose right-hand side is easier to calculate. 
where $S_a(n,l)$ can be recursively calculated via $S_a(n,1)={\Gamma(n-a)}/{\Gamma(1-a)}$, $S_a(n,n)=1$ and $S_a(n+1,l) = (n-al)S_a(n,l)+S_a(n,l-1)$. Multiplying  $S_a(n,l)$ by $a^{-l}$ leads to %Toscano's formula or 
generalized Stirling numbers \citep{charalambides2005combinatorial,csp}.
 Note that when $-ak$ is a nonnegative integer,  $\Gamma(-ak)$ is not well defined but $\Gamma(n-ak)/\Gamma(-ak)=\prod_{i=0}^{n-1}(i-ak)$ is still well defined.  We notice that the generalized NB distribution could  be matched to the the power variance mixture distribution derived in \citet{hougaard1997analysis}, who attributed the key difficulty in applying this distribution to the complicated PMF. 
 %As a byproduct of our analysis, we will show that one may introduce auxiliary variables to make the PMF in (\ref{eq:f_M0}) become fully factorized and hence amenable to posterior simulation. 

The EPPF %of the generalized NB process 
is the ECPF in (\ref{eq:f_Z_M}) divided by the marginal distribution of $n$ in (\ref{eq:f_M}), given by
\begin{align} \label{eq:EPPF1}
 p(\zv\,|\,n,\gamma_0,a,p) &= %\frac{ f(\zv,m\,|\,\gamma_0,a,p)}{\mbox{gNB}(m;\gamma_0,a,p)} = 
 p_n(z_1,\ldots,z_n\,|\,n) % \notag\\ %\frac{f(\zv,m\,|\,\gamma_0,a,p) }{f(m\,|\,G_0,a,p) } =
=\frac{e^{-\frac{\gamma_0}{ap^{a}}} }{ \sum_{k=0}^\infty \frac{1}{k!}{\left(-\frac{\gamma_0}{ap^a}\right)^k}\frac{\Gamma(n-ak)}{\Gamma(-ak)} }  \gamma_0^{l{}}  p^{-al{}} %\prod_{k:\omega_k\in\mathcal{D}}
\prod_{k=1}^{l{}}\frac{\Gamma(n_k-a)}{\Gamma(1-a)}.
\end{align}

Using %Corollary \ref{cor:table}, 
the EPPF  in (\ref{eq:EPPF}) and the identity in (\ref{eq:gStirling}), the conditional distribution of the number of clusters $l$ in a sample of size $n$ %conditioning on the sample size $m$ 
can be expressed as
\begin{align}\label{eq:f_L2}
p_L(l\,|\,n,\gamma_0,a,p)& = \frac{1}{l!}\sum_{*}\frac{n!}{\prod_{k=1}^l n_k!}p(\zv\,|\,n,\gamma_0,a,p)%\notag\\
 =\frac{ \gamma_0^l p^{-al} S_a(n,l)}{e^{\frac{\gamma_0}{ap^{a}}}   \sum_{k=0}^\infty \frac{1}{k!} {\left(\frac{-\gamma_0}{ap^a}\right)^k}  \frac{\Gamma(n-ak)}{\Gamma(-ak)}} ,
\end{align}
which, since $\sum_{l=0}^n p_L(l\,|\,n,\gamma_0,a,p)=1$, further leads to identity 
\beq
e^{\frac{\gamma_0}{ap^{a}}}   \sum_{k=0}^\infty \frac{1}{k!} {\left(\frac{-\gamma_0}{ap^a}\right)^k}  \frac{\Gamma(n-ak)}{\Gamma(-ak)} = \sum_{l=0}^n \gamma_0^l p^{-al} S_a(n,l). \notag
\eeq
Applying this identity on (\ref{eq:f_M}), (\ref{eq:EPPF1}) and (\ref{eq:f_L2}) lead to (\ref{eq:f_M0}),
(\ref{eq:EPPF})  and (\ref{eq:f_L2_0}).

\begin{cor} 
The distribution of the number of clusters in $z_{1:i}$ in a population of size $n$ can be expressed as
\begin{align}\label{eq:l_i}
{p(l_{(i)}\,|\,n,\gamma_0,a,p)}
&=p(l_{(i)}\,|\,i,\gamma_0,a,p)\frac{\sum_{\ell=0}^{i} \gamma_0^\ell p^{-a\ell} S_a(i,\ell)}{\sum_{\ell=0}^n \gamma_0^\ell p^{-a\ell}S_a(n,\ell)}   R_{n,\gamma_0,a,p}(i,l_{(i)}),\notag\\
&=\frac{ \gamma_0^{l_{(i)}}p^{-al_{(i)}}S_a(i,l_{(i)})R_{n,\gamma_0,a,p}(i,l_{(i)})}{\sum_{\ell=0}^n \gamma_0^\ell p^{-a\ell}S_a(n,\ell)}.
\end{align}
\end{cor}
 This can be directly derived using  (\ref{eq:SizeEPPF}) and the relationship between the EPPF and the distribution of the number of clusters. From this PMF, we obtain a useful identity
\beq\notag
 {\sum_{\ell=0}^n \gamma_0^\ell p^{-a\ell}S_a(n,\ell)} = \gamma_0p^{-a}R_{n,\gamma_0,a,p}(1,1),
\eeq
which could be used to calculate the PMF of the generalized NB distribution in (\ref{eq:f_M0}) and the EPPF in (\ref{eq:EPPF}) without the need to compute the generalized Stirling numbers $a^{-l}S_a(n,l)$.

\begin{cor}[Sequential Construction]
Since $p(z_{i+1} \,|\,z_{1:i},n,\gamma_0,a,p)  = \frac{p(z_{1:i+1} \,|\,n,\gamma_0,a,p)} {p(z_{1:i} \,|\,n,\gamma_0,a,p)}$, 
conditioning on the population  size $n$, the sequential prediction rule of the generalized Chinese restaurant sampling formula $\zv\,|\,n\sim\emph{\mbox{gCRSF}}(n,\gamma_0,a,p) $ can be expressed as
 \beq\label{eq:PredictRulej}
P(z_{i+1} = k\,|\,z_{1:i},n,\gamma_0,a,p) =
\begin{cases}\vspace{3mm}
(n_{k,(i)} -a) \frac{R_{n,\gamma_0,a,p}(i+1, ~l_{(i)})}{R_{n,\gamma_0,a,p}(i,~ l_{(i)})}, &  {\mbox{for }} k=1,\ldots,l_{(i)};\\  \gamma_0 p^{-a}\frac{R_{n,\gamma_0,a,p}(i+1, ~l_{(i)}+1)}{R_{n,\gamma_0,a,p}(i, ~l_{(i)})}, & {\mbox{if } }k=l_{(i)}+1; \end{cases}
\eeq
where $i=1,\ldots,n-1$.\end{cor}

With this sequential prediction rule, %similar to a size independent  EPPF, % of $\Pi$, 
we can construct $\Pi_{i+1}$ from $\Pi_i$ in a population of size $n$ by assigning element $(i+1)$ to $A_{z_{i+1}}$.
 When $a=0$, %as expected, we have $$\frac{R_{n,\gamma_0,a,p}(i+1, ~l_{(i)})}{R_{n,\gamma_0,a,p}(i,~ l_{(i)})} = \frac{R_{n,\gamma_0,a,p}(i+1, ~l_{(i)}+1)}{R_{n,\gamma_0,a,p}(i,~ l_{(i)})} = \frac{\Gamma(i+\gamma_0)}{\Gamma(i+1+\gamma_0)} =  \frac{1}{i+\gamma_0},$$  and 
 this sequential prediction rule 
becomes the same as that of a Chinese restaurant process with concentration parameter $\gamma_0$.

\begin{cor} %[Sequence Completion] %\vspace{-2mm}\subsection{Disclosure risk assessment}
The distribution of $z_{i+1\,:\,n}$, given $z_{1:i}$, the population  size $n$, and the model parameters  $\gamma_0$, $a$ and $p$, can be expressed as
\beq \label{eq:extrapolate}
 p(z_{i+1\,:\,n}\,|\,z_{1:i},n,\gamma_0,a,p) %= \frac{f(z_{1:m}|m,\gamma_0,a,p)}{ f(z_{1:i}|m,\gamma_0,a,p)} 
 = \frac{ \gamma_0^{l_{(n)}-l_{(i)}}  p^{-a(l_{(n)}-l_{(i)})}}{ R_{n,\gamma_0,a,p}(i,l_{(i)}) }  \prod_{k=1}^{l_{(i)}} \frac{\Gamma(n_{k,(n)}-a)}{\Gamma(n_{k,(i)}-a)}   \prod_{k=l_{(i+1)}}^{l_{(n)}} \frac{\Gamma(n_{k,(n)}-a)}{\Gamma(1-a)} .
\eeq
%Therefore, given $z_{1:i}$, we can simulate $z_{i+1:m}$, with which we can have a point estimate of the number of sample uniques that are also population uniques as
%$$\sum_{k=1}^{l_{(i)}} \delta(n_{k,(m)}=1, n_{k,(i)}=1).$$ We can collect multiple samples of model parameters $\thetav$, for each of which we collect one or multiple MCMC samples  to estimate 
%$P(PU\,|\, SU,\thetav)$. The final estimation could be calculated with $J$ MCMC samples as 
%
%$$P(PU\,|\, SU)=\sum_{j=1}^J\frac{P(PU\,|\, SU,\thetav^{(j)})}{J}$$
%%
%% and then collect multiple samples 
%%calculate the proportion of sample uniques in $z_{1:i}$ that are also population uniques in  $z_{1:m}$ as
%%\beq
%%P(PU\,|\, SU,\gamma_0,a,p)=\frac{1}{N}\sum_{j=1}^N \frac{\sum_{k=1}^{l_{(i)}} \delta(n^{(j)}_{k,(m)}=1, n_{k,(i)}=1)}{\sum_{k=1}^{l_{(i)}} \delta(n_{k,(i)}=1)}
%%\eeq
%%where $N$ is the total number of collected MCMC samples and $n^{(j)}_{k,(m)}$ is the size of cluster $k$ in the $j$th collected MCMC sample. 
%%
%%\beq
%% f(z_{i+1:m}|z_{1:i},m,\gamma_0,a=0,p) %= \frac{f(z_{1:m}|m,\gamma_0,a,p)}{ f(z_{1:i}|m,\gamma_0,a,p)} 
%% = \frac{ \gamma_0^{l_{(m)}-l_{(i)}} \Gamma(m+\gamma_0) }{ \Gamma(i+\gamma_0) }  \prod_{k=1}^{l_{(i)}} \frac{\Gamma(n_{k,(m)})}{\Gamma(n_{k,(i)})}   \prod_{k=l_{(i+1)}}^{l_{(m)}} {\Gamma(n_{k,(m)})} 
%%\eeq

\end{cor}

\vspace{-4mm}\section{Large $n$ asymptotics for $l_{(n)}$}\label{sec_asym_1}

For $a=0$ it is known from \citet{hollander73} that, as $n\rightarrow+\infty$, $l_{(n)}/\log n$ converges weakly to $\gamma_{0}$. Let us consider the case $a\in(0,1)$. We start by recalling a representation for $\sum_{1\leq l\leq n}(xa)^{l}S_{a}(n,l)$, for any positive $x$. Specifically, let $f_{a}$ denote the density function of a positive stable random variable $X$ with index $a\in(0,1)$, that is $\E[\exp\{-\lambda X\}]=\exp\{-\lambda^{a}\}$. Then, along lines similar to the proof of Proposition 1 in \citet{favaro15}, one may show that
\begin{equation}\label{rap_gen}
\sum_{l=1}^{n}(ax)^{l}S_{a}(n,l)=\exp\{xa\}(xa)^{n/a}\int_{0}^{+\infty}y^{n}\exp\{-(xa)^{1/a}y\}f_{a}(y)dy.
\end{equation}
In order to study the large $n$ asymptotic behavior of $l_{(n)}$, we consider its moment generating function, and we use the representation \eqref{rap_gen}. Specifically, we can write
\begin{align*}
\E[\text{e}^{\lambda l_{(n)}}]&=\sum_{l=0}^{n}\frac{\left(\frac{\text{e}^{\lambda}\gamma_{0}}{p^{a}}\right)^{l}S_{a}(n,l)}{\sum_{l=0}^{n}\left(\frac{\gamma_{0}}{p^{a}}\right)^{l}S_{a}(n,l)}\\
%&=\sum_{l=0}^{n}\frac{\left(\frac{\text{e}^{\lambda}\gamma_{0}}{ap^{a}}\right)^{l}S^{\ast}_{a}(n,l)}{\sum_{l=0}^{n}\left(\frac{\gamma_{0}}{ap^{a}}\right)^{l}S^{\ast}_{a}(n,l)}\\
%&=\frac{\exp\left\{\frac{\text{e}^{\lambda}\gamma_{0}}{ap^{a}}\right\}\left(\frac{\text{e}^{\lambda}\gamma_{0}}{ap^{a}}\right)^{n/a}\int_{0}^{+\infty}y^{n}\exp\left\{-\left(\frac{\text{e}^{\lambda}\gamma_{0}}{ap^{a}}\right)^{1/a}y\right\}f_{a}(y)dy}{\exp\left\{\frac{\gamma_{0}}{ap^{a}}\right\}\left(\frac{\gamma_{0}}{ap^{a}}\right)^{n/a}\int_{0}^{+\infty}y^{n}\exp\left\{-\left(\frac{\gamma_{0}}{ap^{a}}\right)^{1/a}y\right\}f_{a}(y)dy}\\
&=\frac{\exp\left\{\frac{\text{e}^{\lambda}\gamma_{0}}{ap^{a}}\right\}\left(\text{e}^{\lambda}\right)^{n/a}}{\exp\left\{\frac{\gamma_{0}}{ap^{a}}\right\}}\frac{\int_{0}^{+\infty}y^{n}\exp\left\{-\left(\frac{\text{e}^{\lambda}\gamma_{0}}{ap^{a}}\right)^{1/a}y\right\}f_{a}(y)dy}{\int_{0}^{+\infty}y^{n}\exp\left\{-\left(\frac{\gamma_{0}}{ap^{a}}\right)^{1/a}y\right\}f_{a}(y)dy}.
\end{align*}
For large $n$, the ratio of integrals behaves like $\exp\{-n\lambda/a+\lambda\}$. This can be easily verified by using the expression for $f_{a}$, and then solving the integrals. Therefore one obtains $\E[\exp\{{\lambda l_{(n)}}\}]\rightarrow\exp\{\lambda\}\exp\{\gamma_{0}(\exp\{\lambda\}-1)/ap^{a}\}$, as $n\rightarrow+\infty$. This implies that for any $a\in(0,1)$, as $n\rightarrow+\infty$, $l_{(n)}$ converges weakly to $1+X$ where $X$ is a Poisson random variable with parameter $\gamma_{0}/ap^{a}$. Now we consider the case $a=-t$, for $t=1,2,\ldots$ We still use the moment generating function of $l_{(n)}$. Let us define $c_{n}(a)=n^{-a/(1-a)}$, for $a=-t$ with $t=1,2,\ldots$. We can write the moment generating function of $l_{(n)}/c_{n}(-t)$ as
\begin{align*}
\E[\text{e}^{\lambda\frac{l_{(n)}}{c_{n}(-t)}}]&=\sum_{l=0}^{n}\frac{\left(\frac{\text{e}^{\frac{\lambda}{c_{n}}}\gamma_{0}}{(-t)p^{-t}}\right)^{l}S^{\ast}_{-t}(n,l)}{\sum_{l=0}^{n}\left(\frac{\gamma_{0}}{p^{-t}}\right)^{l}S^{\ast}_{-t}(n,l)}\\
&=\sum_{l=0}^{n}\frac{\left(\frac{\text{e}^{\frac{\lambda}{c_{n}(-t)}}\gamma_{0}}{(-t)p^{-t}}\right)^{l}\frac{1}{l!}\sum_{i=0}^{l}(-1)^{i}{l\choose i}\frac{\Gamma(ti+n)}{\Gamma(ti)}}{\sum_{l=0}^{n}\left(\frac{\gamma_{0}}{(-t)p^{-a}}\right)^{l}\frac{1}{l!}\sum_{i=0}^{l}(-1)^{i}{l\choose i}\frac{\Gamma(ti+n)}{\Gamma(ti)}}\\
%&=\sum_{i=0}^{n}\frac{(-1)^{i}(ti)_{n\uparrow1}\sum_{l=i}^{n}\left(\frac{\text{e}^{\frac{\lambda}{c_{n}(-t)}}\gamma_{0}}{(-t)p^{-t}}\right)^{l}\frac{1}{l!}{l\choose i}}{\sum_{i=0}^{n}(-1)^{i}(ti)_{n\uparrow1}\sum_{l=i}^{n}\left(\frac{\gamma_{0}}{(-t)p^{-a}}\right)^{l}\frac{1}{l!}{l\choose i}}\\
&=\sum_{i=0}^{n}\frac{(-1)^{i}\frac{\Gamma(ti+n)}{\Gamma(ti)}\frac{1}{i!}\left(\frac{\text{e}^{\frac{\lambda}{c_{n}(-t)}}\gamma_{0}}{(-t)p^{-t}}\right)^{i}\sum_{l=i}^{n}\left(\frac{\text{e}^{\frac{\lambda}{c_{n}(-t)}}\gamma_{0}}{(-t)p^{-t}}\right)^{l-i}\frac{1}{(l-i)!}}{\sum_{i=0}^{n}(-1)^{i}\frac{\Gamma(ti+n)}{\Gamma(ti)}\frac{1}{i!}\left(\frac{\gamma_{0}}{(-t)p^{-t}}\right)^{i}\sum_{l=i}^{n}\left(\frac{\gamma_{0}}{(-t)p^{-t}}\right)^{l-i}\frac{1}{(l-i)!}}.
\end{align*}
Accordingly, for large $n$ we obtain the following approximated moment generating function
\begin{align*}
\E[\text{e}^{\lambda\frac{L}{c_{n}(-t)}}]&\sim\sum_{i=1}^{n}\frac{\frac{n^{ti}}{i!\Gamma(ti)}\left(\frac{\text{e}^{\frac{\lambda}{n^{t/(t+1)}}}\gamma_{0}}{tp^{-t}}\right)^{i}}{\sum_{i=1}^{n}\frac{n^{ti}}{i!\Gamma(ti)}\left(\frac{\gamma_{0}}{tp^{-t}}\right)^{i}}\\
&\sim\frac{\text{e}^{\frac{\lambda}{n^{t/(t+1)}}}F(-;\frac{t+1}{t},\frac{t+2}{t},\ldots,\frac{t+t-1}{t},2;\frac{\text{e}^{\frac{\lambda}{n^{t/(t+1)}}}\gamma_{0}n^{t}}{t^{t+1}p^{-t}})}{F(-;\frac{t+1}{t},\frac{t+2}{t},\ldots,\frac{t+t-1}{r},2;\frac{\gamma_{0}n^{t}}{t^{t+1}p^{-t}})}
\end{align*}
where $F$ denotes the generalized hypergeometric function. We can make use of asymptotic results for $F$ in Section 5.7 and 5.10 of \citet{luke69} and Section 5.9 of \citet{luke75}. In particular, $\E[\text{e}^{\lambda l_{(n)}/c_{n}(-t)}]\rightarrow\exp\{\lambda(t^{-1}\gamma_{0}p^{t})^{1/(t+1)}\}$. This implies that for any $a=-t$ with $t=1,2,\ldots$, as $n\rightarrow+\infty$, $l_{(n)}/c_{n}(-t)$ converges weakly to %a  degenerate random variable in 
%the constant
$t^{-1}(\gamma_{0}p^{t})^{1/(t+1)}$.

\vspace{-4mm}\section{Large $n$ asymptotics for $M_{i,n}$}\label{sec_asym_2}

For $a=0$ it is known from % Ewens (1972)
\citet{ewens1972sampling}
 that, as $n\rightarrow+\infty$, $M_{i,n}$ converges weakly to a Poisson random variable with parameter $\gamma_{0}/i$. In order to prove the limiting behavior of $M_{i,n}$, for any $a<1$, we make use of the descending factorial moment of order $r$ of $M_{i,n}$. This moment can be easily computed, and it corresponds to 
\begin{align}\label{eq_freq}
&\E\left[\prod_{k=0}^{r-1}(M_{i,n}-k)\right]\\
&\notag\quad=\prod_{k=0}^{ir-1}(n-k)\left[{a\choose i}\right]^{r}\left(-\frac{\gamma_{0}}{p^{a}a}\right)^{r}(-1)^{ir}\frac{\sum_{j=0}^{n-ir}\left(\frac{\gamma_{0}}{p^{a}}\right)^{j}S_{a}(n-ir,j)}{\sum_{j=0}^{n}\left(\frac{\gamma_{0}}{p^{a}}\right)^{j}S_{a}(n,j)}.
\end{align}
Let us consider the case $a\in(0,1)$. As for the case of $l_{(n)}$, we use the representation \eqref{rap_gen}. In particular,
\begin{align*}
&\E\left[\prod_{k=0}^{r-1}(M_{i,n}-k)\right]\\
&\quad=\prod_{k=0}^{ir-1}(n-k)\left[{a\choose i}\right]^{r}\left(-\frac{\gamma_{0}}{p^{a}a}\right)^{r}(-1)^{ir}\frac{\sum_{j=0}^{n-ir}\left(\frac{\gamma_{0}}{p^{a}}\right)^{j}S_{a}(n-ir,j)}{\sum_{j=0}^{n}\left(\frac{\gamma_{0}}{p^{a}}\right)^{j}S_{a}(n,j)}\\
&\quad=\prod_{k=0}^{ir-1}(n-k)\left[{a\choose i}\right]^{r}\left(-\frac{\gamma_{0}}{p^{a}a}\right)^{r}(-1)^{ir}\\
&\quad\quad\times\frac{\left(\frac{\gamma_{0}}{ap^{a}}\right)^{-ir/a}\int_{0}^{+\infty}y^{n-ir}\exp\left\{-\left(\frac{\gamma_{0}}{ap^{a}}\right)^{1/a}y\right\}f_{a}(y)dy}{\int_{0}^{+\infty}y^{n}\exp\left\{-\left(\frac{\gamma_{0}}{ap^{a}}\right)^{1/a}y\right\}f_{a}(y)dy}.\end{align*}
Again, we can use the expression for the $a$-stable density function $f_{a}$ and then solving the integrals in the last expression. In particular, it can be verified the following asymptotics
\begin{displaymath}
\prod_{k=0}^{ir-1}(n-k)\frac{\int_{0}^{+\infty}y^{n-ir}\exp\left\{-\left(\frac{\gamma_{0}}{ap^{a}}\right)^{1/a}y\right\}f_{a}(y)dy}{\int_{0}^{+\infty}y^{n}\exp\left\{-\left(\frac{\gamma_{0}}{ap^{a}}\right)^{1/a}y\right\}f_{a}(y)dy}\rightarrow\left(\frac{\gamma_{0}}{ap^{a}}\right)^{ir/a}
\end{displaymath}
as $n\rightarrow+\infty$. Accordingly, we obtain the following asymptotic descending factorial moments
\begin{align*}
&\E\left[\prod_{k=0}^{r-1}(M_{i,n}-k)\right]\rightarrow\left[{a\choose i}\right]^{r}\left(-\frac{\gamma_{0}}{p^{a}a}\right)^{r}(-1)^{ir}=\left(\frac{a\frac{\Gamma(i-a)}{\Gamma(1-a)}}{i!}\frac{\gamma_{0}}{ap^{a}}\right)^{r}.
\end{align*}
This implies that for any $a\in(0,1)$, as $n\rightarrow+\infty$, $M_{i,n}$ converges weakly to a Poisson random variable with parameter $\Gamma(i-a)\gamma_{0}p^{-a}/i!\Gamma(1-a)$. Now we consider the case $a=-t$, for $t=1,2,\ldots,$. We still use the descending factorial moments. In particular, 
\begin{align*}
&\E\left[\prod_{k=0}^{r-1}(M_{i,n}-k)\right]\\
&\quad=\prod_{k=0}^{ir-1}(n-k)\left[{-t\choose i}\right]^{r}\left(-\frac{\gamma_{0}}{p^{-t}(-t)}\right)^{r}(-1)^{ir}\frac{\sum_{j=0}^{n-ir}\left(\frac{\gamma_{0}}{p^{-t}}\right)^{j}S_{-t}(n-ir,j)}{\sum_{j=0}^{n}\left(\frac{\gamma_{0}}{p^{-t}}\right)^{j}S_{-t}(n,j)}\\
%&\quad=(n)_{ir\downarrow1}\left[{-t\choose i}\right]^{r}\left(-\frac{\gamma_{0}}{p^{-t}(-t)}\right)^{r}(-1)^{ir}\\
%&\quad\quad\times\frac{\sum_{j=0}^{n-ir}\left(\frac{\gamma_{0}}{(-t)p^{-t}}\right)^{j}\frac{1}{j!}\sum_{l=0}^{j}(-1)^{l}{j\choose l}(tl)_{(n-ir)\uparrow1}}{\sum_{j=0}^{n}\left(\frac{\gamma_{0}}{(-t)p^{-t}}\right)^{j}\frac{1}{j!}\sum_{l=0}^{j}(-1)^{l}{j\choose l}(tl)_{n\uparrow1}}\\
&\quad=\prod_{k=0}^{ir-1}(n-k)\left[{-t\choose i}\right]^{r}\left(-\frac{\gamma_{0}}{p^{-t}(-t)}\right)^{r}(-1)^{ir}\\
&\quad\quad\times\frac{\sum_{h=0}^{n-ir}(-1)^{h}\frac{\Gamma(th+n-ir)}{\Gamma(th)}\frac{1}{h!}\left(\frac{\gamma_{0}}{(-t)p^{-t}}\right)^{h}\sum_{j=h}^{n-ir}\left(\frac{\gamma_{0}}{(-t)p^{-t}}\right)^{j-h}\frac{1}{(j-h)!}}{\sum_{h=0}^{n}(-1)^{h}\frac{\Gamma(th+n)}{\Gamma(th)}\frac{1}{h!}\left(\frac{\gamma_{0}}{(-t)p^{-t}}\right)^{h}\sum_{j=h}^{n}\left(\frac{\gamma_{0}}{(-t)p^{-t}}\right)^{j-h}\frac{1}{(j-h)!}}.
\end{align*}
Accordingly, for large $n$ we obtain the following approximated descending factorial moments
\begin{align*}
&\E\left[\prod_{k=0}^{r-1}(M_{i,n}-k)\right]\\
%&\sim(n)_{ir\downarrow1}\left[{-t\choose i}\right]^{r}\left(-\frac{\gamma_{0}}{p^{-t}(-t)}\right)^{r}(-1)^{ir}\\
%&\quad\times n^{-ir}\frac{\sum_{l=0}^{n-ir}\frac{n^{tl}}{l!\Gamma(tl)}\left(\frac{\gamma_{0}}{tp^{-t}}\right)^{l}}{\sum_{l=0}^{n}\frac{n^{tl}}{l!\Gamma(tl)}\left(\frac{\gamma_{0}}{tp^{-t}}\right)^{l}}\\
&\sim\left[{-t\choose i}\right]^{r}\left(-\frac{\gamma_{0}}{p^{-t}(-t)}\right)^{r}(-1)^{ir}\frac{\sum_{h=0}^{n-ir}\frac{n^{th}}{h!\Gamma(th)}\left(\frac{\gamma_{0}}{tp^{-t}}\right)^{h}}{\sum_{h=0}^{n}\frac{n^{th}}{h!\Gamma(th)}\left(\frac{\gamma_{0}}{tp^{-t}}\right)^{i}}\\
&\rightarrow\left[{-t\choose l}\right]^{r}\left(-\frac{\gamma_{0}}{p^{-t}(-t)}\right)^{r}(-1)^{ir}=\left(\frac{\frac{\Gamma(t+i)}{1+t}\gamma_{0}p^{t}}{i!}\right)^{r}.
\end{align*}
This implies that for any $a=-t$ with $t=1,2,\ldots$, as $n\rightarrow+\infty$, $M_{i,n}$ converges weakly to a Poisson random variable with parameter $\Gamma(t+i)\gamma_{0}p^{t}/i!\Gamma(1+t)$.

\vspace{-4mm}\section{MCMC inference}

\subsection{MCMC for the generalized negative binomial process}
For the gNBP, the ECPF in (\ref{eq:f_Z_M}) defines a fully factorized likelihood for $\gamma_0$, $a$ and $p$. We sample $\zv$ using either (\ref{eq:PredictRule}) or (\ref{eq:PredictRulej}).
With a gamma prior $\mbox{Gamma}(e_0,1/f_0)$ placed on $\gamma_0$, we have %the conditional posterior of $\gamma_0$ as
\beqs
(\gamma_0\given -)\sim\mbox{Gamma}\bigg(e_0 + l{},\frac{1}{f_0+ \frac{1-(1-p)^a}{ap^a}}\bigg).
\eeqs
As $a\rightarrow 0$, we have
$
(\gamma_0\given -)\sim\mbox{Gamma}\left(e_0 + l{},\frac{1}{f_0- \ln(1-p)}\right). %\notag
$ This paper sets $e_0=f_0=0.01$.

%\textbf{\emph{Sample $a$}}.
Since $a<1$, we have $\tilde{a}=\frac{1}{1+(1-a)} \in(0,1)$. With a uniform prior placed on $\tilde{a}$ in $(0,1)$ and the likelihood of gNBP in (\ref{eq:f_Z_M}), we use the  griddy-Gibbs sampler \citep{griddygibbs} to
sample $a$ from a discrete distribution
\beq
P(a\given -)\propto e^{-\gamma_0\frac{1-(1-p)^a}{ap^{a}}}
p^{-al{}}
\prod_{k=1}^{l{}} \frac{\Gamma(n_k-a)}{{\Gamma(1-a)} }
%f(\zv,n\,|\,\gamma_0,a,p)
\eeq
over a grid of points $\frac{1}{1+(1-a)}=0.0001,0.0002,\ldots,0.9999$.

%\emph{\textbf{Sample $p$}}. 
We place a uniform prior on $p$ in $(0,1)$.
When $a\rightarrow 0$, the likelihood of the gNBP in (\ref{eq:f_Z_M}) becomes proportional to $p^{m}(1-p)^{\gamma_0}$, thus we have %with a %beta prior $\mbox{Beta}(1,1)$ on $p$, we have 
$
%\lim_{a\rightarrow 0}
(p\given -)\sim\mbox{Beta}(1+n,1+\gamma_0).
$
% the same as that of the NB process count-mixture model \citep{NBP2012}.
When $a\neq 0$,  %with %We %place
%a uniform prior placed on $p$ in $(0,1)$ and the likelihood in (\ref{eq:f_Z_M}),  
we use the  griddy-Gibbs sampler to sample $p$ from a discrete distribution
\beq
P(p\given -)\propto e^{-\gamma_0\frac{1-(1-p)^a}{ap^{a}}}
 p^{n-al{}}
\eeq over %consider
a grid of points $p=0.0001,0.0002,\ldots,0.9999$.

%
%....
%
%
%$$P(unseen speices) = \frac{\gamma_0\frac{1-(1-p)^a}{ap^{a}}- \sum_{i=1}^\infty \delta(m_i>0)  \frac{\Gamma(i-a)\gamma_{0}p^{i-a}}{\Gamma(1-a)i!}}{\gamma_0\frac{1-(1-p)^a}{ap^{a}}} $$
%
%$$\ln(\E[m_i])  = (i-a)\ln(p) + \ln \left(\frac{\Gamma(i-a)\gamma_{0} }{\Gamma(1-a)i!}\right)$$
%
%
\subsection{MCMC for the Pitman-Yor process}

Given the mass parameter $\gamma_0$ and discount parameter $a\in[0,1)$, the EPPF of $(z_1,\ldots,z_i)$ for the Pitman-Yor process  \citep{%perman1992size,pitman1997two,
csp} can be expressed as
\begin{align}
P(z_1,\ldots,z_i\given \gamma_0,a) &= \frac{\Gamma(\gamma_0)}{\Gamma(i+\gamma_0)} \prod_{k=1}^{l_i} \frac{\Gamma(i_k-a)}{\Gamma(1-a)}[\gamma_0+(k-1)a]\notag\\
&= \frac{\Gamma(1+\gamma_0)}{\Gamma(i+\gamma_0)} (1-a)^{l_i}\left[\prod_{k=1}^{l_i} \frac{\Gamma(i_k-a)}{\Gamma(2-a)}\right]\left[ \prod_{k=1}^{l_i-1} (\gamma_0+ka)\right],
\end{align}
where $l_i$ represents the number of clusters in $\{z_{1},\ldots,z_i\}$.
We set in the prior that $\gamma_0\sim\mbox{Gamma}(e_0,1/f_0)$ and $a\sim\mbox{Beta}(1,1)$. 
Following \citet{teh2006bayesian},
with auxiliary variables 
\begin{align}
(p\given i,\gamma_0) &\sim\mbox{Beta}(i-1,\gamma_0+1),\notag\\
 (y_k \given \gamma_0,a)&\sim\mbox{Bernoulli}\left(\frac{\gamma_0}{\gamma_0+ka}\right),~k\in\{1,\ldots,l_i-1\}, \label{eq:PY1}
\end{align}
we sample $\gamma_0$ as
\beq
(\gamma_0\given -) \sim\mbox{Gamma}\left(e_0 + \sum_{k=1}^{l_i-1} y_k,~\frac{1}{f_0- \ln(1-p)}\right),
\eeq
and further with auxiliary variables
\beq
(b_{kj}\given a) \sim\mbox{Bernoulli}\left(\frac{j-1}{j-a}\right),~k\in\{1,\ldots,l_i\},~j\in\{2,\ldots,i_k-1\},
\eeq
we sample $a$ as
\beq
(a\given -)\sim\mbox{Beta}\left(1 +\sum_{k=1}^{l_i-1} (1-y_k), 1+ l + \sum_{k=1}^{l_i} \sum_{j=2}^{i_k-1} (1-b_{kj})\right).
\eeq
We then use the prediction rule of the Pitman-Yor process as 
\beq
P(z_{i+1} = k\given z_1,\ldots,z_i) = \begin{cases} \vspace{3mm}
\displaystyle\frac{i_k-a}{i+\gamma_0} & \text{if } k\in\{1,\ldots,l_i\}, \\ 
\displaystyle\frac{\gamma_0+l_i a}{i+\gamma_0} & \text{if } k=l_i+1.\end{cases}
\label{eq:PYend}
\eeq
to sequentially sample $z_{i+1},\ldots,z_{n}$. Each Gibbs sampling iteration proceeds from 
\eqref{eq:PY1} to \eqref{eq:PYend}.

\end{document}